\newif\ifPreprint \Preprintfalse
\Preprinttrue

\ifPreprint
\documentclass[11pt,a4]{article}
\usepackage[inner=37.8mm,outer=37.8mm,top=25mm,bottom=25mm,includefoot]{geometry}
\renewcommand{\rmdefault}{ptm}
\renewcommand{\sfdefault}{phv}
\usepackage{mathptmx}
\usepackage{authblk}
\usepackage[authoryear]{natbib}
\usepackage{hyperref}
\else % !ifPreprint
\documentclass{jfp}
\fi

\usepackage{robs-thesis-macro-defs}
\usepackage{amsmath}
\usepackage{amssymb}
\usepackage{amsthm}
\usepackage{thmtools}
\usepackage{mathpartir}
\usepackage{multido} % Note that multi-do seems to treat 0 as 1.
\usepackage{xspace} % to fix spacing in macros producing words.
\usepackage{caption}
\usepackage{subcaption}
\usepackage{tikz}
\usepackage{listings}

\input{glyphtounicode}
\graphicspath{{./graphics/}}

\declaretheorem[numberwithin=section]{definition}
\declaretheorem[sibling=definition]{lemma}
\declaretheorem[sibling=definition]{lemmas}
\declaretheorem[sibling=definition]{theorem}
\declaretheorem[sibling=definition]{corollary}
\declaretheorem[sibling=definition]{proposition}
\declaretheorem[numbered=no,name=Reproduced definition]{definition*}
\declaretheorem[numbered=no,name=Reproduced definitions]{definitions*}
\declaretheorem[numberwithin=section,style=definition]{example}

\begin{document}

\newcommand{\ourtitle}{Verified Secure Compilation for Mixed-Sensitivity Concurrent Programs}

\ifPreprint
\title{\vspace{-1em}\ourtitle\thanks{\mycpy}}
% https://tex.stackexchange.com/a/14866
\renewcommand\footnotemark{}
%\journaltitle{Submitted manuscript under review}
% Permitted by publisher for preprints: https://www.cambridge.org/core/services/open-access-policies/open-access-journals/green-open-access-policy-for-journals
% Permitted by copyright owner CSIRO of Robert's contributions for preprints
% Finally, CUP requires CC BY-NC-ND for the accepted manuscript:
% "Accepted Manuscripts can be made accessible under a Creative Commons CC-BY-NC-ND license or equivalent, but not a more permissive license. We do not allow AMs to be made accessible under a CC-BY license, for example."
% (Source: https://www.cambridge.org/core/services/open-access-policies/open-access-journals/green-open-access-policy-for-journals)
\newcommand{\mycpy}{\copyright\ CSIRO (ABN 41 687 119 230), Robert Sison, and Toby Murray 2019-2021; accepted manuscript to appear (excl.~appendices) in CUP \emph{JFP Special Issue on Secure Compilation}; licensed under \href{http://creativecommons.org/licenses/by-nc-nd/4.0/}{CC BY-NC-ND 4.0}}
\else % !ifPreprint
\title{\ourtitle}
% start: from the template
\journaltitle{JFP Special Issue on Secure Compilation}
\cpr{Cambridge University Press}
\doival{10.1017/xxxxx}
\jnlDoiYr{2021}

\lefttitle{\ourtitle}
\righttitle{R. Sison and T. Murray}
\totalpg{\pageref{lastpage01}}
% end: from the template
\fi

\ifPreprint
% https://tex.stackexchange.com/a/370813
\author[$*$ $\dagger$]{Robert Sison}
\author[$*$]{Toby Murray}
\affil[$*$]{\footnotesize School of Computing and Information Systems,
        University of Melbourne, Australia}
\affil[$\dagger$]{\footnotesize CSIRO's Data61 and UNSW Sydney, Australia}
% https://tex.stackexchange.com/a/303689
\affil[ ]{\footnotesize \textit {(e-mail: [firstname].[lastname]@unimelb.edu.au)}}
% https://tex.stackexchange.com/a/2808
\date{\vspace{-6ex}}
\maketitle
% https://tex.stackexchange.com/a/151589
\renewenvironment{abstract}
{\small
 \begin{center}
 \bfseries \abstractname\vspace{-.5em}\vspace{0pt}
 \end{center}
 \relax}

\else % !ifPreprint

\begin{authgrp}
\author{Robert Sison}\textsuperscript{* \textdagger},
\author{Toby Murray}\textsuperscript{*}
\affiliation{\textsuperscript{*}
        School of Computing and Information Systems,\\
        University of Melbourne, Australia\\
        \textsuperscript{\textnormal{\textdagger}}
        CSIRO's Data61 and UNSW Sydney, Australia\\
        (\email{[firstname].[lastname]@unimelb.edu.au})}
\end{authgrp}
\fi

\begin{abstract}
Proving only over source code that programs do not leak sensitive data
leaves a gap between reasoning and reality
that can only be filled by accounting for the behaviour of the compiler.
Furthermore, software does not always have the luxury of limiting itself
to single-threaded computation
with resources statically dedicated to each user
to ensure the confidentiality of their data.
This results in \emph{mixed-sensitivity concurrent programs},
which might reuse memory shared between their threads
to hold data of different sensitivity levels at different times;
for such programs, a compiler must preserve
the \emph{value-dependent} coordination of such \emph{mixed-sensitivity reuse}
despite the impact of \emph{concurrency}.

Here we demonstrate, using Isabelle/HOL,
%that proving preservation of confidentiality
%is feasible for compilers written in functional languages,
that it is feasible to verify that a compiler preserves
\emph{noninterference},
the strictest kind of confidentiality property,
for mixed-sensitivity concurrent programs.
%even when they are to be used to compile programs
%that exploit either of mixed-sensitivity reuse or shared-memory concurrency.
%
First, we present notions of refinement
that preserve a
\emph{concurrent value-dependent} notion of noninterference
that we have designed to support such programs.
As proving noninterference-preserving refinement
can be considerably more complex than the
standard refinements
typically used to verify semantics-preserving compilation,
our notions include a decomposition principle
that separates the semantics-preservation from security-preservation concerns.
Second, we demonstrate that these refinement notions are applicable
to verified secure compilation,
by exercising them on
% robs: Rephrasing suggested by Toby
% an Isabelle function that compiles
a single-pass compiler for
mixed-sensitivity concurrent programs
that synchronise using mutex locks,
from a generic imperative language to
a generic RISC-style assembly language.
Finally, we execute our compiler
on a nontrivial mixed-sensitivity concurrent program
modelling a real-world use case,
thus preserving its source-level noninterference properties
% robs: Reworded 'its' -> 'an' here.
down to an assembly-level model automatically.
All results are formalised and proved
in the Isabelle/HOL interactive proof assistant.

Our work paves the way for more fully featured compilers to
offer verified secure compilation support
to developers of multithreaded software
% robs: Toby suggested rephrasing
%responsible for data of multiple sensitivity levels.
that must handle data of multiple sensitivity levels.
\end{abstract}

\ifPreprint\else % !ifPreprint
\maketitle
\fi

\section{Introduction}
Here we show how to extend secure compilation support to
programs that are designed to %that seek to
%\emph{mixed-sensitivity concurrent programs}, which
%These are programs that feature both (1) concurrency of access to,
%and (2) mixed-sensitivity reuse of memory locations.
address two fundamental problems of scale:
(1) the need to divide work in computer systems that handle information, and
(2) the need to share scarce resources to be able to service every customer
for whom that work is done.
There will always be a program for which that sharing is not abstracted;
that program's responsibility is to implement that sharing in such a way that
it never allows the information of one customer to flow to another.
In this paper, we prove formally %The objective of this paper is to prove
that a compiler does not break that program's responsibility.
%The objective of this paper is to prove that a compiler does not
%break that program's responsibility of preventing
%the information of one customer from flowing to another.

It is well known that program translations of the kind carried out by
compilers can easily break security properties like
confidentiality~\citep{Kaufmann16,Barthe18}.
%Yet source-level reasoning about confidentiality remains
%common~\citep{Murray_SE_18,Mantel_SS_11,Lourenco_Caires_15}.
%
This is especially the case for \emph{mixed-sensitivity concurrent programs}, which feature both:
\begin{itemize}
    \item \emph{Concurrency} of access to memory locations shared between different threads of execution.
    %\item \emph{Shared-memory concurrency},
        A compiler must preserve both %This means a compiler needs both
        (1) the synchronisation that coordinates threads' access to
        shared memory, and
        (2) the absence of any internal timing leaks, to prevent them from
        manifesting as storage leaks~\citep{Volpano98}.

    \item \emph{Mixed-sensitivity reuse} of shared memory
        to hold information of different sensitivity levels at different times.
        %This means a compiler needs
        A compiler must preserve the program functionality
        that coordinates this reuse;
        this implies support for \emph{value-dependent classification} policies,
        which allow the classification of a memory location
        to change dynamically
        depending on values held in other memory locations~\citep{Murray_15}.
        Furthermore, it must do so accounting for the potential impact
        of concurrent access by other threads.
\end{itemize}

Although existing verified compilers
for dialects of mainstream programming languages,
like CompCert \citep{Leroy09} and CakeML \citep{Kumar_MNO_14},
have been proved to preserve program functionality (semantics)
and some
timing-sensitive forms of noninterference~\citep{Barthe20},
% robs: The "prior to ours" doesn't apply
%       if the scope is "for dialects of mainstream PLs".
none %prior to ours presented here
are yet verified to preserve proofs of noninterference
for mixed-sensitivity concurrent programs.
%In doing so our compiler,
Ideally such a compiler,
applied to the threads
of a proved-secure mixed-sensitivity concurrent program,
would yield assembly code that, run concurrently, also composes into
a secure mixed-sensitivity concurrent program.

To this end, here we present notions
of \emph{concurrent value-dependent noninterference}-preserving refinement,
which are compositional
across the threads of mixed-sensitivity concurrent programs.
In these notions,
the usual square-shaped commuting diagram commonly used to depict
(semantics-preserving)
refinement (\autoref{fig:decomp-R})
has been replaced by a \emph{cube} (\autoref{fig:coupling-inv-pres}).
% The additional dimension of this cube
This reflects that it preserves a \emph{2-safety hyperproperty}
\citep{Terauchi05,Clarkson10},
which compares two executions rather than examining a single one.
Our earlier work~\citep{Murray_SPR_16} was the first to make this
observation and to propose a general cube-shaped refinement
property; however other work on verified secure compilation 
targeted towards noninterference preservation
\citep{Barthe18,Barthe20} since made the same observation.
As these cube-shaped properties
are significantly more complicated to prove than standard notions of
semantics-preserving refinement typical in verified
compilation~\citep{Leroy09,Kumar_MNO_14},
we present a principle of
decomposing the cube (\autoref{fig:coupling-inv-pres})
into three separate obligations (\autoref{fig:decomp}):
the first
of these is akin to semantics-preserving refinement, while the rest
prevent the introduction of
any termination- and timing-leaks.
%
%The decomposition principle does not inherently mandate against
%secret-dependent control flow:
A simple comparison of proof effort for a refinement example
%with secret-dependent control flow
(\autoref{fig:refinement-example})
% Superfluous because the figure caption itself cites it thoroughly -robs.
%from %an earlier work
%\citet{Murray_SPR_16}
shows this approach can almost halve its complexity,
and that it is applicable to proofs of
refinement for programs with \emph{secret-dependent control flow}---%
the example pads
an $\ITEg{\var{h}}{\ldots}{\ldots}$ conditional with $\Skip$s, % on a secret $h$,
so as not to introduce a timing leak of $h$.

We then go on to demonstrate that
the decomposition principle we provide
makes our notion of refinement a tractable target
for verified secure compilation.
%
%First, we present a machine-checked formal proof of concurrent value-dependent security preservation, for a proof-of-concept compiler.
%
Our compiler is an executable function in Isabelle/HOL
that translates
mixed-sensitivity concurrent programs
that synchronise using mutex locks,
from a generic imperative While language %(\autoref{sec:source-language})
to a generic RISC-style assembly language. %(\autoref{sec:target-language}).
In particular, it supports the class of programs
that avoid all \emph{implicit flows},
where a secret determines the choice between
two control flow paths with different observable effects,
by disallowing any secret-dependent control flow---%
for example, disallowing $\ITEg{h}{\ldots}{\ldots}$ conditionals
to prevent any timing leaks from $h$.
This is a common approach against implicit flows,
as it avoids any precise source-level reasoning about time.
To preserve confidentiality for programs that take that approach,
we instantiate the decomposition principle so that it enforces
that our compiler does not introduce any new secret-dependent control flow.
%
%In \autoref{chp:wr-compiler},
%We then turn our attention to proof of automatic refinement by our compiler.
%we present our compiler and its formal verification, as an application of the decomposition principle.
Furthermore, as part of satisfying the demands of our refinement notion,
our compiler demonstrates a way of formalising and proving when it is safe for a compiler to perform optimisations in the presence of concurrency.
To ensure that the contents of shared memory locations are preserved under compilation despite potential interference from other threads, our compiler tracks which shared memory locations are free from data races. %\emph{stable} (free from any such interference).
It then makes use of this tracking to avoid redundant loads from ``stable''
(i.e.~race-free) shared variables safely,
that would otherwise be considered unsafe to omit.

Finally, to show that the compiler preserves noninterference
for actual mixed-sensitivity concurrent programs, we execute it
%We then execute our compiler
%This enabled us to execute it
on a real-world use case:
a model of the
software-componentised input-handling regime for the
\emph{Cross Domain Desktop Compositor} \citep{Beaumont_MM_16}, a device
%concurrent program
that enforces information-flow control
over input classified dynamically by a trusted user.
%value-dependently classified input.
We leave treatment on the design and application of per-thread proof techniques
establishing CVDNI for the successive versions of this
model to other works \citep{Murray_SE_18,robs-phd},
and here focus on its CVDNI-preserving compilation---%
expanding on \citet{Sison_Murray_19}, the conference version of this paper.
This yields the first proofs of noninterference %information-flow security
for an assembly-level model
of a nontrivial mixed-sensitivity concurrent program,
demonstrating the power of verified secure compilation
to preserve security properties of compiled code.

The structure of our paper is as follows.
% robs: Moved this to cvdni-notions.
%We begin with an illustrative example of % to give an intuition of
%the kinds of programs for which we are interested in verifying
%secure compilation support (\autoref{sec:example-worker}).
First, we present language-independent notions of noninterference and
its refinement, designed for mixed-sensitivity concurrent programs
(\autoref{sec:cvdni-notions}).
Our attention then turns to preliminaries for our compiler:
the main properties of interest
of the source \WhileLang language it compiles (\autoref{sec:source-language}),
and
of the target \RISCLang language it produces (\autoref{sec:target-language}).
Then, after presenting the details of our compiler and its verification
(\autoref{chp:wr-compiler}), and
the case study to which we apply it (\autoref{chp:cddc}),
we discuss the most closely related work in the area
(\autoref{sec:related-work}), before concluding
%before some concluding remarks
(\autoref{sec:conclusion}).

We expand on
%Our presentation here differs from
the conference version of this paper
\citep{Sison_Murray_19}
as follows:
%in the following ways:
\begin{itemize}
    \item Here we have adapted the noninterference properties
        %we present in \autoref{sec:cvdni-notions}
        %are
        %presented here %We present an adaptation of the security properties
        to support
        %from \citet{Sison_Murray_19}
        %to parameterise over %some
        assumptions %and requirements;
        on initial memory and extra security requirements;
        %and requirements on the security witness;
        %(``$\INITparam$ and $\BISIMREQSparam$'');
        %we need to present these so as
        these will allow us to clarify exactly what
        our compiler is verified to preserve, and for which kinds of programs.

    \item We also in \autoref{sec:cvdni-notions}
        present further preliminaries
        that will allow %for more detailed explanations
        us to explain in greater detail
        the different ways to establish and use proofs
        about a verified compiler
        %in which our compiler can be proved
        to obtain
        %preserves the requirements that make a
        whole-system noninterference
        at the target-language level.
        These include:
        \begin{enumerate}
            \item The side conditions and theorem of
                compositionality for the noninterference properties.
                In \autoref{sec:source-language}, we then
                for the first time
                present the proof,
                whose details were
                elided from \citet{Sison_Murray_19},
                for a noncompositional ``global''
                part of this side condition,
                which is necessary to obtain
                whole-system noninterference from per-thread noninterference
                both at source and target level.

            \item A whole-system refinement theorem, adapted to
                support
                assumptions on initial memory.
                This theorem was alluded to in~\cite{Murray_SPR_16} but,
                until now, has never been formally presented outside of
                the Isabelle/HOL theories.
                %In \autoref{chp:wr-compiler},
                It gives us
                a means to prove preservation
                of whole-system noninterference by the compiler,
                without having
                to re-prove the noncompositional side condition
                at the target-language level.
        \end{enumerate}

\item %Resting on the new details we presented in \autoref{sec:cvdni-notions},
In \autoref{chp:wr-compiler},
we then compare alternative methods of obtaining
whole-system security at RISC level,
%side by side,
that a developer would choose
depending on whether all, or only some
threads are compiled with our compiler.
In contrast, \citet{Sison_Murray_19}
%whose details
stopped after presenting the
application of refinement decomposition principle.

\item In \autoref{chp:cddc},
    the case study to which we apply the compiler is significantly
    expanded, being
%we prove preservation of noninterference for
    a 3-component version of the CDDC input-handling program---%
    closer to a version presented in \citet{Murray_SE_18}---%
    as opposed to the 2-component version
    of \citet{Sison_Murray_19}.
\item Furthermore, we present substantially more details of our case study
    in \autoref{chp:cddc},
    which were mostly elided from \citet{Sison_Murray_19}.
    These include
    formal statements of both the source-level properties preserved
    and the target-level properties obtained,
    alongside explanations of all alternative methods for obtaining the latter
    from the former.

\item Finally, every lemma and theorem we prove
    is presented with a proof sketch or explanation,
    which were largely absent from \citet{Sison_Murray_19}.
\ifPreprint
    We also include here appendices
    with further details of our compiler
    (see Appendices \ref{sec:labels-seq}, \ref{sec:reg-alloc-model},
    and \ref{sec:R_wr-informal}).
\fi % ifPreprint

\end{itemize}
 % from itp19
\label{sec:introduction}

%\section{Concurrent value-dependent noninterference and its preservation}
%\section{Noninterference and its preservation for mixed-sensitivity concurrent programs}
\section{Noninterference and its refinement for mixed-sensitivity concurrent programs}
\label{sec:cvdni-notions}
%\subsection{Concurrent value-dependent noninterference (CVDNI)}

To support mixed-sensitivity concurrent programs,
%composed of worker threads like the example we just presented,
we verify our compiler to preserve
the \emph{concurrent value-dependent noninterference} (CVDNI)
notions
of \citet{Murray_SPR_16}.
% robs: I'm considering reordering the presentation, after I typeset the
%       whole-system refinement theorem.
In this section we present the %language-independent
definitions of CVDNI and its refinement, %preservation,
as we have adapted them from that work's
Isabelle formalisation
\citep{Murray16-Dependent-SIFUM-AFP,Murray16-Refinement-AFP}.
%
% robs: No - the 'extra requirements' part is not in the ITP19 paper proper.
%As in \citet{Sison_Murray_19},
%we present a version of the theory that
In particular, the version of the theory we present here
%is modified %from
%\citep{Murray_SPR_16,Murray16-Dependent-SIFUM-AFP,Murray16-Refinement-AFP}
supports extra customisation of requirements beyond the prior work;
we will need this to parameterise the theory with
initial conditions needed for a compositionality property of our source language
(\autoref{sec:source-language}),
and our compiler's preservation of a ban on secret-dependent control flow
(\autoref{chp:wr-compiler}).
% Revised: Both of these points
% - INIT/EXTRA customisation
% - memory is the same
% were flagged by reviewers as not coming across clearly enough.
Furthermore, it is
simplified to the case where
the shared memory is the same for both the original \emph{abstract}
and the refined \emph{concrete} program---%
refinement adds
no new shared variables.
Later,
we will instantiate this CVDNI theory to have our
compiler's source and target languages
(Sections \ref{sec:source-language}, \ref{sec:target-language})
respectively play the roles of the
abstract and concrete programs' languages in the theory.

We begin by introducing with an illustrative example
the challenges of verifying value-dependent noninterference
in the presence of shared-variable concurrency
(\autoref{sec:example-worker}).
Then we present the per-thread and whole-system
noninterference properties themselves
(\autoref{sec:cvdni-definition}),
followed by the notion of \emph{per-thread refinement}
that preserves the per-thread property
between the two languages
(\autoref{sec:just-cube-reqs}).
As the cube-shapedness of noninterference-preserving refinement diagrams
in general makes them difficult to apply directly to compiler verification,
we present a decomposition principle
(\autoref{sec:just-decomp-reqs})
that we will use to prove CVDNI-preserving refinement for our compiler.
%and discuss its proof impact for a simple example
% robs: Might revisit the way I frame this, after I typeset it.
We then present requirements and a theorem
for \emph{whole-system refinement}
by which we have that
%we will later obtain that
CVDNI-preserving refinement is compositional across the
threads of the program being compiled,
such that it yields the whole-system property at the target language level
(\autoref{sec:sys-refine}).

%\section{Illustrative example of a mixed-sensitivity concurrent worker thread}
\subsection{Illustrative example of a mixed-sensitivity concurrent program}
\label{sec:example-worker}
% Toby: Heading is better this way.
%       Just say "we'll just show you one part of it".
\begin{figure}
    \begin{subfigure}{0.5\textwidth}
{\footnotesize$
\While{\const{TRUE}}{
    \Seq{\myindent{1}\LockAcq{\var{workspace\_lock}}}{
    \Seq{\Whilei{1}{!\var{suspended}}{
    \Seq{\myindent{2}\LockAcq{\var{source\_lock}}}{
    \Seq{\myindent{2}\Assign{\var{workspace}}{\var{source}}}{
        \myindent{2}\text{/* \ldots\ operations on \var{workspace} \ldots */}\\
    \Seq{\ITEi{2}{\var{domain} = \const{LOW}}{
            \myindent{3}\Assign{\var{low\_sink}}{\var{workspace}}
        }{
            \Seq{\myindent{3}\Assign{\var{high\_sink}}{\var{workspace}}}{
            \myindent{3}\Assign{\var{workspace}}{\const{0}}}
        }}{
        \myindent{2}\LockRel{\var{source\_lock}}}}}
    }}{
    \myindent{1}\Seq{\LockRel{\var{workspace\_lock}}}{
    \myindent{1}\Whileg{\var{suspended}}{\Skip}}}}
}
$}
        \caption{Input processing worker thread program}
        \label{fig:example-worker-code}
    \end{subfigure}
    \begin{subfigure}{0.46\textwidth}
        \centering
        \ifPreprint
        \includegraphics[height=0.1\textheight]{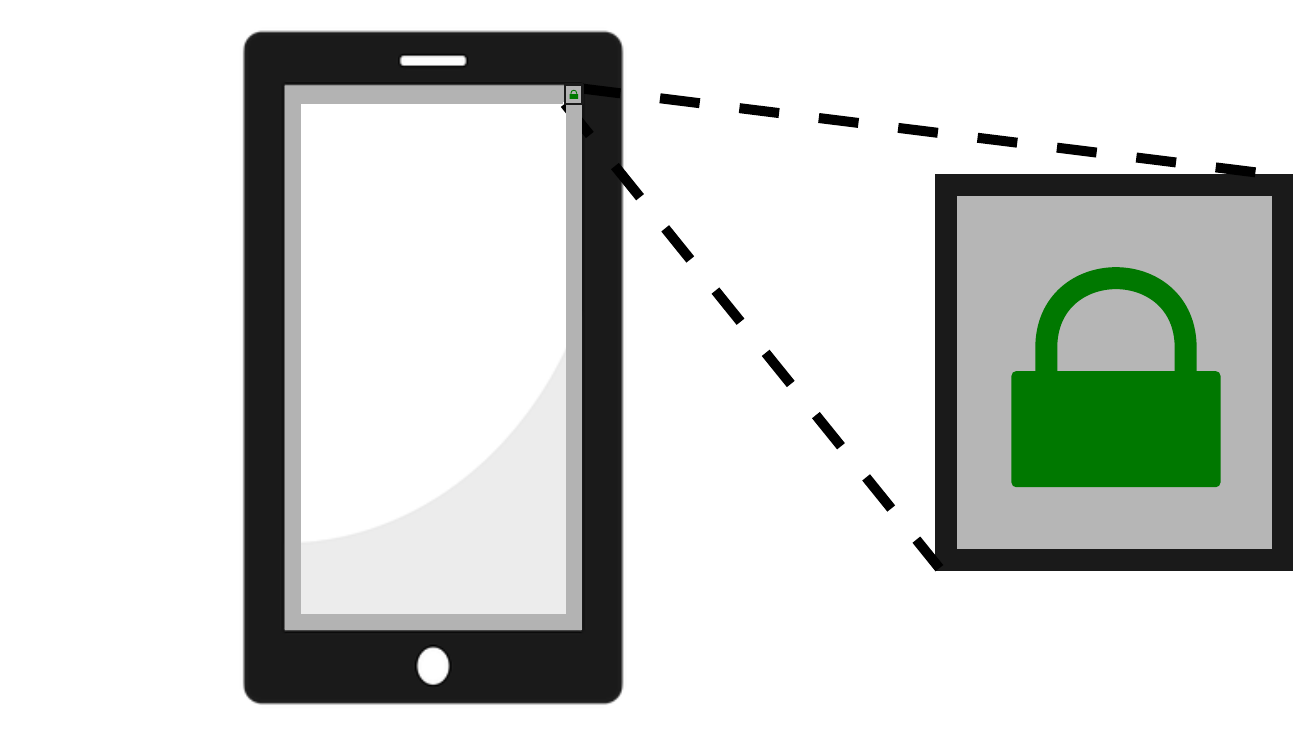}
        \else % !ifPreprint
        \includegraphics[height=0.1\textheight]{smartphone.eps}
        \fi % !ifPreprint
        \caption{The phone providing the $\High$ personality:
                 $\var{domain} \neq \const{LOW}$, and
                 $\var{source}$ is classified $\High$
                 to reflect that the user might type in secrets.}
        \label{fig:example-worker-dma-high}

        \vspace{0.5\baselineskip}

        \ifPreprint
        \includegraphics[height=0.1\textheight]{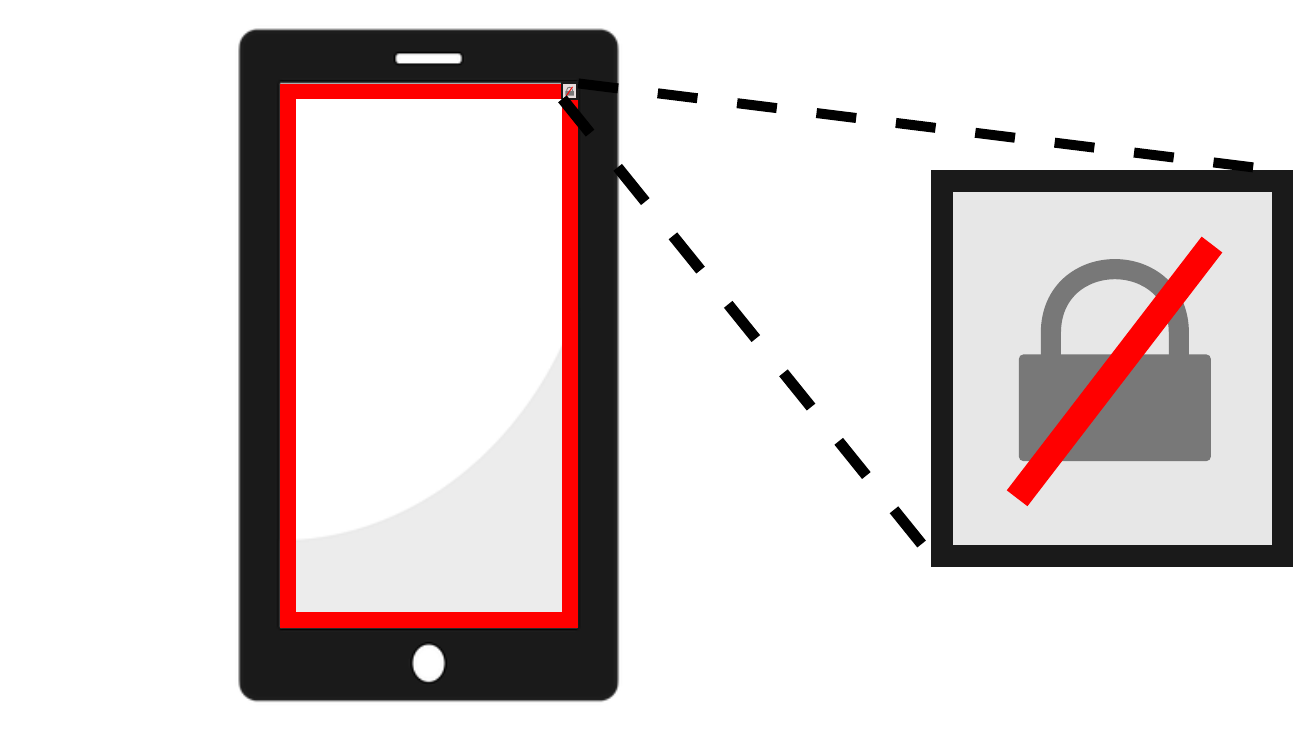}
        \else % !ifPreprint
        \includegraphics[height=0.1\textheight]{smartphone-red.eps}
        \fi % !ifPreprint
        \caption{The phone displaying visual indicators that it is providing the $\Low$ personality:
                 $\var{domain} = \const{LOW}$, and
                 $\var{source}$ is classified $\Low$
                 to reflect that we trust the user not to type in secrets.}
        \label{fig:example-worker-dma-low}
    \end{subfigure}
    \caption{Example: Touchscreen input processing for a dual-personality smartphone. \\ Reproduced from \citet{Sison_Murray_19}.}
    \label{fig:example-worker}
\end{figure}

% specific
Consider the task of verifying a multithreaded system that manages the user interface (UI) for a \emph{dual-personality smartphone}, a phone that provides clearly distinguished user contexts (\emph{personalities}), typically for work versus leisure.
Specifically, our task is to verify that it does not leak \emph{sensitive} information intended only for one of those personalities, which we classify $\High$ (\autoref{fig:example-worker-dma-high}), to locations belonging to the other, which we classify $\Low$ (\autoref{fig:example-worker-dma-low}).

% general
Here and generally throughout this paper, our \emph{attacker model} is
an entity that can read from the system's \emph{untrusted sinks}:
some subset of permanently $\Low$-classified locations not subject to synchronisation.
% specific
In our example, the untrusted sinks may include
WLAN device registers in a hostile environment.

% specific
The smartphone's UI system consists of a number of threads running concurrently with a shared address space;
we aim to verify that, as a whole, this system of threads
satisfies the security requirement.
% general
However, to avoid a state space explosion that is exponential in the number of threads, we must do this \emph{compositionally}:
one thread at a time, then combining the results of these analyses.

% specific
We focus on a particular worker thread (\autoref{fig:example-worker-code}), the one responsible for sending touchscreen input from the \var{source} variable to its intended destination.

The first challenge is that the destination depends on which personality the phone is currently providing, which is indicated by the value of \var{domain}.
This is reflected by the classification of \var{source} being dependent on the value of \var{domain}: \var{source} is classified $\Low$ exactly when $\var{domain} = \const{LOW}$ (where $\const{LOW}$ is a designated constant), and is classified $\High$ otherwise.
Due to this dependency, \var{domain} is known as a \emph{control variable} of \var{source}.

The second challenge is the worker thread runs in a shared address space that might be accessed or modified by other threads, for various purposes.
One of these threads may be responsible for maintaining that $\var{domain} = \const{LOW}$ exactly when the phone indicates it is providing the $\Low$ personality (\autoref{fig:example-worker-dma-low}), so the user knows not to type in anything sensitive.
Another thread may be responsible for assigning $\Assign{\var{suspended}}{\const{TRUE}}$ when the user turns the phone's screen off, to make the worker stop processing touchscreen input.
We may then wish for \var{workspace} to be usable by some other thread---%
for example, processing input from a fingerprint scanner---%
in such a way that it can assume that \var{workspace} no longer contains any sensitive values.

% general
When we analyse one thread like this worker in terms of our compositional security property (\autoref{sec:cvdni-definition}), all the other threads in the system are trusted to do two things:
\begin{enumerate}
    \item They follow a \emph{synchronisation discipline};
        in particular for this example, this is a
        \emph{mutual exclusion (mutex) locking discipline}:
        % specific
        If read- or write-access to a certain variable is governed by a lock,
        each thread may only access the variable in that manner if they hold that lock.
        Mutual exclusion then follows from the semantics of the locking primitives ensuring only one thread may hold a given lock at a time.
    % general
    \item They themselves do not leak values from $\High$-classified locations
        (we refer to such values themselves as $\High$)
        to $\Low$-classified locations that are read-accessible to other threads.
        Note that, here, it is our objective to prove that
        the thread we are analysing can be trusted in the same way.
\end{enumerate}
Even under these assumptions, the concurrency gives rise to some tricky considerations.

% general
First, it is important that no thread in the system
(including the one under analysis) modifies any control variables carelessly.
% specific
For example, writing $\Assign{\var{domain}}{\const{LOW}}$ immediately after the worker reads a $\High$ value from \var{source}, will cause it to leak to \var{low\_sink}.
To prevent this, the worker uses \var{source\_lock}, granting it \emph{exclusive write-access} to \var{source} and \var{domain}.

% general
Furthermore, as noted above,
we may want to ensure that a \emph{non-attacker-observable} location is nevertheless cleared of any sensitive values before being used by another thread.
% specific
In our example, we classify \var{workspace} $\Low$ for the analysis to enforce this when the worker is suspended, but as the worker sometimes uses it to process $\High$ values, it is important to know \var{workspace} is accessible only to the worker during that time.
To ensure this, the worker uses \var{workspace\_lock}, granting it \emph{exclusive read- and write-access} to \var{workspace}.
It is then responsible for clearing it of any $\High$ values by the time it releases that access.
 % from itp19

%\subsection{Compositional noninterference property}
\subsection{Concurrent value-dependent notions of noninterference}
\label{sec:cvdni-definition}
%\subsubsection{Basic elements}
\label{sec:cvdni-basic-elements}
\label{sec:init-extra-params} % redundant label from merged subsection
\label{sec:compositionality-reqs} % redundant label from merged subsection
Having illustrated the challenges with an example,
we now present the definitions of per-thread and whole-system noninterference,
the theorem by which the former composes into the latter, and
%the compositionality requirements demanded as a side-condition to that theorem.
the compositionality side conditions demanded by that theorem.
%(\autoref{sec:compositionality-reqs}).

As proved for each thread, CVDNI
%\emph{concurrent value-dependent noninterference} \citep{Murray_SPR_16} (also abbreviated CVDNI \citep{Sison_Murray_19})
is defined by \citet{Murray_SPR_16} in terms of: % two main elements:
\begin{enumerate}
    \item A binary \emph{strong low-bisimulation (modulo modes)} relation $\mathcal{B}$ between program configurations,
        which serves as witness to CVDNI.
        In the style of other low-bisimulation--based
        noninterference definitions
        \citep{Focardi95,Sabelfeld00,Mantel_SS_11}
        it requires the program configurations it relates to agree on their ``low''-observable portions,
        and demands that lock-step execution preserves that correspondence.
        %(explained in \autoref{sec:verif-concurrency}).
        Furthermore, it is rely--guarantee-style concurrency aware, following \citet{Mantel_SS_11}, but modified to allow value-dependent classifications \citep{Murray_15} for mixed-sensitivity reuse (see next point). %explained in \autoref{sec:verif-reuse}).
    \item A \emph{classification} function $\dmaFunc$ that determines the ``low''-observable portion of a program configuration, thus affecting $\mathcal{B}$'s requirements.
        The innovation of $\dmaFunc$,
        as parameterised first by \citet{Murray_15}
        and then by \citet{Murray_SPR_16}
        as reproduced here,
        is that $\dmaFunc$ can depend on values in the program configuration itself, thus expressing dynamic and not just static classifications.
        %(explained by \autoref{sec:verif-reuse}).
\end{enumerate}

The theory is parameterised over the type of values~$\Val$, a finite set of shared variables~$\Var$, and a \emph{deterministic evaluation step semantics} $\EvalStep$ between \emph{local configurations} of a thread in a concurrent program.
Each local configuration is a triple $\LocalConf{\tps}{\mds}{\mem}$:
\begin{itemize}
    \item $\tps \oftype \Tps$ is the \emph{thread-private state}, which the theory will consider to be permanently inaccessible to the attacker and not shared with the other threads.
        Note that, due to this inaccessibility, we allow the user of the theory to parameterise the type $\Tps$, and we do not impose any particular structure on it.
    \item $\mds \oftype \Mode \Rightarrow \Var\ \set$
        is the \emph{(assume--guarantee) mode state}, which is ghost state associating each of
      $\Mode \defineq \{\mathbf{AsmNoW},\mathbf{AsmNoRW},\mathbf{GuarNoW},\mathbf{GuarNoRW}\}$
      with a set of shared variables.
      Intuitively, it identifies the set of variables for which the thread currently \textbf{As}su\textbf{m}es it possesses (or \textbf{Guar}antees it respects) exclusive permission to \textbf{W}rite (or \textbf{R}ead and \textbf{W}rite), granted (or obligated) for those variables typically by some synchronisation scheme.
      %\footnote{There is, strictly speaking, nothing in this formalism to stop the sets of any two different modes from overlapping, but my instantiations of this theory for this thesis will in practice prevent any such overlaps from occurring
      %(via restrictions on parameters in \autoref{sec:locking-restrictions}, and via invariants proved in \autoref{sec:global-compatibility-locks}).}
    This facilitates compositional, rely--guarantee-style reasoning about such access \citep{Jones:phd,Mantel_SS_11}. % (as explained in \autoref{sec:verif-concurrency}).

        For example, when our worker thread (of \autoref{fig:example-worker-code})
        holds \var{source\_lock}, it \emph{assumes that no other threads write} to \var{source} or its control variable \var{domain}
        (i.e.~$\{\var{source}, \var{domain}\} \subseteq \mds\ \mathbf{AsmNoW}$),
        otherwise it \emph{guarantees it does not write} to them
        ($\mathbf{GuarNoW}$).
      Similarly, when it holds \var{workspace\_lock}
      it assumes that no other threads \emph{read or write} to \var{workspace}
        (i.e.~$\var{workspace} \in \mds\ \mathbf{AsmNoRW}$),
        and at all other times it makes the corresponding guarantee
        ($\mathbf{GuarNoRW}$).

    \item $\mem \oftype \Mem$ is \emph{shared memory} considered potentially accessible to the attacker and other threads.
        To make what is accessible amenable to analysis, we impose the structure $\Mem \defineq \Var \Rightarrow \Val$, a total map from shared variable names to values.
\end{itemize}

The theory is then further parameterised by the value-dependent classification function $\dmaFunc \oftype \Mem \Rightarrow \Var \Rightarrow \{\High, \Low\}$, inducing a function $\Cvars \oftype \Var \Rightarrow \Var\ \set$ that returns all the control variables of a given variable. % on which its classifications depend.
In our worker thread example, $\dmaApp{\mem}{x}$ gives:
\begin{itemize}
    \item $\High$ when $x$ is \var{high\_sink}, meaning \var{high\_sink} is classified $\High$ at all times.
    \item when $x$ is \var{source}: $\Low$ if $\mem\ \var{domain}=\const{LOW}$, and $\High$ otherwise.
    \item $\Low$ for all other variables $x$, meaning they are classified $\Low$ at all times.
\end{itemize}

The set $\Cset = \{y\ |\ \exists x.\ y \in \Cvars\ x\}$ is then defined to contain all control variables in the system.
Thus in our worker thread example, $\Cvars\ \var{source} = \{\var{domain}\}$ and $\Cset = \{\var{domain}\}$.

With these parameters having been set, we can now define
notions of
\emph{observational equivalence}---underpinning noninterference properties---%
that are value dependent.

The notion of observational equivalence of memories,
used by the
\emph{whole-system} noninterference property
to quantify over initial state pairs,
%to be proved for the entire concurrent program,
is as follows:
% It's not actually "typical", because it's value dependent.
%in typical fashion for noninterference properties,
%such that
%(cf.~the non-modulo modes version $\LowEqOp$ in \autoref{sec:verif-reuse})
Variables that are value-dependently classified $\Low$
\emph{according to both memories}
are required to have the same value \emph{in both memories}.
Formally, as defined originally
by \citet{Murray_15}:

\begin{definition} [Low-equivalent memories] \label{def:low_eq-mem}
\label{def:low-eq-mem}
%\begin{align*}
\[
    \LowEq{\mem_1}{\mem_2}\ \defineq\ \forall x.\ \dmaApp{\mem_1}{x} = \Low\ \longrightarrow\ \mem_1\ x = \mem_2\ x
\]
%\end{align*}
\end{definition}

\noindent
Note that the asymmetry of \autoref{def:low-eq-mem}
(also \autoref{def:low_mds_eq-mem} to follow)
referring only to $\mem_1$ is resolved by
%\footnote{Note the asymmetry of referring only to $\mem_1$ is resolved by furthermore requiring the classification function $\dmaFunc$ to statically (i.e.~always, regardless of the memory state) classify all control variables as $\Low$.}
requiring the classification function
$\dmaFunc$ to classify
all control variables as $\Low$ \emph{statically}---%
that is, $\Low$ \emph{always},
regardless of the memory state
(cf.~our restriction \autoref{thm:no-H-locks} on the classification of state
used to implement locks,
later in \autoref{sec:locking-restrictions}).

%The notion of observational indistinguishability used for the noninterference property is defined over memories
%(cf.~the non-modulo modes version $\LowEqOp$ in \autoref{sec:verif-reuse})

To support compositionality for concurrent programs, however,
the equivalence notion for the \emph{per-thread} noninterference property
%---for compositional analysis of each thread---%
%the ``low'' portion demanded to be equal by the analysis
is relaxed to be \emph{modulo modes}
in the style of \citet{Mantel_SS_11}:
%(as described in \autoref{sec:verif-concurrency})
Here, $\Low$-classified non-control variables $x \notin \Cset$
are only required to have the same value
if they are assumed to be \emph{readable} by other threads
according to the mode state.
(Control variables $x \in \Cset$ are excluded from that relaxation,
and are \emph{always} required to be equal.)
%In contrast, control variables ($x \in \Cset$) are always treated as observables.
%, and non-control variables ($x \notin \Cset$) are subjected to that requirement if they are both classified $\Low$ and considered $\Readable$ in the current memory and mode state.
Defined more formally, again as originally by \citet{Murray_15}:%
\footnote{Logical operator precedence here is just as in Isabelle/HOL---from most tightly to least: $\land, \lor, \longrightarrow$.}

\begin{definition} [Readability of variable $x$, according to mode state $\mds$] \label{def:readable}
    \[
        \Readable\ \mds\ x\ \defineq\ x \notin \mds\ \mathbf{AsmNoRW}
    \]
\end{definition}

\begin{definition} [Low-equivalence of memories, modulo the mode state $\mds$] \label{def:low_mds_eq-mem}
\begin{align*}
    & \LowEqModulo{\mds}{\mem_1}{\mem_2}\ \defineq \\
    & \qquad \forall x.\ x \in \Cset\ \lor\ \dmaApp{\mem_{1}}{x} = \Low\ \land\ \Readable\ \mds\ x\ \longrightarrow \
      \mem_1\ x = \mem_2\ x
\end{align*}
\end{definition}

\noindent Moreover, we will use notation
$\lc_1 \LCLowEqModuloMdsOp \lc_2$
from \citet{Sison_Murray_19}
to lift \autoref{def:low_mds_eq-mem}
% This qualifier applied to a re-definition for Covern logic's support for
% general assume-guarantee conditions.
%this and subsequent definitions
%of $\LowEqModuloOp{\mds}$
to local program configurations, asserting also that the
local configurations $\lc_1$ and $\lc_2$ have
the same assume--guarantee mode state.
Additionally, we will use notation $\LCSameMds{\lc_1}{\lc_2}$ to denote (only) that $\lc_1$ and $\lc_2$ have the same assume--guarantee mode state.

Thus, intuitively, the user of the theory should model the
permanent untrusted output sinks, of their whole concurrent program, as variables for which $\dmaFunc$ \emph{always returns $\Low$}, ungoverned by any synchronisation scheme that the attacker cannot be trusted to follow.
In our worker example program (of \autoref{fig:example-worker-code}),
\var{low\_sink} is untrusted permanently in this way, but \var{workspace} is untrusted only when unlocked.

We now have almost enough definitions to state the per-thread compositional security property. % (deferred to \autoref{sec:init-extra-params}).
This property will assert the existence of a witness \emph{bisimulation}
relation
$\mathcal{B}$ for every possible observationally equivalent pair of starting configurations.
Specifically, this witness relation must be a \emph{strong low-bisimulation (modulo modes)}
(denoted by $\StrongLowBisimMM\ \mathcal{B}$), % formalised by \autoref{def:strong-low-bisim-mm})
meaning that it must satisfy the following three conditions:
\begin{enumerate}
    \item It must maintain observational indistinguishability by requiring that
all configuration pairs it relates (i.e.~$(\lc_1, \lc_2) \in \mathcal{B}$)
that have the same mode state ($\LCSameMds{\lc_1}{\lc_2}$),
are low-equivalent modulo modes ($\LCLowEqModuloMds{\lc_1}{\lc_2}$).

    \item Furthermore, it must be a \emph{bisimulation} by being symmetric (denoted by $\Sym\ \mathcal{B}$) and \emph{progressing to itself}:
Any step taken by one of the configurations ($\lc_1 \EvalStep \lc_1'$)
must be matched by some step taken by the configuration related to it ($\lc_2 \EvalStep \lc_2'$),
so the destinations remain related % by $\mathcal{B}$
(i.e.~$(\lc_1', \lc_2') \in \mathcal{B}$)
and modes-equal ($\LCSameMds{\lc_1'}{\lc_2'}$).

    \item Finally,
        %$\mathcal{B}$
        it must be \emph{closed under globally consistent changes}
        made to memory by other threads
        (denoted by $\CgConsistent\ \mathcal{B}$)---%
        that is, changes that preserve low-equivalence and are permitted by the current mode state $\mds$.
        Specifically, other threads are permitted to change either of variable~$x$'s value or its classification only when~$x$ is considered \emph{writable} by the current mode state (denoted by $\Writable\ \mds\ x$, \autoref{def:writable}).
        This is the most crucial element of the per-thread CVDNI property itself that ensures its compositionality for concurrent programs.
\end{enumerate}

\noindent These requirements are formalised by \autoref{def:strong-low-bisim-mm}, using Definitions \ref{def:writable} and \ref{def:cg-consistent}:

\begin{restatable} [Strong low bisimulation, modulo modes] {definition} {defstronglowbisimmm}
\label{def:strong-low-bisim-mm}%
\begin{align*}
    & \StrongLowBisimMM\ \mathcal{B}\ \defineq \ 
      \CgConsistent\ \mathcal{B}\ \land \ 
      \Sym\ \mathcal{B}\ \land \\
    & \qquad (\forall \lc_1\ \lc_2.\ (\lc_1, \lc_2) \in \mathcal{B}\ \land \ 
      \LCSameMds{\lc_1}{\lc_2} \longrightarrow \\
    & \qquad \qquad \LCLowEqModuloMds{\lc_1}{\lc_2}\ \land \\
    & \qquad \qquad (\forall \lc_1'.\ \lc_1 \EvalStep \lc_1' \longrightarrow \ 
             (\exists \lc_2'.\ \lc_2 \EvalStep \lc_2'\ \land \ 
              \LCSameMds{\lc_1'}{\lc_2'}\ \land \ 
              (\lc_1', \lc_2') \in \mathcal{B})))
\end{align*}
\end{restatable}

\begin{definition} [Writability of variable $x$, according to mode state $\mds$] \label{def:writable}
    \[
        \Writable\ \mds\ x\ \defineq\ x \notin \mds\ \mathbf{AsmNoW}\ \land\ x \notin \mds\ \mathbf{AsmNoRW}
    \]
\end{definition}

\begin{definition} [Closedness under globally consistent changes] \label{def:cg-consistent}
\begin{align*}
    & \CgConsistent\ \mathcal{B}\ \defineq \ 
      \forall \tps_1\ \mem_1\ \tps_2\ \mem_2\ \mds. \\
    & \qquad (\LocalConf{\tps_1}{\mds}{\mem_1},
              \LocalConf{\tps_2}{\mds}{\mem_2}) \in \mathcal{B}\ \longrightarrow \\
    & \qquad (\forall \mem_1'\ \mem_2'.\ 
              (\forall x.\ (\mem_1\ x \ne \mem_1'\ x\ \lor \ 
                            \mem_2\ x \ne \mem_2'\ x\ \lor \\
    & \qquad \ \ \dmaApp{\mem_1}{x} \ne \dmaApp{\mem_1'}{x}) \ 
                 \longrightarrow\ \Writable\ \mds\ x)\ \land \ 
                 \LowEqModulo{\mds}{\mem_1'}{\mem_2'}\ \longrightarrow \\
    & \qquad (\LocalConf{\tps_1}{\mds}{\mem_1'},
              \LocalConf{\tps_2}{\mds}{\mem_2'}) \in \mathcal{B})
\end{align*}
\end{definition}

\noindent
Note that, to prevent unnecessary proof effort,
$\StrongLowBisimMM$ assumes instead of asserting
the initial modes-equality ($\LCSameMds{\lc_1}{\lc_2}$),
as the security property that will use $\StrongLowBisimMM$
will take responsibility for asserting it
(to follow, in \autoref{def:com-secure}).

We now present definitions of the CVDNI security properties
that differ from those published in \citet{Murray_SPR_16}
and our conference paper \citet{Sison_Murray_19},
in that they
allow two additional forms of customisation
as parameters to the theory,
necessary for a fuller written presentation of the
formal verification of our compiler:
\begin{enumerate}
    \item Initialisation requirements for the system, in the form of a predicate over shared memory called $\INITparam$. % $\oftype \Mem \Rightarrow \bool$.

        The per-thread and whole-system security properties are \emph{relaxed} such that they only quantify over initial shared memories that obey this predicate.

    \item Extra requirements to be imposed on top of strong low-bisimulation modulo modes, in the form of a predicate over bisimulation relations called $\BISIMREQSparam$.

        The per-thread security property is \emph{strengthened} to impose these additional requirements on any candidate security witness.

\end{enumerate}
\noindent
When dropped from each of the names of the properties
``$\ComSecure$'' and ``$\SysSecure$'' soon to be introduced,
$\INITparam$ and $\BISIMREQSparam$ default to $\AlwaysTrue$;
in that case, the definitions of those properties will then simplify
to their original versions
as presented in \citet{Murray_SPR_16,Sison_Murray_19}.

The per-thread security property is then as follows:

\begin{definition} [Per-thread compositional security, with $\INITparam, \BISIMREQSparam$ requirements] \label{def:com-secure}
\begin{align*}
    & \ComSecureParam{\INITparam}{\BISIMREQSparam}\ (\tps, \mds)\ \defineq \ 
      \forall \mem_1\ \mem_2. \\
    & \qquad \LowEqModulo{\mds}{\mem_1}{\mem_2}\ \land\
             \INITparam\ \mem_1\ \land\
             \INITparam\ \mem_2\ \longrightarrow\\
    & \qquad \qquad (\exists \mathcal{B}.\ 
              \StrongLowBisimMM\ \mathcal{B}\ \land \
              \BISIMREQSparam\ \mathcal{B}\ \land \\
    & \qquad \qquad \qquad (\LocalConf{\tps}{\mds}{\mem_1},
               \LocalConf{\tps}{\mds}{\mem_2}) \in \mathcal{B})
\end{align*}
\end{definition}

We have proved in Isabelle/HOL that the compositionality theorem of \citet{Murray_SPR_16} holds regardless of the $\INITparam, \BISIMREQSparam$ chosen---%
in short, the $\INITparam$ condition relaxes the goal sufficiently to relax each of its assumptions the same way, and the $\BISIMREQSparam$ requirement only strengthens its assumptions.
Subject to some ``sound mode use'' side conditions (to be discussed soon),
%(which in this thesis I will call ``compositionality requirements'', deferred to \autoref{sec:compositionality-reqs}),
it gives us that the parallel composition
$\cms \oftype (\Tps \times (\Mode \Rightarrow \Var\ \set))\ \listtype$
of
$\ComSecure$ program threads will itself be a concurrent program
that enforces ``$\SysSecure$'',
a system-wide value-dependent noninterference property.
Here, the $\toSet$ operator returns the set of all the elements in a given list:

\begin{restatable} [Compositionality of $\ComSecureParam{\INITparam}{\BISIMREQSparam}$] {theorem} {thmcomsecurecomposes}
\label{thm:com-secure-composes}%
\[
\inferrule{
 \forall (\tps, \mds)\in \toSet\ \cms.\ \ComSecureParam{\INITparam}{\BISIMREQSparam}\ (\tps, \mds) \\\\
 \forall \mem.\ \INITparam\ \mem\ \longrightarrow\ \SoundModeUse\ (\cms, \mem)
 } {
 \SysSecureParam{\INITparam}\ \cms
 }
\]
\end{restatable}

We first introduce the elements of
this whole-system property ``$\SysSecure$'', before defining it formally
(to follow, in \autoref{def:sys-secure}).

% Redundant with above.
%\noindent This property $\SysSecure \oftype (\Tps \times \Mode)\ \listtype$ is a judgement over a list of all the per-thread parts of an initial global configuration.
From all low-equivalent pairs of initial memories
that both satisfy the $\INITparam$ conditions,
this whole-system property ``$\SysSecure$''
asserts a form of low-equality between
all global configuration pairs that are reachable via evaluation
$\EvalSched{\sched}$
%of both initial configurations
to \emph{the same} fixed schedule $\sched$, for all such finite lists $\sched$
giving an order of steps of execution from each thread:
\begin{align*}
    &  \gc \EvalSched{[]} \gc'\ \defineq\ (\gc = \gc') \\
    & \gc \EvalSched{n . ns} \gc'\ \defineq\ 
     (\exists \gc''.\ \gc \EvalSchedStep{n} \gc''\ \land\ 
    \gc'' \EvalSched{ns} \gc')
\end{align*}
Here $[]$ is an empty list, $\_ . \_$ is the cons operator,
and $\EvalSchedStep{n}$ means
the $n$th thread in the global configuration takes one step.
% A bit superfluous. -robs.
%---%
%details on $\EvalSchedStep{n}$ are unchanged from \citet{Murray_SPR_16,Murray16-Dependent-SIFUM-AFP}.

In always comparing pairs of runs executing against the same schedule,
the property models the class of schedulers whose decisions never depend on
any secrets.
Consequently, this excludes schedulers that are specialised,
in the manner of \citet{Barthe07_multi},
to actively monitor the sensitivity level of each thread's control flow,
%when the control flow of a thread has
so as to intervene and avoid interleaving it with others
when it has become dependent on secrets.
Instead, the CVDNI theory puts the onus on the developer of the program
to prove that any branching on secret conditionals does not lead to
timing-sensitive flows of the secret as discernible via low-classified sinks
accessible to other threads in the system.
%that any secret-dependent control flow does not lead to any
%timing sensitive secret-dependent behaviour when measured at the granularity
%of the points at which a context switch can occur.
%On the other hand, as
%locking state determines mode state and
Note that, as CVDNI prohibits mode state from ever becoming secret dependent,
it will implicitly prohibit any leaks into parts of memory with which
the mode state is directly associated---%
in \autoref{sec:source-language}, we will need to prohibit leaks into
the memory we use to implement mutex locks, for this reason.
%secret-dependent locking state at all.

The special form of low-equality applied by the whole-system property is one that
is modified from \autoref{def:low_eq-mem},
so that it only
requires each $\Low$-classified non-control variable $x \notin \Cset$
to be of equal value in both global configurations
if the mode states of \emph{all threads} consider $x$ to be
$\Readable$ (\autoref{def:readable}).
Furthermore, the property ensures that paired global configurations
continue to agree on the number of threads in the system,
and on the mode states for all threads,
written
%---here,
%by asserting
$\AllSameMds{\cms_1'}{\cms_2'}\ \defineq\ (\mapmdsOf{\cms_1'} = \mapmdsOf{\cms_2'})$,
where the syntax ``$\mapmdsOf{\cms}$'' denotes the mapped projection
that extracts a list
%$\mdss$
of mode states from a list $\cms$ of
$\Tps \times (\Mode \Rightarrow \Var\ \set)$ pairs.
Finally, we will use
syntax $\cms[i]$ to denote the $i$th element in list $\cms$.

This whole-system noninterference property, written formally, is then as follows:
%and $\ReadableAnyMds$ is a predicate that checks if a variable is assumed to be $\Readable$ (\autoref{def:readable}) by any thread.

\begin{definition} [Whole-system value-dependent security, with $\INITparam$ requirements]
\label{def:sys-secure}
\begin{align*}
    & \SysSecureParam{\INITparam}\ \cms\ \defineq \ 
      \forall \mem_1\ \mem_2. \\
    & \INITparam\ \mem_1\ \land\ \INITparam\ \mem_2\ \land\ \LowEq{\mem_1}{\mem_2} \longrightarrow \\
    & \qquad (\forall \sched\ \cms_1'\ \mem_1'.\ 
             (\cms, \mem_1) \EvalSched{\sched} (\cms_1', \mem_1') \longrightarrow \\
    & \qquad \qquad (\exists \cms_2'\ \mem_2'.\ 
                    (\cms, \mem_2) \EvalSched{\sched} (\cms_2', \mem_2'))\ \land \\
    & \qquad \qquad (\forall \cms_2'\ \mem_2'.\ 
                    (\cms, \mem_2) \EvalSched{\sched} (\cms_2', \mem_2')\ \longrightarrow \\
    %& \qquad \qquad \qquad \AllSameMds{\cms_1'}{\cms_2'}\ \land \\
    %& \qquad \qquad \qquad \length\ \cms_1' = \length\ \cms_2'\ \land\ \mapmdsOf{\cms_1'} = \mapmdsOf{\cms_2'} \ \land \\
    & \qquad \qquad \qquad \length\ \cms_1' = \length\ \cms_2'\ \land\ \AllSameMds{\cms_1'}{\cms_2'}\ \land \\
    & \qquad \qquad \qquad (\forall x.\ x \in \Cset\ \lor\ 
                           \dmaApp{\mem_{1}'}{x} = \Low\ \land\\
                           %\ReadableAnyMds\ \cms_1'\ x\
    & \qquad \qquad \qquad \qquad
                           (\forall i < \length\ \cms_1'.\ \Readable\ \listderef{\cms_1'}{i}\ x) \
                           \longrightarrow \
                                \mem_1'\ x = \mem_2'\ x)))
\end{align*}
\end{definition}

 % from thesis
%\subsection{Requirements for compositionality}
%\subsubsection{Compositionality requirements}
Finally,
%Briefly, we mention that
we must note that
the use of assume--guarantee reasoning to obtain the compositionality of
the per-thread property
in the style of \citet{Mantel_SS_11}
gives rise to requirements justifying the soundness of that reasoning;
requirements that we will prove our compiler to preserve.
For CVDNI, these are summed up by the ``$\SoundModeUse$'' side condition of \autoref{thm:com-secure-composes}, which consists of a ``local'' and a ``global'' part:

\begin{definition} [Sound mode use side-condition] \label{def:sound-mode-use}
\begin{align*}
    & \SoundModeUse\ (\cms, \mem) \defineq \\
    & \qquad (\forall \cm \in \toSet\ \cms.\ \LocalModeUse\ (\cm, \mem))\ \land \\
    & \qquad \GlobalModeUse\ (\cms, \mem)
\end{align*}
\end{definition}

First, all threads must each obey a \emph{local mode compliance} requirement.
This says that for all reachable local configurations of the program,
at no point will the thread violate any of its own guarantees not to access
a particular location in the shared state,
which implies also not accessing any of its control variables.
We leave precise definitions for ``$\LocReach$'' and ``\DoesntRorWText''
to this paper's Isabelle/HOL supplement material,%
\ifPreprint
\footnote{The Isabelle/HOL theories are available at \url{http://covern.org/jfpsc.html}.}
\else
\ 
\fi
but mention here that it is the \DoesntRorWText assertions that enforce that
%for this paper, as those have not changed since
%\citet{Murray_SPR_16,Murray16-Dependent-SIFUM-AFP}:
any guarantees not to access some variable $x$
will effectively apply also to all of $x$'s control variables:

\begin{restatable} [Local mode compliance] {definition} {deflocalmodecompliance}
\label{def:local-mode-compliance}%
\begin{align*}
    & \LocalModeUse\ \lc \defineq \\
    & \forall c\ \mds\ \mem.\ \LocalConf{c}{\mds}{\mem} \in
                 \LocReach\ \lc\ \longrightarrow\ 
                 \RespectsOwnGuars\ (c, \mds) \\
    & \textnormal{where} \\
    & \qquad \RespectsOwnGuars\ (c, \mds)\ \defineq \\
    & \qquad \qquad (\forall x.\ (x \in \mds\ \GuarNoRW \longrightarrow \DoesntRW\ c\ x)\ \land \\
    & \qquad \qquad \qquad (x \in \mds\ \GuarNoW \longrightarrow \DoesntW\ c\ x))
\end{align*}
\end{restatable}

Then, all threads must together obey a \emph{global modes compatibility} requirement.
This requirement says that the threads' mode states in all reachable global configurations of the concurrent program (the ``$\ReachableMdsLists$'') are \emph{compatible}%
%($\CompatibleModes$)%
---that is, if any one thread assumes a particular location will not be accessed for writing or reading, then all other threads must be guaranteeing not to access that location for the same purpose:

\begin{restatable} [Global modes compatibility] {definition} {defglobalmodescompatibility}
\label{def:global-modes-compatibility}%
\begin{align*}
    & \GlobalModeUse\ \gc\ \defineq \ 
      \forall \mdss \in \ReachableMdsLists\ \gc.\ \CompatibleModes\ \mdss \\
    & \textnormal{where} \\
    & \qquad \ReachableMdsLists\ \gc\ \defineq \\
    & \qquad \qquad \{\mdss\ |\ \exists \cms'\ \mem'\ \sched.\ 
            \gc \EvalSched{\sched} (\cms', \mem')\ \land \ 
            \map\ \mdsOf{\cms'} = \mdss\} \\
    & \qquad \CompatibleModes\ \mdss\ \defineq \ 
      \forall i\ x.\ i < \length\ \mdss \longrightarrow \\
    & \qquad \qquad (x \in \listderef{\mdss}{i}\ \AsmNoRW \longrightarrow \\
    & \qquad \qquad \qquad (\forall j < \length\ \mdss.\ j \neq i \longrightarrow
             x \in \listderef{\mdss}{j}\ \GuarNoRW))\ \land \\
    & \qquad \qquad (x \in \listderef{\mdss}{i}\ \AsmNoW \longrightarrow \\
    & \qquad \qquad \qquad (\forall j < \length\ \mdss.\ j \neq i \longrightarrow
                    x \in \listderef{\mdss}{j}\ \GuarNoW))
\end{align*}
\end{restatable}

Note that this global modes compatibility requirement
is \emph{not compositional};
consequently,
instead of obliging the program developer to prove it for the source programs
to be fed to our compiler, we will prove it
as an invariant maintained by the execution semantics
of our source language---particularly, by its synchronisation primitives
(see \autoref{sec:source-language}).

For more details and precise presentations of all the definitions we have
adapted
from \citet{Murray_SPR_16,Murray16-Dependent-SIFUM-AFP} to enable
the compiler verification work
described in this paper,
%please refer to Section III-2(a) of \citet{Murray_SPR_16}, and to its Isabelle formalisation \citep{Murray16-Dependent-SIFUM-AFP}.
please refer to the Isabelle/HOL formalisation in our supplement material.
 % from thesis

%\subsection{Requirements for per-thread CVDNI preservation}
%\subsubsection{Original notion with cube-shaped diagram}
%\subsubsection{General cube-shaped refinement formulation}
%\subsection{CVDNI-preserving refinement}
\subsection{Cube-shaped refinement for preserving noninterference}
\label{sec:decomp-reqs} % badly named label from thesis
\label{sec:just-cube-reqs}
Proof of \emph{CVDNI-preserving refinement}
(also \emph{security-preserving} or \emph{secure refinement}), for
a single-threaded program that will be run as
a thread of a concurrent program, requires the user of the theory to nominate two binary relations (both illustrated by \autoref{fig:refinement-example}):
\begin{enumerate}
    \item A \emph{refinement relation} $\mathcal{R}$ relating local configurations of the abstract program to local configurations of the concrete program:
        Abstract must simulate concrete, in a sense typical of much other work on program refinement, including compiler verification.
    \item A \emph{concrete coupling invariant} $\mathcal{I}$ that allows us to use $\mathcal{B}$ and $\mathcal{R}$ to build a new strong low-bisimulation (modulo modes) for the concrete program, by discarding pairs of local configurations \emph{after the refinement} that should not be reached in the same number of evaluation steps.
        It thereby witnesses that any changes a refinement (or compiler) might make to the execution time do not introduce any timing channels.
\end{enumerate}

% robs: JFP requests that graphics be supplied as .eps or .ps (FIXME)
\begin{figure}
    \begin{subfigure}[c]{0.34\textwidth}
    \ifPreprint
    \input{graphics/tikz-refinement-example-abs.tex}
    \else % !ifPreprint
    \includegraphics{tikz-refinement-example-abs.eps}
    \fi % !ifPreprint
    \caption{Abstract \IfKw-conditional. \\
        Relation $\mathcal{R}$ pairs configurations of this program
        with configurations of the program
        in \autoref{fig:refinement-example-conc}
        that are of the same-shaded region.}
        \label{fig:refinement-example-abs}
    \end{subfigure}
    \hfill
    \begin{subfigure}[c]{0.61\textwidth}
    \ifPreprint
    \input{graphics/tikz-refinement-example-conc.tex}
    \else % !ifPreprint
    \includegraphics{tikz-refinement-example-conc.eps}
    \fi % !ifPreprint
    \caption{Concrete \IfKw-conditional.
        Relation $\mathcal{I}$ pairs configurations of this program as shown by the dashed lines.}
        \label{fig:refinement-example-conc}
    \end{subfigure}
    %\vspace{-1\baselineskip}
    \newcommand{\figRefinementExample}{Excerpts from a CVDNI-preserving refinement example with secret-dependent control flow}
    \caption[\figRefinementExample]{\figRefinementExample:
        $h$ contains a secret, $y$ and $z$ contain zero,
        and $x$ is an untrusted sink.
        Reproduced from \citet{Sison_Murray_19}---%
        the example is originally from \citet{Murray_SPR_16}.} %, where it was used to compare the impact of the decomposition principle on proof effort.
    %\caption{Excerpts from refinement example \cite{Murray_SPR_16} used to compare proof effort (in \autoref{sec:case-study-decomp}).}
    \label{fig:refinement-example}
\end{figure}

The essence of the proof technique is to require that a number of conditions---analogous to those for $\StrongLowBisimMM$ (\autoref{def:strong-low-bisim-mm})---be imposed on the nominated $\mathcal{R}$ and $\mathcal{I}$, in relation to a given witness relation $\mathcal{B}$ establishing $\ComSecure$ (\autoref{def:com-secure}) for the abstract program.
The definitions to follow are adapted from \citet{Murray_SPR_16} Section V, as we presented in \citet{Sison_Murray_19}---for better readability, a simplified version in which no new shared variables are added by the refinement.
Consequently, we use the notation $\LCSameMdsMemOp$ to denote that two local configurations have equal mode state and memory, regardless of whether relating configurations of the same or differing languages.
% In case we want to switch to the new naming convention earlier.
%Also for better readability, from here onwards we will use $\lcA{}, \lcC{}$ instead of $\lc_{A}$, $\lc_{C}$ as the names for local configuration variables when comparing abstract and concrete executions.

Regarding the maintenance of modes equivalence and observational equivalence across the relation, the restrictions on refinement are tighter than those that were applied to $\StrongLowBisimMM$, in that $\mathcal{R}$ is required to preserve the shared memory in its entirety:
\begin{definition} [Preservation of modes and memory] \label{def:preserves_modes_mem}
    \[
      \PreservesMM\ \mathcal{R}\ \defineq \ 
      \forall \lc_{A}\ \lc_{C}. \ 
      (\lc_{A}, \lc_{C}) \in \mathcal{R} \longrightarrow \ 
      \LCSameMdsMem{\lc_A}{\lc_C}
    \]
\end{definition}
% In case we want to switch to the new naming convention earlier.
%\begin{definition} [Preservation of modes and memory] \label{def:preserves_modes_mem}
%    \[
%      \PreservesMM\ \mathcal{R}\ \defineq \ 
%      \forall \lcA{}\ \lcC{}. \ 
%      (\lcA{}, \lcC{}) \in \mathcal{R} \longrightarrow \ 
%      \LCSameMdsMem{\lcA{}}{\lcC{}}
%    \]
%\end{definition}

Regarding the closedness under changes by other threads that ensures compositionality for concurrency, on $\mathcal{I}$ we again impose
$\CgConsistent$ (\autoref{def:cg-consistent}) from \autoref{sec:cvdni-definition}.
However, in the case of $\mathcal{R}$, we instead impose
``$\ClosedOthers$'', a simplification of $\CgConsistent$ that considers only environmental actions that affect the memories on both sides of the relation identically.
Furthermore, $\ClosedOthers$ ensures equality of \emph{all} shared variables, not just those judged observable.
Defined formally:
\begin{definition} [Closedness of refinements under changes by others] \label{def:closed-others}
\begin{align*}
    & \ClosedOthers\ \mathcal{R}\ \defineq \ 
      \forall \tps_A\ \tps_C\ \mds\ \mem\ \mem'. \\
    & \qquad (\LocalConfAbs{\tps_A}{\mds}{\mem},
              \LocalConfConc{\tps_C}{\mds}{\mem}) \in \mathcal{R})\ \land \\
    & \qquad (\forall x.\ (\mem\ x \ne mem'\ x\ \lor \ 
                           \dmaApp{\mem}{x} \ne \dmaApp{\mem'}{x}) \ 
                        \longrightarrow \ 
                        \Writable\ \mds\ x)\ \longrightarrow \\
    & \qquad (\LocalConfAbs{\tps_A}{\mds}{\mem'},
              \LocalConfConc{\tps_C}{\mds}{\mem'}) \in \mathcal{R})
\end{align*}
\end{definition}

The final major---and hardest---requirement for confidentiality preservation is to prove $\mathcal{R}$ and $\mathcal{I}$ closed simultaneously under the pairwise executions of the concrete and abstract programs, using a cube-shaped ``refinement and coupling invariant preservation'' diagram ($\CouplingInvPres$, depicted in \autoref{fig:coupling-inv-pres}), whose edges are configuration pairs in $\mathcal{B}$, $\mathcal{R}$, and $\mathcal{I}$.
(Reducing its difficulty is the focus of the decomposition principle in \autoref{sec:just-decomp-reqs}.)

% robs: JFP requests that graphics be supplied as .eps or .ps (FIXME)
\begin{figure}
    \ifPreprint
    % https://tex.stackexchange.com/a/304606
\makeatletter
\@ifundefined{ifPreprint}{\newif\ifPreprint \Preprintfalse}{}
\makeatother

\ifPreprint\else % !ifPreprint
% https://tex.stackexchange.com/a/499110
\documentclass[margin=4pt,varwidth]{standalone}
\renewcommand{\rmdefault}{ptm}
\renewcommand{\sfdefault}{phv}
\usepackage{mathptmx}
\usepackage{../robs-thesis-macro-defs}
\usepackage{amsmath}
\usepackage{amssymb}
\usepackage{multido}
\usepackage{xspace}
\usepackage{tikz}
\usepackage{subcaption}

\begin{document}

% When building an .eps from this, we need a dummy figure for it to sit in;
% when we input it for the preprint, we'll do so from inside the real one.
\begin{figure}
\fi

\begin{subfigure}[l]{0.5\textwidth}
{\footnotesize
\begin{align*}
    & \CouplingInvPres\ \mathcal{B}\ \mathcal{R}\ \mathcal{I}\ \defineq \\
    & \forall \lcA{1}\ \lcC{1}. \ 
      (\lcA{1}, \lcC{1}) \in \mathcal{R}\ \longrightarrow \\
    & \quad (\forall \lcC{1}'. \ 
             \lcC{1} \EvalStepConc \lcC{1}'\ \longrightarrow \\
    & \qquad (\exists n\ \lcA{1}'.\ 
             \lcA{1} \NEvalStepAbs{n} \lcA{1}'\ \land\
             (\lcA{1}', \lcC{1}') \in \mathcal{R}\ \land \\
    & \qquad \quad (\forall \lcA{2}\ \lcC{2}\ \lcA{2}'.\ 
                     (\lcA{1}, \lcA{2}) \in \mathcal{B}\ \land\ 
                     \LCSameMds{\lcA{1}}{\lcA{2}}\ \land \\
    & \qquad \qquad (\lcA{2}, \lcC{2}) \in \mathcal{R}\ \land \ 
                    (\lcC{1}, \lcC{2}) \in \mathcal{I}\ \land \ 
                    \LCSameMds{\lcC{1}}{\lcC{2}}\ \land \\
    & \qquad \qquad \lcA{2} \NEvalStepAbs{n} \lcA{2}'\ \land\ 
                    \LCSameMds{\lcA{1}'}{\lcA{2}'}\ \longrightarrow \\
    & \qquad \qquad \quad (\exists \lcC{2}'.\ 
                           \lcC{2} \EvalStepConc \lcC{2}'\ \land\ 
                           \LCSameMds{\lcC{1}'}{\lcC{2}'}\ \land\\
    & \qquad \qquad \qquad (\lcA{2}', \lcC{2}') \in \mathcal{R}\ \land \ 
                           (\lcC{1}', \lcC{2}') \in \mathcal{I}))))
\end{align*}}
\end{subfigure}
\begin{subfigure}[l]{0.5\textwidth}

% Reproduced from our CSF'16 paper repo. Originally by Toby Murray
% NB: scale ensures desired spacing
\begin{tikzpicture}[scale=0.7]
    \node at (0,0) {\footnotesize{$\lcA{1}$}};
    \draw [->, dashed] (0.4,0) -- (3.6,0);;
    \node at (2,-0.3) {\footnotesize$n$};
    \node at (4,0) {\footnotesize{$\lcA{1}'$}};

    \node at (0.5,1) {\footnotesize{$\lcA{2}$}};
    \draw [->] (0.9,1) -- (4.1,1);;
    \node at (2.5,0.7) {\footnotesize$n$};
    \node at (4.5,1) {\footnotesize{$\lcA{2}'$}};

    \node at (0.25,0.5) {$\mathcal{B}$};
    \node at (4.25,0.5) {$\mathcal{B}$};

    \node at (0.25,-3.5) {$\mathcal{I}$};
    \node at (4.25,-3.5) {$\mathcal{I}$};

    \node at (-1.85,0.5) {\begin{minipage}{2cm}\centering\ \ \ \ \  {\em abstract execution}\end{minipage}};

    \node at (-1.85,-3.5) {\begin{minipage}{2cm}\centering\ \ \ \ \ {\em concrete execution}\end{minipage}};

    \draw (0,-0.5) -- (0,-1.5);
    \node at (0,-2) {$\mathcal{R}$};
    \draw (0,-2.5) -- (0,-3.5);

    \draw (0.5,0.5) -- (0.5,-0.5);
    \node at (0.5,-1) {$\mathcal{R}$};
    \draw (0.5,-1.5) -- (0.5,-2.5);

    \draw [dashed] (4,-0.5) -- (4,-1.5);
    \node at (4,-2) {$\mathcal{R}$};
    \draw [dashed] (4,-2.5) -- (4,-3.5);

    \draw [dashed] (4.5,0.5) -- (4.5,-0.5);
    \node at (4.5,-1) {$\mathcal{R}$};
    \draw [dashed] (4.5,-1.5) -- (4.5,-2.5);

    \node at (0,-4) {\footnotesize{$\lcC{1}$}};
    \draw [->] (0.4,-4) -- (3.6,-4);;
    \node at (2,-4.3) {\footnotesize$1$};
    \node at (4,-4) {\footnotesize{$\lcC{1}'$}};

    \node at (0.5,-3) {\footnotesize{$\lcC{2}$}};
    \draw [->, dashed] (0.9,-3) -- (4.1,-3);;
    \node at (2.5,-3.3) {\footnotesize$1$};
    \node at (4.5,-3) {\footnotesize{$\lcC{2}'$}};
\end{tikzpicture}
\end{subfigure}

\ifPreprint\else % !ifPreprint
\end{figure}
\end{document}
\fi
    \else % !ifPreprint
    \includegraphics{tikz-coupling-inv-pres.eps}
    \fi % !ifPreprint
    \vspace{-1\baselineskip}
    \newcommand{\figCubePres}{Definition and graphical depiction of refinement preservation obligation}
    %refinement (and coupling invariant) preservation for $\SecureRefinement$ (Def. \ref{def:secure-refinement})}
    \caption[\figCubePres]{\figCubePres\ for $\SecureRefinement$ (\autoref{def:secure-refinement}).
    Reproduced from \citet{Sison_Murray_19}---the definition is a simplified restatement of its original formalisation in \citet{Murray_SPR_16}.}
    \label{fig:coupling-inv-pres}
\end{figure}

All that then remains is for the nominated concrete coupling invariant $\mathcal{I}$ to be symmetric ($\Sym\ \mathcal{I}$), and the predicate $\SecureRefinement$ puts together all the requirements:

\begin{definition} [Requirements for confidentiality-preserving secure refinement]
\label{def:secure-refinement}
\begin{align*}
    \SecureRefinement\ \mathcal{B}\ \mathcal{R}\ \mathcal{I}\ \defineq\ 
  & \PreservesMM\ \mathcal{R}\ \land \ 
    \ClosedOthers\ \mathcal{R}\ \land \\
  & \CgConsistent\ \mathcal{I}\ \land \ 
    \Sym\ \mathcal{I}\ \land \ 
    \CouplingInvPres\ \mathcal{B}\ \mathcal{R}\ \mathcal{I}
\end{align*}
\end{definition}

The soundness theorem for confidentiality-preserving refinement by \citet{Murray_SPR_16} then gives us that,
under these conditions,
the concrete relation ``$\BCofApplied$'', derived from
a witness $\StrongLowBisimMM$ relation $\mathcal{B}$,
refinement relation $\mathcal{R}$,
and coupling invariant $\mathcal{I}$,
is itself a witness $\StrongLowBisimMM$ for the concrete program.
For readability, from here onwards we will use $\lcA{1}, \lcC{1}, \ldots$ instead of $\lc_{1A}$, $\lc_{1C}, \ldots$ for local configuration variables when comparing abstract and concrete executions simultaneously:

\begin{definition} [Concrete bisimulation relation derived from $\mathcal{B}, \mathcal{R}$ and $\mathcal{I}$]
\label{def:concrete-bisim}
\begin{align*}
\BCofApplied\ \defineq\ 
\{(\lcC{1}, \lcC{2})\ |\ 
    & \exists \lcA{1}\ \lcA{2}.\ 
      (\lcA{1}, \lcC{1}) \in \mathcal{R}\ \land \ 
      (\lcA{2}, \lcC{2}) \in \mathcal{R}\ \land \\
    & (\lcA{1}, \lcA{2}) \in \mathcal{B}\ \land \ 
                  \LCLowEqModuloMds{\lcC{1}}{\lcC{2}}\ \land \ 
                  (\lcC{1}, \lcC{2}) \in \mathcal{I}\}
\end{align*}
\end{definition}

\begin{theorem} [Preservation of $\StrongLowBisimMM$ by $\SecureRefinement$] \label{thm:strong-low-bisim-mm-preserved}
\[
\inferrule{
    \StrongLowBisimMM\ \mathcal{B}
    \and
    \SecureRefinement\ \mathcal{B}\ \mathcal{R}\ \mathcal{I}
} {
    \StrongLowBisimMM\ (\BCofApplied)
}
\]
\end{theorem}

 % from thesis

%\subsubsection{Decomposition principle and proof effort comparison}
%\subsection{Refinement decomposition principle and its proof impact}
\subsection{Decomposition principle and its impact on refinement proofs}
\label{sec:just-decomp-reqs}
\label{sec:h-branch-example} % redundant label from merged subsection
%\subsubsection{Decomposition principle}

%Recent work, published in \citet{Sison_Murray_19} alongside (but outside the scope of) \autoref{chp:wr-compiler}'s contributions,
We now present, as we first did in \citet{Sison_Murray_19}, an alternative way
to prove $\SecureRefinement$ (\autoref{def:secure-refinement})
that obviates the need to use the cube-shaped, two-sided refinement obligation
%$\CouplingInvPres\ \mathcal{B}\ \mathcal{R}\ \mathcal{I}$
(depicted by \autoref{fig:coupling-inv-pres}), by decomposing its concerns into:
\begin{enumerate}
    \item Proving $\mathcal{R}$ closed %under concrete--abstract pairwise execution,
        %under the pairwise executions of the concrete and abstract programs alone,
        using a square-shaped simulation diagram (depicted by \autoref{fig:decomp-R}) akin to the \emph{backward simulations} commonly used to prove semantics-preserving refinement by compilers (e.g.~for CompCert \citep{Leroy09}), and
    \item Security-focused proof obligations
        (depicted by Figures \ref{fig:decomp-abs-steps}, \ref{fig:decomp-I}),
        separable from the square-shaped simulation,
        that %together
        prevent the introduction of \emph{timing leaks}, \emph{termination leaks},
        and secret-dependent differences in assume--guarantee mode state. %smaller and more separable obligations, %gathered together under the side-condition predicate $\SimplerRefinementSafe$ (see \autoref{def:simpler-refinement-safe}).
\end{enumerate}

The decomposition
%(specified by \autoref{def:secure-refinement-simpler})
requires the verifier to nominate a new parameter, called $\abssteps$ or the \emph{pacing function}.
Its role is to dictate the pace of the square-shaped simulation
%that proves refinement,
by specifying the number of abstract steps that ought to be taken for one concrete step,
as depicted by \autoref{fig:decomp-R}.
%as follows (also depicted by \autoref{fig:decomp-R}): % for each given abstract-concrete local configuration pair related by $\mathcal{R}$:
Deferring the security-focused side conditions
(``$\SimplerRefinementSafe$'')
to afterwards, the decomposition principle is then defined formally as follows:

%\begin{definition} [Decomposed requirements for $\SecureRefinement$]
\begin{definition} [Decomposition principle for $\SecureRefinement$]
\label{def:secure-refinement-simpler}
\begin{align*}
    & \SecureRefineSimpler\ \mathcal{B}\ \mathcal{R}\ \mathcal{I}\ \abssteps\ \defineq \\
    & \qquad \PreservesMM\ \mathcal{R}\ \land \ 
      \ClosedOthers\ \mathcal{R}\ \land \ 
      \CgConsistent\ \mathcal{I}\ \land \ 
      \Sym\ \mathcal{I}\ \land \\
    & \qquad \SimplerRefinementSafe\ \mathcal{B}\ \mathcal{R}\ \mathcal{I}\ \abssteps\ \land \ 
      (\forall \lcA{}\ \lcC{}.\ (\lcA{}, \lcC{}) \in \mathcal{R} \longrightarrow \\
    & \qquad \qquad (\forall \lcC{}'.\ \lcC{} \EvalStepConc \lcC{}' \longrightarrow \ 
             (\exists \lcA{}'.\ \lcA{} \NEvalStepAbs{(\abssteps\ \lcA{}\ \lcC{})} \lcA{}'\ \land \ (\lcA{}', \lcC{}') \in \mathcal{R})))
\end{align*}
\end{definition}

The aforementioned side conditions on all refinement parameters, depicted by Figures \ref{fig:decomp-abs-steps}, \ref{fig:decomp-I}, are then defined formally under the predicate $\SimplerRefinementSafe$ as follows:

\begin{figure}
    \centering
    % All three subfigures are adapted by Robert Sison from
    % coupling_invariant.tex from our CSF'16 paper repo,
    % originally by Toby Murray. -robs.
    \begin{subfigure}[b]{0.3\textwidth}
        \ifPreprint
        % https://tex.stackexchange.com/a/304606
\makeatletter
\@ifundefined{ifPreprint}{\newif\ifPreprint \Preprintfalse}{}
\makeatother

\ifPreprint\else % !ifPreprint
\documentclass{standalone}
\renewcommand{\rmdefault}{ptm}
\renewcommand{\sfdefault}{phv}
\usepackage{mathptmx}
\usepackage{../robs-thesis-macro-defs}
\usepackage{amsmath}
\usepackage{tikz}

\begin{document}
\fi

% NB: scale ensures desired spacing
\begin{tikzpicture}[scale=0.7]
    \node at (0,0) {\footnotesize{$\lcA{}$}};
    \draw [->, dashed] (0.4,0) -- (3.8,0);;
    \node at (2.1,-0.3) {\footnotesize{$\abssteps\ \lcA{}\ \lcC{}$}};
    \node at (4.2,0) {\footnotesize{$\lcA{}'$}};

    \draw (0,-0.5) -- (0,-1.5);
    \node at (0,-2) {$\mathcal{R}$};
    \draw (0,-2.5) -- (0,-3.5);

    \draw [dashed] (4.2,-0.5) -- (4.2,-1.5);
    \node at (4.2,-2) {$\mathcal{R}$};
    \draw [dashed] (4.2,-2.5) -- (4.2,-3.5);

    \node at (0,-4) {\footnotesize{$\lcC{}$}};
    \draw [->] (0.4,-4) -- (3.8,-4);;
    \node at (2,-4.4) {\footnotesize$1$};
    \node at (4.2,-4) {\footnotesize{$\lcC{}'$}};
\end{tikzpicture}

\ifPreprint\else % !ifPreprint
\end{document}
\fi
        \else % !ifPreprint
        \includegraphics{tikz-decomp-R.eps}
        \fi % !ifPreprint
        \caption{Refinement preservation for relation $\mathcal{R}$ under program execution paced by $\abssteps$.
        (Part of \autoref{def:secure-refinement-simpler}.)}
        \label{fig:decomp-R}
    \end{subfigure}\hfill
    \begin{subfigure}[b]{0.31\textwidth}
        \ifPreprint
        % https://tex.stackexchange.com/a/304606
\makeatletter
\@ifundefined{ifPreprint}{\newif\ifPreprint \Preprintfalse}{}
\makeatother

\ifPreprint\else % !ifPreprint
\documentclass{standalone}
\renewcommand{\rmdefault}{ptm}
\renewcommand{\sfdefault}{phv}
\usepackage{mathptmx}
\usepackage{../robs-thesis-macro-defs}
\usepackage{amsmath}
\usepackage{tikz}

\begin{document}
\fi

% NB: scale ensures desired spacing
\begin{tikzpicture}[scale=0.7]
    \node at (0,0) {\footnotesize{$\lcA{1}$}};
    \draw [-, dashed] (0.4,0) -- (1.2,0);;
    \node at (2.8,0) {\footnotesize{$\abssteps\ \lcA{1}\ \lcC{1}$}};
    \node at (2.8,0.5) {\footnotesize{=}};

    \node at (0.5,1) {\footnotesize{$\lcA{2}$}};
    \draw [-, dashed] (0.9,1) -- (1.7,1);;
    \node at (3.3,1) {\footnotesize{$\abssteps\ \lcA{2}\ \lcC{2}$}};

    \node at (0.25,0.5) {$\mathcal{B}$};

    \node at (0.25,-3.5) {$\mathcal{I}$};

    \draw (0,-0.5) -- (0,-1.5);
    \node at (0,-2) {$\mathcal{R}$};
    \draw (0,-2.5) -- (0,-3.5);

    \draw (0.5,0.5) -- (0.5,-0.5);
    \node at (0.5,-1) {$\mathcal{R}$};
    \draw (0.5,-1.5) -- (0.5,-2.5);

    \node at (0,-4) {\footnotesize{$\lcC{1}$}};
    \draw [-, dashed] (0.4,-4) -- (1.2,-4);;
    \node at (2.1,-4) {\footnotesize{\textsf{stops} $\lcC{1}$}};
    \node at (2.1,-3.5) {\footnotesize{=}};

    \node at (0.5,-3) {\footnotesize{$\lcC{2}$}};
    \draw [-, dashed] (0.9,-3) -- (1.7,-3);;
    \node at (2.6,-3) {\footnotesize{\textsf{stops} $\lcC{2}$}};
\end{tikzpicture}

\ifPreprint\else % !ifPreprint
\end{document}
\fi
        \else % !ifPreprint
        \includegraphics{tikz-decomp-abs-steps.eps}
        \fi % !ifPreprint
        \caption{Consistency of pacing and stopping behaviour, to prevent timing and termination leaks.
        (Part of \autoref{def:simpler-refinement-safe}.)}
        \label{fig:decomp-abs-steps}
    \end{subfigure}\hfill
    \begin{subfigure}[b]{0.3\textwidth}
        \ifPreprint
        % https://tex.stackexchange.com/a/304606
\makeatletter
\@ifundefined{ifPreprint}{\newif\ifPreprint \Preprintfalse}{}
\makeatother

\ifPreprint\else % !ifPreprint
\documentclass{standalone}
\renewcommand{\rmdefault}{ptm}
\renewcommand{\sfdefault}{phv}
\usepackage{mathptmx}
\usepackage{../robs-thesis-macro-defs}
\usepackage{amsmath}
\usepackage{tikz}

\begin{document}
\fi

% NB: scale ensures desired spacing
\begin{tikzpicture}[scale=0.7]
    \node at (0,0) {\footnotesize{$\lcA{1}$}};

    \node at (0.5,1) {\footnotesize{$\lcA{2}$}};

    \node at (0.25,0.5) {$\mathcal{B}$};

    \node at (0.25,-3.5) {$\mathcal{I}$};
    \node at (4.2,-3.5) {$\mathcal{I}$};
    \draw [dashed] (4.2,-3.5) circle [x radius=1.1, y radius=0.6, rotate=60];;

    \draw (0,-0.5) -- (0,-1.5);
    \node at (0,-2) {$\mathcal{R}$};
    \draw (0,-2.5) -- (0,-3.5);

    \draw (0.5,0.5) -- (0.5,-0.5);
    \node at (0.5,-1) {$\mathcal{R}$};
    \draw (0.5,-1.5) -- (0.5,-2.5);

    \node at (0,-4) {\footnotesize{$\lcC{1}$}};
    \draw [->] (0.4,-4) -- (3.7,-4);;
    \node at (2,-4.3) {\footnotesize$1$};
    \node at (4.1,-4) {\footnotesize{$\lcC{1}'$}};

    \node at (0.5,-3) {\footnotesize{$\lcC{2}$}};
    \draw [->] (0.9,-3) -- (4.1,-3);;
    \node at (2.5,-3.3) {\footnotesize$1$};
    \node at (4.5,-3) {\footnotesize{$\lcC{2}'$}};
\end{tikzpicture}

\ifPreprint\else % !ifPreprint
\end{document}
\fi
        \else % !ifPreprint
        \includegraphics{tikz-decomp-I.eps}
        \fi % !ifPreprint
        \caption{Closedness of the coupling invariant relation $\mathcal{I}$ under lockstep program execution.
        (Part of \autoref{def:simpler-refinement-safe}.)}
        \label{fig:decomp-I}
    \end{subfigure}
    \newcommand{\figDecompPres}{Graphical depictions of decomposed refinement preservation obligations}
    \caption[\figDecompPres]{\figDecompPres. \\ % \ (\autoref{def:simpler-refinement-safe}).
    Reproduced from \citet{Sison_Murray_19}.}
    \label{fig:decomp}
\end{figure}

%\begin{definition} [Side conditions for $\SecureRefinement$ decomposition]
\begin{definition} [Security-focused side conditions for decomposition principle]
\label{def:simpler-refinement-safe}
\begin{align*}
    & \SimplerRefinementSafe\ \mathcal{B}\ \mathcal{R}\ \mathcal{I}\ \abssteps\ \defineq \ 
      \forall \lcA{1}\ \lcA{2}\ \lcC{1}\ \lcC{2}. \ 
      (\lcA{1}, \lcA{2}) \in \mathcal{B}\ \land \\
    & \LCSameMds{\lcA{1}}{\lcA{2}}\ \land\ 
      (\lcA{1}, \lcC{1}) \in \mathcal{R}\ \land\ 
      (\lcA{2}, \lcC{2}) \in \mathcal{R}\ \land\ 
      (\lcC{1}, \lcC{2}) \in \mathcal{I}\ \land\ 
      \LCSameMds{\lcC{1}}{\lcC{2}} \\
    & \longrightarrow\ \stops\ \lcC{1} = \stops\ \lcC{2}\ \land\ 
             \abssteps\ \lcA{1}\ \lcC{1} = \abssteps\ \lcA{2}\ \lcC{2}\ \land \\
    & \quad\ \ \ \ (\forall \lcC{1}'\ \lcC{2}'. \ 
             \lcC{1} \EvalStepConc \lcC{1}'\ \land \ 
             \lcC{2} \EvalStepConc \lcC{2}'\ \longrightarrow \ 
             (\lcC{1}', \lcC{2}') \in \mathcal{I}\ \land \ 
             \LCSameMds{\lcC{1}'}{\lcC{2}'})
\end{align*}
\end{definition}

\noindent The intuitive meanings of the side conditions in \autoref{def:simpler-refinement-safe} are:
\begin{itemize}
    \item $\stops\ \lcC{1} = \stops\ \lcC{2}$ ensures that the refinement has not introduced any termination leaks, by asserting \emph{consistent stopping behaviour} for $\mathcal{I}$-related concrete program configurations, which we know to be observationally indistinguishable.
    \item $\abssteps\ \lcA{1}\ \lcC{1} = \abssteps\ \lcA{2}\ \lcC{2}$ ensures that the refinement has not introduced any timing leaks, by asserting \emph{consistency of the pace of the refinement} for $\mathcal{R}$-related program configurations, which we again know to be observationally indistinguishable.
    \item The final $\forall$-quantified clause asserts $\mathcal{I}$'s suitability as a coupling invariant, in that it must remain \emph{closed under lockstep evaluation} of the concrete program configurations it relates.
        Furthermore it must \emph{maintain mode state equality} with each lockstep evaluation, which ensures that the refinement has not introduced any inconsistencies in the memory access assumptions and guarantees needed for the concurrent compositionality of the property.
\end{itemize}
Note that the $\mathcal{B}$- and $\mathcal{R}$-edges in \autoref{fig:decomp-I} may capture useful
facts about a particular program verification technique and compiler (respectively), so their availability as assumptions is intended to reduce greatly the effort needed to specify a coupling invariant $\mathcal{I}$ and prove it satisfies the condition.

Assuming the fulfilment of all the decomposed requirements,
% robs: Actually whoops - we didn't! Only the extended (arXiv) version did,
% so this remark is a mistake in my thesis.
%\citet{Sison_Murray_19} provided a proof that they are a sound method for establishing
we obtain that they are a sound method for establishing
secure refinement of the per-thread confidentiality property, as desired:
%\begin{theorem} [Soundness of $\SecureRefineSimpler$]
\begin{theorem} [Soundness of the decomposition principle]
\label{thm:secure-refinement-simpler-sound}
\begin{align*}
\SecureRefineSimpler\ \mathcal{B}\ \mathcal{R}\ \mathcal{I}\ \abssteps\ \implies \SecureRefinement\ \mathcal{B}\ \mathcal{R}\ \mathcal{I}
\end{align*}
\end{theorem}
\begin{proof}
    The only obligation for $\SecureRefinement$ (\autoref{def:secure-refinement}) not obtained immediately from $\SecureRefineSimpler$ (\autoref{def:secure-refinement-simpler}) is the cube-shaped $\CouplingInvPres$ (\autoref{fig:coupling-inv-pres}).
    We discharge this as follows:

    The front face of the cube is just ordinary square-shaped refinement preservation (depicted in \autoref{fig:decomp-R}), given to us by $\SecureRefineSimpler$:
    that a single concrete step from $\lcC{1}$ is simulated by $n$ abstract steps from $\lcA{1}$, where $n$ is given by $\abssteps$.

    We are then obliged to prove a simulation in the other direction (the back face of the cube), that $n$ abstract steps from all configurations $\lcA{2}$ related by $\mathcal{B}$ to $\lcA{1}$ are simulated by some concrete step from $\lcC{2}$ related by $\mathcal{R}$ to $\lcA{2}$ and by $\mathcal{I}$ to $\lcC{1}$.

    Here, we lean on the determinism of the abstract program's evaluation semantics (required by the theory) to flip the direction of simulation, knowing that $n$ abstract steps from $\lcA{2}$, simulating a single concrete step from $\lcC{2}$, could only be the very same $n$ abstract steps from $\lcA{2}$ that we were required to consider.
    This allows us to obtain that simulation by using, once again,
    the square-shaped refinement preservation (\autoref{fig:decomp-R})
    given to us by $\SecureRefineSimpler$.

    Consistency of refinement pacing and stopping behaviour (depicted in \autoref{fig:decomp-abs-steps}) given by $\SimplerRefinementSafe$ (\autoref{def:simpler-refinement-safe}) then respectively ensure that $n$ (via $\abssteps$) is the correct number of abstract steps to consider, and that there will indeed be a concrete step from $\lcC{2}$ to drive the matching simulation step.

    Finally, the remainder of $\SimplerRefinementSafe$ (depicted in \autoref{fig:decomp-I}) discharges the requirement of closedness and modes-equality maintenance of $\mathcal{I}$ under lockstep execution, demanded by the bottom face of the cube.
\end{proof}
 % from thesis
%\subsubsection{Proof effort comparison}
To demonstrate how the decomposition principle reduces proof complexity and effort, we returned to the example program refinement discussed in Section V-E of \citet{Murray_SPR_16} and proved in its Isabelle formalisation \citep{Murray16-Refinement-AFP},
an excerpt of which is shown in \autoref{fig:refinement-example}.
The abstract program (9 imperative commands) branches on a sensitive value, and executes a single atomic expression assignment in each branch.
Its refinement (to 16 commands) models expansion of the expressions into multiple steps, resolving a timing disparity between the two branches by padding with $\Skip$.

We use proof size as a proxy for proof effort, since the former
is known to be strongly linearly correlated with the latter~\citep{Staples_JAMKK_14}.
Formalised in Isabelle/HOL as \texttt{EgHighBranchRevC.thy} \citep{Murray16-Refinement-AFP}, the proof line count for that theory stood at about 4.6K lines of definitions
and proof, of which approx.\ 3.6K line were proofs.
Adapting the proof instead to use the decomposition principle
($\SecureRefineSimpler$, \autoref{def:secure-refinement-simpler}),
the proof line count drops from 3.6K to approx.\ 2K, a 44\% reduction.
Regarding definition changes, the new proof makes less than 10 lines of adaptations to a coupling invariant and pacing function used by the old proof, and adds about 30 lines worth of new helper definitions, for use with the decomposition principle.
The rest of the theory and its external dependencies remain in common between the two versions.

As would be expected, the bulk of the deletions are from the full cube-shaped refinement diagram proof (\autoref{fig:coupling-inv-pres}) of $\SecureRefinement$ (\autoref{def:secure-refinement}) for the refinement relation.
The surviving parts of that proof just become the square-shaped refinement diagram proof (\autoref{fig:decomp-R}) of the decomposition principle
(\autoref{def:secure-refinement-simpler}),
without much modification.
The deletions are replaced by newly added proofs of
the decomposition principle's more security-focused side conditions
(\autoref{def:simpler-refinement-safe},
depicted by Figures \ref{fig:decomp-abs-steps}, \ref{fig:decomp-I}).
 % from itp19

\subsection{Compositional whole-system secure refinement}
%\subsection{Requirements for whole-system CVDNI preservation}
% Toby: Placeholder for reproducing whole-system refinement theory
%       from CSF'16 isa-afp entry.
\label{sec:sys-refine}
We now present the whole-system refinement theorem
from \citet{Murray_SPR_16,Murray16-Refinement-AFP},
which we adapt here to support
the specification of $\INITparam$ requirements
(as in $\ComSecureParam{\INITparam}{\BISIMREQSparam}$, \autoref{def:com-secure}),
and simplify to the case of
refinements that add no shared variables.

%The whole-system refinement theorem
%from \citet{Murray_SPR_16,Murray16-Refinement-AFP}
The main usefulness of this theorem is that, %Its main usefulness is that,
beyond demanding $\SecureRefinement$
(\autoref{def:secure-refinement}),
which dealt with the preservation of
per-thread security as witnessed by a $\StrongLowBisimMM$
(\autoref{def:strong-low-bisim-mm}),
it deals additionally with the preservation of the
$\SoundModeUse$ side conditions (\autoref{def:sound-mode-use})
that will be demanded by the compositionality theorem for CVDNI
(\autoref{thm:com-secure-composes})
at the target language level.

Notably, although it imposes the requirement for the refinement
to preserve the ``local'' part of $\SoundModeUse$
(\autoref{def:local-mode-compliance}),
%$\PreservesLocalCompliance$
it \emph{automatically} preserves the non-compositional
``global'' part of this side condition
(\autoref{def:global-modes-compatibility})
%global modes compatibility side condition.
as a consequence of the requirements imposed by
the per-thread secure refinements. % requirements.
Thus, our source-level proof of the global condition %'s invariance
(see \autoref{sec:source-language})
will be sufficient,
and %so, beyond the aforementioned requirements,
there will be no need for us to prove anything extra about our compiler
for it to preserve that to the target-language level.
%our compiler to preserve it to target level.

We now present the requirements and theorem for whole-system refinement formally.

First, in addition to the per-thread refinement notion
$\SecureRefinement$ (\autoref{def:secure-refinement})
that we addressed in
Sections \ref{sec:just-cube-reqs} and \ref{sec:just-decomp-reqs},
our whole-system refinement theorem will % ask the compiler developer
%to prove that
require that
the refinement relation $\mathcal{R}$
established by the compiler additionally
preserves the compositional local mode compliance property for each thread.
Here, ``$\RespectsOwnGuars$''
%$\PreservesLocalCompliance$
is from \autoref{def:local-mode-compliance}:

\begin{definition} [Refinement $\mathcal{R}$ preserves local mode compliance]
\begin{align*}
  & \PreservesLocalCompliance\ \mathcal{R}\ \defineq \ 
    \forall \tps_A\ \mds_A\ \mem_A\ \tps_C\ \mds_C\ \mem_C. \\
  & \qquad \RespectsOwnGuars\ (\tps_A, \mds_A)
    \ \land \\
  & \qquad (\LocalConfAbs{\tps_A}{\mds_A}{\mem_A},
            \LocalConfConc{\tps_C}{\mds_C}{\mem_C}) \in \mathcal{R}
           \ \longrightarrow \\
  & \qquad \qquad \RespectsOwnGuars\ (\tps_C, \mds_C))
\end{align*}
\end{definition}

We define a new ``compositional refinement'' predicate
to capture all per-thread requirements that will be demanded
by our compositional whole-system refinement theorem.
This bundles together
$\PreservesLocalCompliance$ and
$\SecureRefinement$ % (\autoref{def:secure-refinement})
so as to preserve the $\StrongLowBisimMM$ relations
(\autoref{def:strong-low-bisim-mm})
that witness
noninterference for each thread of the abstract program.
Alongside all these requirements just described,
it also requires %imposes a simple requirement that ensures that
the concrete coupling invariant $\mathcal{I}$ to cover
all possible initial memory pairs
that are low-equal modulo modes (\autoref{def:low_mds_eq-mem})
and satisfy the $\INITparam_C$ conditions
that will parameterise the target language-level CVDNI property:

\begin{definition} [Requirements for compositional whole-system refinement]
\label{def:sys-refine-reqs}
\begin{align*}
  & \SysRefineReqs\ \mathcal{B}\ \mathcal{R}\ \mathcal{I}\ \defineq \\
  & \qquad
    \SecureRefinement\ \mathcal{B}\ \mathcal{R}\ \mathcal{I}\ \land \ 
    \StrongLowBisimMM\ \mathcal{B}\ \land \\
    % Note: "sym I" is here for the Isabelle definition, but
    %       it is bundled into \SecureRefinement for the paper.
  & \qquad
    \PreservesLocalCompliance\ \mathcal{R}\ \land \\
  & \qquad
    (\forall \tps_C\ \mds\ \mem_1\ \mem_2.\ 
    \LowEqModulo{\mds}{\mem_1}{\mem_2}\ \land\
             \INITparam_C\ \mem_1\ \land\
             \INITparam_C\ \mem_2\ \longrightarrow\\
  & \qquad \qquad (\LocalConfConc{\tps_C}{\mds}{\mem_1},
                   \LocalConfConc{\tps_C}{\mds}{\mem_2}) \in \mathcal{I})
\end{align*}
\end{definition}

With these requirements, we prove using Isabelle/HOL that
a whole-system refinement theorem,
proved originally by \citet{Murray_SPR_16,Murray16-Refinement-AFP},
can be adapted to support the specification of $\INITparam$ requirements
on initial memory at both abstract- and concrete-level.
(As with \autoref{thm:com-secure-composes},
the relaxation of the goal by $\INITparam_C$
is enough to permit the relaxations of its assumptions
by $\INITparam_C, \INITparam_A$.)
First we will state the theorem, then we will explain it, line-by-line.
This theorem proves
that abstract-level $\SoundModeUse$ (including its global part)
%by a $\SysSecure$ (\autoref{def:sys-secure})
by a system of secure mixed-sensitivity concurrent program threads
(i.e.~list $\cms_A$, as witnessed by bisimulations $\Bs$ for each thread)
is sufficient for a set of per-thread secure refinements
(in terms of the lists $\Bs, \Rs, \Is$ of bisimulation, refinement, and concrete coupling invariant relations for each thread, respectively)
to yield a concrete-level secure concurrent program
(i.e.~list $\cms_C$ that satisfies $\SysSecure$, \autoref{def:sys-secure}):

% Isabelle name: refined_prog_secure_init_simplified_for_print
%                as proved for locale sifum_refinement_same_mem_init
% Proof is in lemma refined_prog_secure_init_explicit for that locale.
% (NB: Ignore the one proved for locale sifum_refinement_init)
%\begin{theorem} [Whole-system secure refinement]
%\begin{theorem} [Compositionality of secure refinement]
\begin{theorem} [Whole-system compositionality of per-thread secure refinement]
\label{thm:sys-refine}
\begin{gather*}
\begin{align*}
    & (\forall \mem.\ \INITparam_{C}\ \mem\ \longrightarrow\ \INITparam_{A}\ \mem)
    \ \land \\
    & (\forall \mem.\ \INITparam_{A}\ \mem\ \longrightarrow\ \SoundModeUse\ (\cms_{A}, \mem))
    \ \land \\
    & \length\ \cms_{A}\,=\,\length\ \Bs\,=\,\length\ \Rs\,=\,\length\ \Is\,=\,\length\ \cms_{C}
    \ \land \\
    & (\forall i < \length\ \cms_{C}. \\
    & \qquad \SysRefineReqs\ \listderef{\Bs}{i}\ 
                            \listderef{\Rs}{i}\ 
                            \listderef{\Is}{i}
    \ \land \\
    & \qquad (\forall \mem.\ \INITparam_{C}\ \mem\ \longrightarrow
                ((\listderef{\cms_{A}}{i},\mem),
                 (\listderef{\cms_{C}}{i},\mem)) \in \listderef{\Rs}{i})
    \ \land \\
    & \qquad (\forall \mem_1\ \mem_2.\ 
              \INITparam_{A}\ \mem_1\ \land\ 
              \INITparam_{A}\ \mem_2\ \land\ 
              \LowEqModulo{(\snd\ \listderef{\cms_{C}}{i})}{\mem_1}{\mem_2}
    \ \longrightarrow \\
    & \qquad \qquad ((\listderef{\cms_{A}}{i},\mem_1),
                 (\listderef{\cms_{A}}{i},\mem_2)) \in \listderef{\Bs}{i}))
\end{align*} \\
\cline{1-2}
\SysSecureParam{\INITparam_C}\ \cms_{C}
\end{gather*}
\end{theorem}

The premises of this theorem can be understood as follows:
\begin{itemize}
    \item
        $(\forall \mem.\ \INITparam_{C}\ \mem\ \longrightarrow\ \INITparam_{A}\ \mem)$: \\
        The concrete-level ``$\INITparam_C$'' initial condition
        must be no weaker than
        the abstract-level ``$\INITparam_A$'' one.
    \item
        $(\forall \mem.\ \INITparam_{A}\ \mem\ \longrightarrow\ \SoundModeUse\ (\cms_{A}, \mem))$: \\
        For the abstract program, $\SoundModeUse$
        (\autoref{def:sound-mode-use})
        must hold for all possible initial memories.
    \item
        The lists of initial thread-private and mode states
        at abstract and concrete level (resp.~$\cms_A, \cms_C$),
        and lists of bisimulation, refinement, and
        concrete coupling invariant relations
        (resp.~$\Bs,\Rs,\Is$)
        must all be for the same number of threads.
    \item
        Then, for all threads $i$ in the system:
        \begin{itemize}
            \item The relations
                $\mathcal{B},\mathcal{R},\mathcal{I}$
                for thread $i$ must meet the requirements for
                ``compositional whole-system refinement''
                (\autoref{def:sys-refine-reqs}).
            \item The refinement relation $\mathcal{R}$
                for thread $i$
                must hold initially, i.e.\
                cover its initial thread-private and mode states
                at concrete and abstract level
                (resp.~$\cms_C, \cms_A$),
                for all initial memories that satisfy the concrete
                $\INITparam_C$ requirement.
            \item The abstract bisimulation relation $\mathcal{B}$
                for thread $i$ must hold initially, i.e.\ 
                must relate its initial thread-private and mode state
                to itself, for all pairs of memories that
                are low-equal modulo that mode state,
                and that both satisfy the abstract $\INITparam_A$ requirement.
        \end{itemize}
\end{itemize}

Given all these assumptions, \autoref{thm:sys-refine}
%the whole-system theorem just described
yields a whole-system noninterference property
$\SysSecureParam{\INITparam_C}$ %$\ \cms_{C}
for the resulting
concurrent program (with the list of
initial thread-private and mode states $\cms_C$)
that assumes that the initial memory satisfies $\INITparam_C$.
 % new, presenting simplifications from csf16

\section{Source language: \WhileLang with mutex locks}
% Toby: Note that you can put any technique here, see CSF'16/EuroS&P'18/Thesis
% (Maybe not so much CSF'16, because of the improvements on it.)
\label{sec:source-language}

%In this section we present our compiler's source language, \WhileLang with mutex locks,
In this section, we give a focused presentation
of our compiler's source language,
centered on its properties
that enable \emph{the composition}
of per-thread proofs of CVDNI-preserving refinement to the compiler's
target \RISCLang language.
Our Isabelle/HOL supplement provides full formalisations of its semantics, and
of instances of all
per-thread proof techniques
for proving CVDNI itself
(developed for \WhileLang
by \citet{robs-phd,Murray_SPR_16,Murray16-Dependent-SIFUM-AFP}).

\WhileLang with mutex locks (hereafter \WhileLang) is a generic imperative language with support for conditional looping, consisting of the commands $\cmd$ over arithmetic expressions $\expr$:
\[
\begin{array}{r@{\ }l}
\expr \deftype & n\ |\ v\ |\ \expr\,\oplus\,\expr \\
\cmd \deftype & \Skip\ |\ \Seqg{\cmd}{\cmd}\ |\ \ITEg{\expr}{\cmd}{\cmd}\ |\ \\
         & \Whileg{\expr}{\cmd}\ |\ \Assign{v}{\expr}\ |\ \Stop\ |\ \\
         & \LockAcq{k}\ |\ \LockRel{k}
\end{array}
\]

%The generic type $\expr$ used for both arithmetic and boolean expressions
The language is parameterised over
shared program-variable identifiers
$v \oftype \Var$,
shared lock-variable identifiers
$k \oftype \Lock$,
constant values
$n \oftype \Val$,
and binary arithmetic operators
$\oplus \oftype \Val \Rightarrow \Val \Rightarrow \Val$ that
each have a big-step evaluation semantics;
these induce a big-step evaluation semantics for $\expr$ as a whole.
The commands $\cmd$ then have a small-step operational semantics,
% Overkill back-citation -robs.
%largely unchanged from \citet{Murray_SPR_16,Murray16-Dependent-SIFUM-AFP},
wherein
$\Skip$ and variable assignment $\Assign{v}{\expr}$ execute in one step to
$\Stop$ (which itself does not step to anything);
conditional branch $\IfKw$ steps to the appropriate $\cmd$ depending on
whether its expression evaluates to zero;
and conditional loop $\WhileKw$ steps to an $\IfKw$-conditional between either
(1) the loop body sequenced with a repetition of the $\WhileKw$ command,
or (2) $\Stop$.
Finally the sequential command $\Seqg{c_1}{c_2}$ executes to $c_2$
when $c_1$ executes to $\Stop$,
and to $\Seqg{c_1'}{c_2}$ ($c_1'$ being $c_1$'s destination) otherwise.
Of these aforementioned commands,
only variable assignments can modify the shared memory
(program-variables only),
and none can directly modify the mode state or lock-variables.
%The formal semantics of %the remaining
%all \WhileLang commands
%are available in full detail
%in the supplement Isabelle/HOL material for this paper.
%%we leave to \citet{Murray_SPR_16,Murray16-Dependent-SIFUM-AFP}.

%\WhileLang leaves users free to supply as parameters
%the types and big-step semantics
%of arithmetic and boolean expressions (resp.~$\aexpr, \bexpr$);
%over a set of shared program-variable identifiers $v \oftype \Var$ and a type of constant values $\Val$.
%Here, we
%will fix both %$\aexpr$ and $\bexpr$ to

We will give special
focus to the addition to the \WhileLang language of
%For this version of \WhileLang,
the mutex synchronisation primitives $\LockAcq{k}$ and $\LockRel{k}$, which
are the sole means of modifying lock variables and mode state.
These replace both the ad-hoc mode annotations and the $\Await{v}$ synchronisation primitive that were previously offered for \WhileLang by \citet{Murray_SPR_16}.
After briefly noting here how \WhileLang instantiates the underlying theory from
\autoref{sec:cvdni-notions},
we will present these new primitives' operational semantics,
which depends on the program developer supplying details of the locking discipline as a parameter (\autoref{sec:locking-discipline}) subject to some restrictions (\autoref{sec:locking-restrictions}).
We will then prove that $\GlobalModeUse$
(\autoref{def:global-modes-compatibility})
%part of $\SoundModeUse$ (\autoref{def:sound-mode-use})
%side condition to
%Theorems \ref{thm:com-secure-composes} and \ref{thm:sys-refine}
is invariant for systems of \WhileLang programs
running concurrently (\autoref{sec:global-compatibility-locks}),
subject to some initial conditions (\autoref{sec:init-conds-global-compat}).
Discharging this once-off noncompositional proof obligation
is crucial in enabling both
composition of per-thread noninterference properties
(using \autoref{thm:com-secure-composes}),
and compositional whole-system secure refinement of noninterference
%as we described in \autoref{sec:sys-refine},
down to \RISCLang by our compiler
(using \autoref{thm:sys-refine}).

\WhileLang instantiates the concurrent value-dependent noninterference theory described in \autoref{sec:cvdni-definition}.
%Again in common with \citet{Murray_SPR_16},
% Overkill back-citation removed -robs.
%As in \citet{Murray_SPR_16},
This instantiation assumes that the underlying concurrent execution model
(e.g.\ operating system, scheduler)
for the \WhileLang language prevents threads from seeing
each others' current program location.
Thus the \WhileLang program command $c \oftype \cmd$ being executed
(understood as the current program location)
is modelled as the thread-private state
of the local configuration triple: $\LocalConfWhile{c}{\mds}{\mem}$.
(The subscript $_\defined{w}$ distinguishes
\WhileLang program triples
from \RISCLang ones, which are subscripted $_\defined{r}$.)

To ease formalisation of $\LockAcq{k}$ and $\LockRel{k}$,
we instantiate the shared $\mem \oftype \Mem$ type
as a total mapping from a sum type to values $\Val$.
This sum type, with constructors $\LockCons, \VarCons$,
distinguishes
lock-variable identifiers $k \oftype \Lock$
(which can only be read or written by the lock primitives)
from program-variable identifiers $v \oftype \Var$
(which can be read or written by the rest of the commands).
In Isabelle/HOL's datatype notation, this is:
\[
    \Mem\ \defineq\ (\LockCons\ \Lock\ |\ \VarCons\ \Var) \Rightarrow \Val
\]

\noindent
For readability, we will elide this distinction between $\Lock$ and $\Var$---or applications of their constructors $\LockCons$ and $\VarCons$---%
from the presentation
whenever clear from the context.
 % from thesis
\subsection{Locking discipline and its semantics}
\label{sec:locking-discipline}
% NB: there is a subsection heading hidden in locking-discipline
% Point 1 -robs.
%We assume that the \texttt{While} language program being compiled follows a certain locking discipline.
%Adherence to this locking discipline is verified using a local compliance check (\autoref{sec:local-compliance-locks}).

The program developer
%user of the theory
provides the details of the program's locking discipline in the form of a \emph{lock interpretation} parameter
$\lockinterp \oftype \Lock \Rightarrow (\Var\ \set \times \Var\ \set)$,
% robs: I must word it the following way because if nobody is holding lock l
% which is responsible for NoW for variable x, then they all guarantee not
% to write to x. However, they may all still read from x.
which gives for each lock the two non-overlapping sets of program-variables over which acquiring the lock grants exclusive permission to write, (resp.)~read and write.
For readability, %clarity,
this presentation will elide $\lockinterp$ from the arguments of definitions,
and %also
use the notation $\NoWvarsFunc,\NoRWvarsFunc \oftype \Lock \Rightarrow \Var\ \set$ to refer
%directly
to its $\fst$ and $\snd$ projection.

% Point 2 -robs.
Alongside encoding the mutex primitives' usual effect on control flow---most crucially, $\LockAcq{k}$ should refuse to proceed meaningfully if the lock $k$ is already held---we will now specify for them an evaluation semantics that furthermore encodes the permissions implied by the locking discipline, as assumptions and guarantees expressed in the mode state.
This semantics assumes that, initially, no locks are held, and all threads are
making guarantees not to access the variables they govern
(conditions we will define formally in
\autoref{sec:init-conds-global-compat}).
%Later (in \autoref{sec:}), we will define initial conditions
%wherein all locks begin unlocked,
%and their guarantees are in the mode state for all threads.
%and all threads have the guarantees for all locks.

The following two helpers specify how acquiring (resp.~releasing) a lock affects the mode state under a given lock interpretation $\lockinterp$.
When a thread acquires a lock it gains more assumptions, and makes fewer guarantees about the region of memory concerned:

\begin{definition} [Impact on mode state $\mds$ of acquiring lock $k$]
\label{def:lock-acq-mds}
\begin{align*}
\LockAcqUpd{\mds}{k}\ \defineq\ 
\lambda\ m.\ \caseof{m}\
       & \GuarNoW \Rightarrow \mds\ \GuarNoW - \NoWvars{k} \\
    |\ & \AsmNoW \Rightarrow \mds\ \AsmNoW\ \cup\ \NoWvars{k} \\
    |\ & \GuarNoRW \Rightarrow \mds\ \GuarNoRW - \NoRWvars{k} \\
    |\ & \AsmNoRW \Rightarrow \mds\ \AsmNoRW\ \cup\ \NoRWvars{k}
\end{align*}
\end{definition}

The converse occurs when releasing a lock: the thread drops the assumptions it was making about that region of memory, and once again makes guarantees not to access it.

\begin{definition} [Impact on mode state $\mds$ of releasing lock $k$]
\label{def:lock-rel-mds}
\begin{align*}
\LockRelUpd{\mds}{k}\ \defineq\ 
\lambda\ m.\ \caseof{m}\ 
       & \GuarNoW \Rightarrow \mds\ \GuarNoW\ \cup\ \NoWvars{k} \\
    |\ & \AsmNoW \Rightarrow \mds\ \AsmNoW - \NoWvars{k} \\
    |\ & \GuarNoRW \Rightarrow \mds\ \GuarNoRW\ \cup\ \NoRWvars{k} \\
    |\ & \AsmNoRW \Rightarrow \mds\ \AsmNoRW - \NoRWvars{k}
\end{align*}
\end{definition}

The operational semantics for $\LockAcq{k}$ is then given by two rules: \EvalLockAcq when lock $k$ is available,
and \EvalLockSpin when it is already held.
For these, we use predicate $\EvalLockFunc \oftype \Val \Rightarrow \bool$ with designated constants $\locktrue,\lockfalse \oftype \Val$ to indicate that the lock is, resp.~is not held---i.e.~$\EvalLockFunc(\locktrue) = \True$, and $\EvalLockFunc(\lockfalse) = \False$.\footnote{All three of $\EvalLockFunc,\locktrue,\lockfalse$ are parameters that are set by the user of the theory, with the proviso that their choice of parameters satisfy that $\EvalLockFunc(\locktrue)$ and $\neg \EvalLockFunc(\lockfalse)$ hold as required.}

Apart from impacting the mode state as already specified (by \autoref{def:lock-acq-mds}), attempting to acquire an available lock will succeed in the usual manner, setting the lock-variable to the designated constant ($\locktrue$) to prevent subsequent lock acquisition attempts:
\[
\inferrule{
 \neg \EvalLock{mem}{(\LockCons\ k)}
 \and
 \mem' = \mem[\LockCons\ k \mapsto \locktrue] \\\\
 \and
 \mds' = \LockAcqUpd{\mds}{k}
 } {
 \LocalConfWhile{\LockAcq{k}}{\mds}{\mem}
 \EvalStepWhile
 \LocalConfWhile{\Stop}{\mds'}{\mem'}
 }\ \EvalLockAcq
\]

Attempting to acquire an already-held lock results in a stuttering evaluation step:
\[
\inferrule{
 \EvalLock{mem}{(\LockCons\ k)}
 } {
 \LocalConfWhile{\LockAcq{k}}{\mds}{\mem}
 \EvalStepWhile
 %\LocalConfWhile{\LockAcq{k}}{\mds'}{\mem'}
 \LocalConfWhile{\LockAcq{k}}{\mds}{\mem}
 }\ \EvalLockSpin
\]

Then, the operational semantics for $\LockRel{k}$ is given by two rules, of which only one, $\EvalLockRel$, will ever be used by programs that follow locking discipline.
%(according to the local mode compliance check to be presented in \autoref{sec:local-compliance-locks}).
This rule requires that the mode state $\mds$ is consistent with the present thread having previously acquired the lock $k$:
In short, it should have all the assumptions, but none of the guarantees, associated with the variables governed by the lock.
To specify this, we define the following helper:

\begin{definition} [Mode state is consistent with holding a lock $k$]
\label{def:lock-held-mds-correct}
\begin{align*}
 & \LockHeldMdsCorrect\ \mds\ k\ \defineq\ \\
 & \quad \forall x.\ (x \in\ \NoWvars{k}\ \longrightarrow\ 
              x \notin \mds\ \GuarNoW\ \land\ x \in \mds\ \AsmNoW)\ \land \\
 & \quad\ \quad\ (x \in\ \NoRWvars{k}\ \longrightarrow\ 
              x \notin \mds\ \GuarNoRW\ \land\ x \in \mds\ \AsmNoRW)
\end{align*}
\end{definition}

With that condition satisfied, the $\EvalLockRel$ rule specifies that an $\LockRel{k}$ will proceed successfully, to enact lock release on the memory and mode state as expected: % accordingly:
\[
\inferrule{
 \LockHeldMdsCorrect\ \mds\ k
 \and
 \mem' = \mem[\LockCons\ k \mapsto \lockfalse] \\\\
 \and
 \mds' = \LockRelUpd{\mds}{k}
 } {
 \LocalConfWhile{\LockRel{k}}{\mds}{\mem}
 \EvalStepWhile
 \LocalConfWhile{\Stop}{\mds'}{\mem'}
 }\ \EvalLockRel
\]

To ensure that the \WhileLang evaluation semantics is defined for all possible configurations, the \EvalLockInvalid rule defines a stuttering evaluation step for attempts to $\LockRel{k}$ that violate the locking discipline due to not having previously acquired the lock $k$:
%Program developers can rely on a local mode compliance check
%%(see \ComplyRuleUnlock rule,
%%in \autoref{sec:local-compliance-new-rules})
%to reject programs that misbehave in attempting to do this;
%details are relegated to \citet{robs-phd}.
\[
\inferrule{
 \neg\ \LockHeldMdsCorrect\ \mds\ k %\\\\
 } {
 \LocalConfWhile{\LockRel{k}}{\mds}{\mem}
 \EvalStepWhile
 %\LocalConfWhile{\LockRel{k}}{\mds'}{\mem'}
 \LocalConfWhile{\LockRel{k}}{\mds}{\mem}
 }\ \EvalLockInvalid
\]

As mode state is nominally a form of ghost state, having the operational semantics appear to depend on it in this manner is rather unusual.
To remove the semantics' reliance on ghost state, the program developer must use a check for $\LocalModeUse$ (\autoref{def:local-mode-compliance}) that
%However, as the local mode compliance check will
only ever admits programs that satisfy the $\LockHeldMdsCorrect$ check whenever attempting to $\LockRel{k}$.
% TM's wording:
For such programs, the operational semantics is equivalent to one that (1) omits the $\LockHeldMdsCorrect$ check from the $\EvalLockRel$ rule, and (2) omits the \EvalLockInvalid rule from the \WhileLang-language semantics entirely.
%for such programs, evaluation according to \EvalLockInvalid is not reachable---thus eliminating any effective run-time dependency on the mode state.
An example of such a check is included in our Isabelle/HOL supplement.
%its adaptation to \WhileLang with mutex locks is presented by \citet{robs-phd}.

\subsection{Restrictions on locking disciplines}
\label{sec:locking-restrictions}

Here we lay out some cleanliness conditions on locking disciplines,
giving particular focus to those relevant
to our locking semantics (\autoref{sec:locking-discipline}),
and %that will be relevant
to our verification efforts for \WhileLang's global compositionality property
(\autoref{sec:global-compatibility-locks})
and our compiler (\autoref{chp:wr-compiler}).

% This used to be "3" on the list.
Of these, only one
is a hard consequence of the underlying
CVDNI theory we presented in \autoref{sec:cvdni-notions}:
The per-thread CVDNI property
$\ComSecure$ (\autoref{def:com-secure})
effectively compels us
to enforce that secrets are never allowed to leak into the locking state.
Otherwise, mode state would become tainted
upon any attempt to acquire a lock whose status is secret,
which would violate $\ComSecure$'s requirement
that modes-equality must be maintained at all times
(note the $\LCSameMdsOp$ enforced by $\StrongLowBisimMM$,
\autoref{def:strong-low-bisim-mm}).
To ensure that $\ComSecure$ will always treat the
locking state as an untrusted sink,
we impose the following requirement on the $\dmaFunc$ parameter
supplied by the program developer:

\begin{proposition} [$\dmaFunc$ must permanently assign $\Low$ classification to all lock-variables $k$]
\label{thm:no-H-locks}
% Isabelle name: no_H_locks
%   Imposed by
%       - RISCCompilation (locale risc_compilation)
%         USED by
%          - R_preserves_simple_evB_eq_B_pc_mds
%            when current_instB (RISC) is LockAcq or LockRel.
%            This in turn is used by
%              simpler_refinement_safe_R_simple_evB_eq_B
%            \autoref{thm:simpler-refinement-safe_wr}
%         ONLY.
\[
    \forall k\ \mem.\ \dmaApp{\mem}{(\LockCons\ k)} = \Low
\]
\end{proposition}

% Still to do: 5 and 6
%Next we have made a number of convenient simplifications.
The remaining restrictions are consequences of various
simplifications of convenience. % that we have made.

% This used to be "1" on the list.
First, note that the type signature of the $\lockinterp$ parameter
(given in \autoref{sec:locking-discipline})
only allows locks to govern program variables,
not other locks.
%this simplifies our reasoning in
%\autoref{sec:global-compatibility-locks}.
We justify this simplification with the fact that
if some lock $k$ governed lock $k'$,
then $k$ would already have to be held whenever acquiring $k'$---%
otherwise, the change to $k'$ would violate
a no-write assumption implied by the locking discipline.
This, however, would make $k'$ entirely redundant with $k$.

% This used to be "5" on the list.
% Should we move this to the previous section,
% to where we introduce the lock-acq/rel semantics?
Second, the lock acquisition and release semantics
we gave in \autoref{sec:locking-discipline}
is rather simplified, in that
releasing a lock will drop
the assumptions of all its variables
from the mode state,
even if another lock for that variable is still held!
Thus, we signal that it
only works for disciplines wherein
no more than one lock governs each program variable,
%Therefore we must assert the following, %We must assert this, because
by asserting:

\begin{proposition} [No variable can be managed by more than one lock]
\label{thm:lone-lock-per-var}
\begin{align*}
    & \forall v\ k.\ v \in \NoWvars{k}\ \cup\ \NoRWvars{k}\ \longrightarrow \\
    & \qquad (\forall k'.\ v \in \NoWvars{k'}\ \cup\ \NoRWvars{k'}\ \longrightarrow\ k' = k)
\end{align*}
\end{proposition}

\noindent
We believe that it would be feasible to
relax \autoref{thm:lone-lock-per-var}, by generalising \WhileLang's
locking semantics to allow disciplines %locking disciplines wherein,
wherein %for example, 
multiple locks must be held to access a given variable.
% This belabors the point a bit. -robs.
%This would require implementing more sophisticated bookkeeping
%to ensure the mode state of a variable is only impacted by the release
%of the last lock governing it.
%
To satisfy CVDNI-preserving refinement
(particularly \autoref{def:preserves_modes_mem}),
a compiler would need to %will then need to preserve the operations on the
preserve the lock memory operations that
% The classification doesn't actually matter, here,
% because preserves_modes_mem is agnostic of the classification. -robs.
%statically $\Low$-classified lock variables (due to \autoref{thm:no-H-locks})
%lock variables that would
implement the more sophisticated bookkeeping needed,
as ours does for the current, much simpler locking semantics.

% This used to be "6" on the list.
Next, we assume that the program developer has not specified any
``vacuous'' locks
(i.e.~ones that govern no variables),
and that all locks grant at most one of $\AsmNoW$ or $\AsmNoRW$
(i.e.~not both)
on any given variable.
These two assumptions allow us to exclude various pathological cases from
our reasoning in \autoref{sec:global-compatibility-locks}
and \autoref{chp:wr-compiler}, respectively:

\begin{proposition} [Every lock governs access to some variable]
\label{thm:no-vacuous-locks}
% Isabelle name: no_vacuous_locks
%   Imposed by
%       - GloballySoundWhileLockUse (locale sifum_global_modes)
%         USED by
%          - locks_held_consistent
%            (that lock_held_mds_correct, lock_held_mds_not_correct
%             cannot be true simultaneously;
%             used by lock_acq_preserves_correctness
%                     lock_rel_preserves_correctness
%                     non_lock_commands_preserve_correctness)
%         ONLY.
%       - GloballySoundRISCLockUse (locale risc_global_modes), USED
%         ^^^ Exactly the same as for the WhileLang case.
\[
    \forall k.\ \NoWvars{k}\ \cup\ \NoRWvars{k} \neq \emptyset
\]
\end{proposition}

% This used to be "7" on the list.
%The two lock interpretation sets for a given lock are never overlapping:
\begin{proposition}[The lock interpretation sets for any given lock $k$ do not overlap]
\label{thm:lock-interp-no-overlap}
% Isabelle name: lock_interp_no_overlap
%   Imposed by
%       - RISCCompilation (locale risc_compilation)
%         USED by
%            This is the only place that needs a reference to this:
%            vvvvvvvvvvvvvvvvvvvvvvvvvvvvvvvvvvvvvvvvvvvvvvvvvvvvv
%          - no_locks_acquired_C0_mds0_init_reqs
%            (that with no locks acquired,
%             compiled_cmd_init_reqs holds)
%            \autoref{thm:init-conds-consistent}
%            Note: It's only needed for this to satisfy lock_consistent_mds,
%            a condition we don't even seem to be talking about!
%            ^^^^^^^^^^^^^^^^^^^^^^^^^^^^^^^^^^^^^^^^^^^^^^^^^^^^^
%          - eval_r_preserves_lock_consistent_mds
%            (used by reach_r_preserves_lock_consistent_mds, in turn
%             used by if_c1, if_c2 cases of R_preservation lemma)
%          - lock_acq and lock_rel case of R_preservation lemma
%            \autoref{thm:R_wr-refinement-abs-steps_wr}
%            Note: I managed to remove the dependency
%            of R_preservation and eval_r_preserves_lock_consistent_mds
%            on lock_interp_no_overlap!
%         ONLY.
%   Note that it is enforced by the lock_rel_type rule in
%       - WhileLockTypeSystem (locale sifum_types)
%   and then used (see name "S_no_dups") in tyenv_sec_lock_rel
\[
    \forall k.\ \NoWvars{k}\ \cap\ \NoRWvars{k} = \emptyset
\]
\end{proposition}

The final two restrictions simplify the possible interactions between
locks and control variables:
%in some of the compiler verification proofs of \autoref{chp:wr-compiler}.
%They play a more major role in simplifications to
%%In addition, they enable some simplifications to
%the specific proof techniques we used to obtain per-thread CVDNI
%and $\LocalModeUse$ (\autoref{def:local-mode-compliance})
%for our case study program of \autoref{chp:cddc};
%%to give our compiler of \autoref{chp:wr-compiler} something to preserve;
%however, we relegate details on that %of their impact on those proof techniques
%to \citet{robs-phd}.
%
We disallow locks from being control variables, and require variables to be
governed by the same lock as their control variables.
In particular, they will help us establish
(in \autoref{sec:compiler-outputs})
%be used in a stand-alone proof establishing
that the compiler produces programs that satisfy
$\LocalModeUse$.

First, recall we mentioned
that, as part of $\LocalModeUse$ (\autoref{def:local-mode-compliance}),
the \DoesntRorWText assertions
entail that any guarantees not to access some variable $v$ will effectively apply also to all of $v$'s control variables.
Disallowing lock-variables from being control variables %will thus simplify the reasoning for $\LocalModeUse$ in \autoref{sec:local-compliance-locks},
thus %simplifies the proof of soundness of the local compliance check,
ensures that $\LockAcq{k}$ and $\LockRel{k}$,
because they only access lock-variable $k$,
cannot violate \DoesntRorWText for any program-variables:

% This used to be "2" on the list.
\begin{proposition} [Lock-variables $k$ cannot be control variables]
\label{thm:no-locks-in-C}
% Isabelle name: no_locks_in_C
%   Imposed by
%       - LocallySoundWhileLockUse (locale sifum_modes),
%         (NB: This is that locale's only assume-requirement)
%         USED by
%          - lock_acq_doesnt_modify_progvars
%          - lock_acq_doesnt_read_progvars
%          - lock_rel_doesnt_modify_progvars
%          - lock_rel_doesnt_read_progvars
%         ONLY.
%       - RISCCompilation (locale risc_compilation)
%         USED by
%            Its only use in the theorems of this paper:
%            vvvvvvvvvvvvvvvvvvvvvvvvvvvvvvvvvvvvvvvvvvvvvvvvvvvvv
%          - lock_acq_doesnt_modify_vars (RISC)
%          - lock_acq_doesnt_read_vars (RISC)
%          - lock_rel_doesnt_modify_vars (RISC)
%          - lock_rel_doesnt_read_vars (RISC)
%            These four are then used by the LockAcq/Rel cases of:
%              establishes_local_guarantee_compliance_R
%              \autoref{thm:Rwr-respects-own-guarantees}
%            ^^^^^^^^^^^^^^^^^^^^^^^^^^^^^^^^^^^^^^^^^^^^^^^^^^^^^
%          - Note: I managed to remove the dependency of
%              R_preserves_simple_evB_eq_B_pc_mds
%            on no_locks_in_C - it doesn't need it!
%            This in turn was used by
%              simpler_refinement_safe_R_simple_evB_eq_B
%            \autoref{thm:simpler-refinement-safe_wr}
%         ONLY.
%   Note it can be derived from C_vars_correct, as it is in CSF'16 by
%       - WhileLockTypeSystem (locale sifum_types), USED
%         USED by
%          - lock_acq and lock_rel cases of R_typed_step
%         ONLY.
\[
    \forall k.\ (\LockCons\ k) \notin \Cset
\]
\end{proposition}

Finally, requiring variables to be
%variables are always
governed by the same lock as their control variables
effectively
ensures they %variables and their control variables are always locked
are always locked simultaneously.
Apart from making it easier for %to %helping prove compliance with the corresponding
programs to satisfy
%comply with the guarantees demanded by
$\LocalModeUse$,
% This is not really the reason that helps local-mode-compliance,
% but we might as well mention it. -robs.
this also
naturally prevents leaks caused by other threads %from %violations of CVDNI
changing a variable's classification to $\Low$ when it
still contains $\High$ data:
%and ensures that not holding a lock simultaneously implies
%the same guarantees on variables and their control variables,
%as will be demanded by $\LocalModeUse$:

% This used to be "4" on the list.
% Appears to be relevant to preservation of locally sound mode use.
%This restriction simplifies matters in a number of proofs of
%\autoref{chp:wr-compiler}
%by ensuring that variables and their control variables are always locked simultaneously:
%a lock-variable $k$ governing access to a program-variable $v$ must govern the same kind of access to all of $v$'s control variables:
\begin{proposition} [Variables are always governed by the same lock as their control variables]
\label{thm:lock-interp-C-vars}
% Isabelle name: lock_interp_C_vars
%   Imposed (NEW since CSF'16) by
%       - WhileLockTypeSystem (locale sifum_types), USED
%         USED by
%          - types_stable_mode_update
%         ONLY.
%       - RISCCompilation (locale risc_compilation)
%         USED by
%            This refers to detail elided from the paper:
%            vvvvvvvvvvvvvvvvvvvvvvvvvvvvvvvvvvvvvvvvvvvvvvvvvvvvv
%          - control_vars_mds
%            (that in a lock_consistent_mds, control vars must have
%             the same modes as the vars they're controlling)
%          - compiled_expr_respects_guarantees
%            (that the RISC output of a compiled expression
%             respects its own guarantees;
%             used in the guise of control_vars_mds)
%            ^^^^^^^^^^^^^^^^^^^^^^^^^^^^^^^^^^^^^^^^^^^^^^^^^^^^^
%            -----------------------------------------------------
%            This is the only place that needs a reference to this:
%            vvvvvvvvvvvvvvvvvvvvvvvvvvvvvvvvvvvvvvvvvvvvvvvvvvvvv
%          - establishes_local_guarantee_compliance_R
%            \autoref{thm:Rwr-respects-own-guarantees}
%            (that a RISC configuration in the refinement relation
%             respects its own guarantees;
%             used to derive contradictions in the assign_store case)
%            ^^^^^^^^^^^^^^^^^^^^^^^^^^^^^^^^^^^^^^^^^^^^^^^^^^^^^
\begin{align*}
    \forall c\ v\ k.\ \VarCons\ c \in \Cvars\ (\VarCons\ v) \longrightarrow\ 
    &(c \in \NoWvars{k} = v \in \NoWvars{k})\ \land \\
    &(c \in \NoRWvars{k} = v \in \NoRWvars{k})
\end{align*}
\end{proposition}

 % from thesis
% robs: I think we need the more long-winded section names (than in thesis)
%       because in JFP, the reader gets no per-paper ToC to orient them.
\subsection{Proof of global modes compatibility as an invariant}
\label{sec:global-compatibility-locks}
\label{sec:invariant-global-compat} % redundant label from merged subsection
% Toby: Pitch global soundness details as one (of many potential)
%       methods of getting whole-system proof?
This section will present proof that $\GlobalModeUse$ (\autoref{def:global-modes-compatibility})
holds as an invariant
for concurrent \WhileLang programs
(\autoref{sec:invariant-global-compat})
when initialised to have no locks held (\autoref{sec:init-conds-global-compat}).
Consequently, it is sufficient for a developer to use
a local compliance check \citep{robs-phd}
%of \autoref{sec:local-compliance-locks}
%to establish $\LocalModeUse$
%is enough to for a developer
to obtain the $\SoundModeUse$ condition (\autoref{def:sound-mode-use}) needed for per-thread security proofs to be compositional via
%per-thread security to compose into whole-system security via
\autoref{thm:com-secure-composes}.

%The global component of $\SoundModeUse$
%short for global compatibility of assume--guarantee modes.
Recall from \autoref{sec:cvdni-definition} that
%$\GlobalModeUse$
this compatibility requirement
formalises that for all reachable global configurations of a concurrent program,
%a condition
%called $\CompatibleModes$
%must hold whereby
any assumptions made by any of the threads must be met by corresponding guarantees made by all of the other threads:
%---i.e.~the condition just described, formalised as $\CompatibleModes$ below, must be invariant:
%Formally, all of this was defined as follows:

\defglobalmodescompatibility*

The approach to establish $\GlobalModeUse$ here will be to define three \emph{mode management requirements} that taken together imply $\CompatibleModes$, and to prove them invariant for concurrent \WhileLang programs when initialised such that they hold to begin with.

The first of these pertains to variables whose access is governed by some lock, according to the locking discipline.
To define it,
we need, alongside $\LockHeldMdsCorrect$ (\autoref{def:lock-held-mds-correct}) from \autoref{sec:locking-discipline},
a predicate that specifies the correct mode state
for \emph{not} holding a lock $k$:
%$\LockNotHeldMdsCorrect\ \mds\ k$, which we will use to specify that
%whenever a thread is \emph{not} holding a given lock $k$, its mode state $\mds$
It should make all of the guarantees, and have none of the assumptions associated with the variables governed by $k$.%
\footnote{Note that this not merely the negation of $\LockHeldMdsCorrect\ \mds\ k$ (\autoref{def:lock-held-mds-correct})!}
Stated formally:
%it is necessary also to specify when a mode state is consistent with \emph{not} holding a lock $k$.

\begin{definition} [Mode state is consistent with \emph{not} holding a lock $k$]
\label{def:lock-not-held-mds-correct}
\begin{align*}
 & \LockNotHeldMdsCorrect\ \mds\ k\ \defineq\ \\
 & \quad \forall x.\ (x \in \NoWvars{k}\ \longrightarrow\ 
                x \in \mds\ \GuarNoW\ \land\ x \notin \mds\ \AsmNoW)\ \land \\
 & \quad\ \quad\ (x \in \NoRWvars{k}\ \longrightarrow\ 
                x \in \mds\ \GuarNoRW\ \land\ x \notin \mds\ \AsmNoRW)
\end{align*}
\end{definition}

\noindent
Note that our simplifying exclusion of ``vacuous'' locks
(\autoref{thm:no-vacuous-locks})
%assumption that no locks are vacuous
ensures we never have to deal with a case where
$\LockHeldMdsCorrect\ \mds\ k$ and $\LockNotHeldMdsCorrect\ \mds\ k$
hold simultaneously.

The requirement on global configurations regarding these lock-managed variables is then as follows:
If and only if a given lock is held by anybody, then exactly one thread has a mode state consistent with holding it; furthermore, all other threads will have a mode state consistent with not holding it.
Formally, with
$\mdssOf{\gc}\,\defineq\,\map\ \mdsOf{({\cmsOf{\gc}})}$:

% Isabelle name: lock_managed_vars_mds_mem_correct
\begin{definition} [Lock-managed variable modes are compatible with memory]
\label{def:lock-managed-vars-mds-mem-correct}
\begin{align*}
    &\LockManagedVarsMdsMemCorrect\ \gc\ \defineq\ \\
    &\qquad \forall k.\ \IfThen{\EvalLock{(\memOf{\gc})}{k}} \\
    &\qquad \qquad \exists!i.\ i < \length\ (\cmsOf{\gc})\ \land \\
    &\qquad \qquad \qquad \LockHeldMdsCorrect\ \listderef{(\mdssOf{\gc})}{i}\ k\ \land \\
    &\qquad \qquad \qquad (\forall j < \length\ (\cmsOf{\gc}).\ i \neq j \longrightarrow \\
    &\qquad \qquad \qquad \qquad \LockNotHeldMdsCorrect\ \listderef{(\mdssOf{\gc})}{j}\ k) \\
    &\qquad \qquad \Else\ \forall i < \length\ (\cmsOf{\gc}). \\
    &\qquad \qquad \qquad \LockNotHeldMdsCorrect\ \listderef{(\mdssOf{\gc})}{i}\ k
\end{align*}
\end{definition}

The second requirement pertains to variables whose access is entirely ungoverned by any locks in the locking discipline.
For these we specify a more direct check that if any thread in the global configuration has an assumption about access to any of these variables, then all other threads must be providing the corresponding guarantee to that assumption:

% Isabelle name: unmanaged_var_modes_compatible
\begin{definition} [Unmanaged variable modes are compatible]
\label{def:unmanaged-var-modes-compat}
\begin{align*}
    &\UnmanagedVarModesCompatible\ \gc\ \defineq\ 
      \forall i\ x.\ i < \length\ (\mdssOf{\gc}) \longrightarrow \\
    & \qquad (x \notin \bigcup_{k\oftype\Lock} \NoRWvars{k} \longrightarrow \\
    & \qquad \qquad (x \in \listderef{(\mdssOf{\gc})}{i}\ \AsmNoRW \longrightarrow \\
    & \qquad \qquad \qquad (\forall j < \length\ (\mdssOf{\gc}).\ j \neq i \longrightarrow
             x \in \listderef{(\mdssOf{\gc})}{j}\ \GuarNoRW)))\ \land \\
    & \qquad (x \notin \bigcup_{k\oftype\Lock} \NoWvars{k} \longrightarrow \\
    & \qquad \qquad (x \in \listderef{(\mdssOf{\gc})}{i}\ \AsmNoW \longrightarrow \\
    & \qquad \qquad \qquad (\forall j < \length\ (\mdssOf{\gc}).\ j \neq i \longrightarrow
                    x \in \listderef{(\mdssOf{\gc})}{j}\ \GuarNoW)))
\end{align*}
\end{definition}

Also proved invariant is a third, minor property that enforces globally that no assumptions or guarantees are ever recorded regarding access to lock-variables:

\begin{definition} [No assumptions and guarantees on lock variables]
\label{def:no-lock-mds-gc}
\begin{align*}
    \NoLockMds\ \mds\ &\defineq\ \forall l\ m.\ \LockCons\ l \notin \mds\ m \\
    \NoLockMdsGc\ \gc\ &\defineq\ \forall \mds \in \toSet\ (\mdssOf{\gc}).\ \NoLockMds\ \mds
\end{align*}
\end{definition}

\noindent This follows trivially from
%the restriction
(1) our simplification (discussed in \autoref{sec:locking-restrictions})
%that the lock interpretation parameter
only to allow
%permit
locks to protect access to program variables and not other locks, and
(2) the resulting fact that no \WhileLang primitives ever touch any mode state pertaining to lock variables.
Thus, further details on this third management requirement will be elided.

We then have straightforwardly from their definitions that  together, these three mode management requirements imply compatible modes
%$\CompatibleModes$
for a given global configuration:
% Isabelle name: management_requirements_ensure_compatibility
\begin{lemma} [Management requirements ensure compatibility]
\label{thm:management-requirements-ensure-compatibility}
\[
\inferrule{
    \LockManagedVarsMdsMemCorrect\ \gc
    \and
    \UnmanagedVarModesCompatible\ \gc \\
    \and
    \NoLockMdsGc\ \gc
} {
    \CompatibleModes\ (\mdssOf{\gc})
}
\]
\end{lemma}

Proofs of invariance then proceed by induction over the single-step evaluation semantics of an arbitrary thread taking a step to progress the system to a new global configuration.

For the first management requirement (\autoref{def:lock-managed-vars-mds-mem-correct}):

% Isabelle name: eval_preserves_correctness
\begin{lemma} [Single-step preservation of $\LockManagedVarsMdsMemCorrect$]
\[
\label{thm:single-step-managed}
\inferrule{
    \LockManagedVarsMdsMemCorrect\ (\cms, \mem) \\
    \and
    \LocalConfWhile{c_i}{\mds_i}{\mem} \EvalStepWhile
        \LocalConfWhile{c_i'}{\mds_i'}{\mem'}
    \and
    i < \length\ \cms \\
    \and
    \cms' = \listsubst{\cms}{i}{(c_i',\mds_i')}
    \and
    \listderef{\cms}{i} = (c_i, \mds_i)
} {
    \LockManagedVarsMdsMemCorrect\ (\cms', \mem')
}
\]
\end{lemma}
\begin{proof}
    By induction over the single-threaded evaluation semantics of the program at index $i$ that is taking a step.

    $\LockAcq{k}$ preserves the property because it only allows a thread to set lock $k$'s memory if it is not already set -- it would then become the single unique thread whose mode state is consistent with holding $k$.
    Otherwise, the mode states and memory remain unchanged.

    Similarly, $\LockRel{k}$ preserves the property because its only possible change is to unset lock $k$'s memory, and return the unique thread holding lock $k$ to a mode state consistent with not holding $k$.

    The other \WhileLang commands preserve the property because they do not touch the mode state nor any lock-variables.
\end{proof}

For the second management requirement (\autoref{def:unmanaged-var-modes-compat}):

% Isabelle name: eval_preserves_unmanaged_var_modes_compatibility
\begin{lemma} [Single-step preservation of $\UnmanagedVarModesCompatible$]
\label{thm:single-step-unmanaged}
\[
\inferrule{
    \UnmanagedVarModesCompatible\ (\cms, \mem) \\
    \and
    \LocalConfWhile{c_i}{\mds_i}{\mem} \EvalStepWhile
        \LocalConfWhile{c_i'}{\mds_i'}{\mem'}
    \and
    i < \length\ \cms \\
    \and
    \cms' = \listsubst{\cms}{i}{(c_i',\mds_i')}
    \and
    \listderef{\cms}{i} = (c_i, \mds_i)
} {
    \UnmanagedVarModesCompatible\ (\cms', \mem')
}
\]
\end{lemma}
\begin{proof}
    Again, by induction over the single-threaded evaluation semantics of the program at index $i$ that is taking a step.

    We prove and use lemmas that $\LockAcq{k}$ and $\LockRel{k}$ do not touch any mode state pertaining to variables that are unmanaged by any locks, and that the remaining \WhileLang commands do not touch the mode state at all.
    Therefore evaluation steps cannot possibly have any effect on the compatibility of modes on these variables.
\end{proof}

These single-step evaluation results lift easily to invariance results over the global multi-step evaluation semantics quantified over arbitrary schedules.
These invariance results, % (the full details of which we leave to the Isabelle formalisation),
with the fact that the management requirements ensure compatibility (\autoref{thm:management-requirements-ensure-compatibility}), yield in a straightforward manner the desired global compatibility invariance theorem:

\begin{theorem} [Mode management requirements ensure global compatibility]
\label{thm:soundness-mode-management}
\[
\inferrule{
    \LockManagedVarsMdsMemCorrect\ \gc
    \and
    \UnmanagedVarModesCompatible\ \gc \\
    \and
    \NoLockMdsGc\ \gc
} {
    \GlobalModeUse\ \gc
}
\]
\end{theorem}
 % from thesis
\subsection{Initial conditions ensuring global modes compatibility}
\label{sec:init-conds-global-compat}
%Having defined requirements under which global modes compatibility is invariant for \WhileLang,
We now define conditions on memory and mode state consistent with no locks being held, and show that initialising a system under these conditions is enough to satisfy the global compatibility part (\autoref{def:global-modes-compatibility}) of the $\SoundModeUse$ side condition (\autoref{def:sound-mode-use}) of the compositionality theorem for our security property (\autoref{thm:com-secure-composes}).
%to satisfy the requirements we just showed (in \autoref{sec:global-compatibility-locks}) ensure global modes compatibility for \WhileLang.

We define the following predicate for initial memory:

\begin{definition} [A requirement for initial memory that no locks are held]
\label{def:no-locks-held}
\[
    \NoLocksHeld\ \mem\ \defineq\ \forall k.\ \neg \EvalLock{\mem}{k}
\]
\end{definition}

We then define an initial mode state $\InitMds \oftype \Mode \Rightarrow \Var\ \set$ that provides all guarantees demanded by the lock interpretation parameters $\NoWvarsFunc,\NoRWvarsFunc$ (described in \autoref{sec:locking-discipline}) for all lock variables in the system, and makes no assumptions:

\begin{definition} [Initial mode state $\InitMds$]
\label{def:init-mds}
\begin{align*}
\InitMds \defineq\ 
\lambda\ m.\ \caseof{m}\
       & \GuarNoW \Rightarrow \bigcup_{k\oftype\Lock} \NoWvars{k} \\
    |\ & \GuarNoRW \Rightarrow \bigcup_{k\oftype\Lock} \NoRWvars{k} \\
    |\ & \AsmNoW \Rightarrow \emptyset \\
    |\ & \AsmNoRW \Rightarrow \emptyset
\end{align*}
\end{definition}

We are then able to show that these conditions are enough to satisfy the requirements we just showed (in \autoref{sec:global-compatibility-locks}) ensure global modes compatibility for \WhileLang:

% Isabelle name: no_locks_empty_mds_management_conds
% Isabelle name: no_locks_empty_mds_global_soundness
\begin{lemma} [Initialising with $\NoLocksHeld, \InitMds$ ensures global modes compatibility]
\label{thm:init-conds-global-compat}
\[
\inferrule{
    \NoLocksHeld\ \mem \and
    \forall (c, \mds) \in \toSet\ \cms.\ \mds = \InitMds
} {
    \GlobalModeUse\ (\cms, \mem)
}
\]
\end{lemma}
\begin{proof}
    \autoref{thm:soundness-mode-management} obliges us to show that the mode management conditions (Definitions \ref{def:lock-managed-vars-mds-mem-correct}, \ref{def:unmanaged-var-modes-compat}, and \ref{def:no-lock-mds-gc}) hold.
    This follows straightforwardly from all the relevant definitions.
\end{proof}
 % from thesis

\section{Target language: \RISCLang with mutex locks}
\label{sec:target-language}
Here we introduce \RISCLang with mutex locks (hereafter \RISCLang),
the target of
our compiler.
%the \Covern \WRCompiler.
%The \WRCompiler targets \RISCLang with mutex locks (hereafter \RISCLang),
This is
a generic RISC-style assembly language
based on
the \RISCLang architecture
targeted by
the compilation scheme of
\citet{Tedesco16}.
A \RISCLang program text
is a list of \RISCLang instructions $I$, each optionally associated with a label:

\[
\begin{array}{r@{\ }l}
I \deftype & [l :] B \\
B \deftype & \Load\ r\ v\ |\ \Store\ v\ r\ |\ \Jmp\ l\ |\ \Jz\ l\ r\ |\ \Nop \\
      & \MoveK\ r\ n\ |\ \MoveR\ r\ r\ |\ \Op\ \oplus\ r\ r \\
      & \RISCLockAcq\ k\ |\ \RISCLockRel\ k
\end{array}
\]

Here we fix the types of the constant values $n \oftype \Val$, binary arithmetic operators
$\oplus \oftype \Val \Rightarrow \Val \Rightarrow \Val$,
shared program variables $v \oftype \Var$, and shared lock variables $k \oftype \Lock$ to be the same as those for the source \WhileLang language being compiled.
Thus, the only new types here compared to \autoref{sec:source-language}
are for the register identifiers $r \oftype \Reg$,
and labels $l \oftype \Lab$.

%Apart from this notational adaptation to the configuration triple format for CVDNI proofs,
%Apart from the new $\RISCLockAcq$ and $\RISCLockRel$ instructions,
\RISCLang has a small-step operational semantics that is
largely unchanged from %that of
%evaluation semantics follows that of
%the \RISCLang target architecture of
\citet{Tedesco16}, in that
each step updates a distinguished \emph{program counter} register,
which captures the current thread's program location
as an index into its \RISCLang program text.
%on which it is based.
%to which we relegate further details.
The instructions
%This includes
$\MoveK$, $\MoveR$,
%and direct-addressing
$\Load$, and $\Store$,
for moving values to and between the registers and shared memory,
and the ``no-op'' instruction $\Nop$,
all increment the program counter;
the ``jump if zero'' instruction $\Jz\ l\ r$
updates it to
the index of the instruction at $l$ if $r$ contains zero (else increments it);
the unconditional $\Jmp\ l$ does so unconditionally.

Modifying this instruction set from \citet{Tedesco16},
we then customise the $\Op$ instruction, and add %$\ \oplus\ r\ r$,
$\RISCLockAcq\ k$ and $\RISCLockRel\ k$ instructions,
to have semantics mirroring those of $\oplus$, % each $\oplus$ binary operation,
$\LockAcq{k}$, and $\LockRel{k}$ from \WhileLang respectively.
Whereas the \RISCLang equivalents for the
\EvalLockAcq and
\EvalLockRel
evaluation rules
(described by \autoref{sec:locking-discipline} for the \WhileLang language)
increment the program counter,
those for
\EvalLockSpin and
\EvalLockInvalid
leave it unchanged.
There is no \RISCLang evaluation rule that changes the program text.

Although it has only direct-addressing $\Load$ and $\Store$ instructions,
our \RISCLang target language is adequate for implementing
all features of \WhileLang present in \autoref{sec:source-language},
with the big-step semantics of $\expr$
replaced by small-step operations on registers.
We relegate \RISCLang's full formal semantics
%the \RISCLang language
to this paper's supplement Isabelle/HOL material.

%Here, we add lock-based synchronisation operations
%$\RISCLockAcq\ k$ and $\RISCLockRel\ k$,
%which we give
%the same operational semantics on shared memory and mode state
%as the $\LockAcq{k}$ and $\LockRel{k}$ primitives from \WhileLang
%(\autoref{sec:source-language}).
%
Our defining %Our addition of % the lock-based synchronisation operations
$\RISCLockAcq\ k$ and $\RISCLockRel\ k$
to have the same operational semantics on shared memory and mode state
as \WhileLang's $\LockAcq{k}$ and $\LockRel{k}$
has two consequences:
\begin{itemize}
    \item Our compiler will expect the program developer to supply
        the details of the locking discipline
        for the \WhileLang program being compiled,
        so as to be able to ensure that the \RISCLang program it produces
        follows the same discipline.
    \item %Then, owing to having the same semantics
        %on shared memory and mode state as for the \WhileLang language,
        We then have that
        global compatibility is invariant for \RISCLang execution,
        by a near-identical argument to the one we presented in
        \autoref{sec:invariant-global-compat},
        when initialised with the conditions we presented in
        \autoref{sec:init-conds-global-compat}.
        This presents one option for obtaining \RISCLang-level
        composition of per-thread noninterference properties;
        however, invoking it directly will not be necessary
        when using the compositional whole-system secure refinement
        method of \autoref{sec:sys-refine}.
        %which preserves the corresponding \WhileLang-level result automatically.
        (The alternative options and their application
        will be demonstrated further, respectively
        in \autoref{sec:compiler-outputs}, \autoref{sec:case-study-wr}.)
\end{itemize}

As for \WhileLang in \autoref{sec:source-language}, we instantiate
here for \RISCLang
the CVDNI theory of \citet{Murray_SPR_16}
as recalled in \autoref{sec:cvdni-definition},
assuming that the underlying concurrency model (e.g.~OS, scheduler etc.)~prevents one thread from reading the program text of another.
For \RISCLang, we furthermore assume that the context switching mechanism
ensures effectively that no thread can read or interfere with the contents
of the registers (including the program counter)
when active for another thread.
Based on these assumptions, we model all three of the program counter register's value $\pc \oftype \nat$, \RISCLang program text $P \oftype I\ \listtype$, and register bank $\regs \oftype \Reg \Rightarrow \Val$, as thread-private state in the local configuration triple:
$\LocalConfRISC{\pc}{P}{\regs}{\mds}{\mem}$.
(We use the subscript $_\defined{r}$ to distinguish \RISCLang triples.)
 % from thesis

%\section{Noninterference-preserving compiler for mixed-sensitivity concurrent programs}
\section{Verified secure compiler for mixed-sensitivity concurrent \WhileLang programs}
\label{chp:wr-compiler}

This section presents the \Covern \WRCompiler:
the first compiler proved to preserve proofs of noninterference
for mixed-sensitivity concurrent programs.
By using assume--guarantee modes \citep{Mantel_SS_11} %Murray_SPR_16
and the decomposition principle of \autoref{sec:just-decomp-reqs}
to prove it introduces (resp.)~%
no race conditions or timing leaks,
we demonstrate the \emph{applicability to compiler verification}
of the CVDNI-preserving refinement notion of \autoref{sec:just-cube-reqs}
originally posed by \citet{Murray_SPR_16}.
Here the decomposition principle (\autoref{fig:decomp})
is crucial because, in separating the concern of preventing new timing leaks,
it avoids directly having to prove
the cube-shaped refinement diagram (\autoref{fig:coupling-inv-pres})
arising from its need to preserve
a 2-safety hyperproperty \citep{Terauchi05,Clarkson10}.

To preserve security for mixed-sensitivity concurrent programs,
CVDNI-preserving refinement demands small-step preservation
of \emph{the contents of all shared memory locations}
including those that control value-dependent classifications
and implement locks.
As it is unusual for verified compilers
to make such promises,
%in terms of the contents of memory
%(as opposed to the operations on them),
we show that a valid approach is
to take advantage of CVDNI's assume--guarantee framework to:
\begin{enumerate}
    \item \emph{test and preserve any absence of race conditions}
        implied (via the framework)
        by mutex lock-based synchronisation of access to such locations,
        and then
    \item \emph{use this absence of race conditions} to establish
        the small-step preservation of their contents demanded for
        security-preserving refinement.
\end{enumerate}
\noindent
In doing so, we prove that some optimisations the \WRCompiler
performs with its knowledge of the locking discipline---%
it avoids unnecessary $\Load$s
and recalculation of common subexpressions over shared memory when locked---%
are safe to allow without violating CVDNI.

In preserving CVDNI, the \WRCompiler
preserves security proofs that are produced by
the program verification techniques of \citet{robs-phd}
for \WhileLang with mutex locks,
which in turn were adapted from
\citet{Murray_SPR_16,Murray16-Dependent-SIFUM-AFP}.
We will present such an application of our compiler, to a
case study program verified using these techniques,
in \autoref{chp:cddc}.

\autoref{sec:compiler-impl} will focus on the \WRCompiler's particular
adaptations to CVDNI %concurrent value-dependent noninterference
(beyond the fault-resilient noninterference
targeted by the original compilation scheme of \citet{Tedesco16}),
%These are %These adaptations come
in the form of
static checks and invariants
that (resp.)~test for and maintain
the absence of race conditions on lock-protected shared variables.
\autoref{sec:wr-compiler-no-hbranch} formalises a ban,
preserved by the \WRCompiler,
on secret-dependent control flow.
\autoref{sec:proof-cvdni-compilation} then presents formal proof
(structured by our decomposition principle of \autoref{sec:just-decomp-reqs})
that the \WRCompiler implements CVDNI-preserving refinement.
\autoref{sec:compiler-outputs} ultimately presents proofs of overall security preservation results useful to users of the \WRCompiler:
Namely, it can be used either to preserve security down to \RISCLang for an entire concurrent \WhileLang-language program, or to preserve the per-thread security for threads that will be run alongside others written directly in the \RISCLang-language.

\subsection{Preserving race-free expression evaluation}
\label{sec:compiler-impl}
\label{sec:tracking-regs-stability} % redundant label from thesis
% NB: there are subsection headings hidden in compiler-impl
Recall from \autoref{sec:decomp-reqs} that CVDNI-preserving refinement
\citep{Murray_SPR_16}
demands that all shared memory contents be preserved,
between each target-
and source-language configuration that it relates.
This is security critical for mixed-sensitivity concurrent programs, as it ensures that any future influence of those contents on value-dependent classifications (via control variables) or readability by other threads (in the case of the \WhileLang and \RISCLang languages, via lock variables) is preserved.

The \WRCompiler's approach to preserving the contents of shared memory
is to ensure:
\begin{enumerate}
\item That values calculated by expressions are \emph{preserved}
by compilation---that is, they have the same value
when written back to shared memory (or conditionally branched on)
by the \RISCLang program,
as they did in the original \WhileLang program; and
\item That expression evaluation is \emph{race-free}---that is,
free of any race conditions
with other threads that would render the calculated expression inaccurate.

To this end,
the \WRCompiler requires
of the original \WhileLang program
that whenever each thread
attempts to evaluate an expression, it must hold locks ensuring the stability of \emph{all} variables referenced by the expression.
\end{enumerate}

\noindent
Thus, its knowledge and enforcement of the locking discipline is crucial, not
only to show that its optimisations preserve CVDNI,
but that \emph{any meaningful operation} over shared memory preserves it.
It therefore tests for \emph{and rejects} programs that exhibit potential race
conditions due to their failure to follow locking discipline---%
these result in a failed compilation.

The \WRCompiler tracks two kinds of information to achieve these outcomes:
the contents of registers as expressions over shared variables,
and assumptions on access to variables by other threads.
The structures the \WRCompiler uses to do this are, respectively:
\begin{itemize}
    \item A \emph{register record}
        $\Phi \oftype \RegrecType \defineq \Reg \rightharpoonup \expr$.
        This draws inspiration from
        that used by the compilation scheme of \citet{Tedesco16}
        (originally of type $\Reg \rightharpoonup \Var$)
        to avoid generating unnecessary $\Load$ instructions
        to registers that already contain a variable;
        in addition, here we extend it to track entire expressions on shared variables.

    \item An \emph{assumption record}
        $\stableS \oftype \AsmrecType \defineq (\Var\ \set \times \Var\ \set)$
        that
        %like the security type system of \autoref{chp:add-locks},
        tracks
which variables at a given point in the source \WhileLang program are ``stable''
due to having, respectively, an $\AsmNoW$ or $\AsmNoRW$ assumption.
\end{itemize}

The \WRCompiler's main function $\CompileCmd$ then
outputs every register--assumption record pair (or \emph{compilation record}) $C = (\Phi, \mathcal{S}) \oftype \Comprec \defineq \RegrecType \times \AsmrecType$
associated with the program state \emph{before} execution of each instruction in
the output \RISCLang program.\footnote{For readability, we will use $\Regrec$, $\Asmrec$ to denote a $\Comprec$'s (resp.)~$\fst$, $\snd$ projections.}
A typical invocation to compile some $c \oftype cmd$ takes
an \emph{initial compilation record} $C$,
and returns
the $\Comprec$-annotated \RISCLang program $\PCs \oftype (I \times \Comprec)\ list$
(i.e.~$\mapfst\ \PCs$ recovers an
unannotated \RISCLang text),
and a \emph{final compilation record} $C'$:

\begin{example} [Example invocation of the \Covern \WRCompiler]
\label{eg:compile-cmd-invocation}
\[
  (\PCs, l', \nl', C', \failed) = \CompileCmd\ C\ l\ \nl\ c
\]
\end{example}

The remainder of this section will focus on formal properties of the compilation records output alongside each \RISCLang text:
\autoref{sec:stability-vs-data-races} will elaborate on checks enforced on input programs with the help of $\AsmrecType$s, and
\autoref{sec:proof-regrecs-stable} will present a resulting property
that $\RegrecType$s track stable expressions,
needed to prove security preservation (in \autoref{sec:proof-cvdni-compilation}).

Remaining details (e.g.~$l,l',nl,nl'$ for label allocation) will be relegated to
\ifPreprint appendices.
\else appendices that we provide as supplement material.
\fi
We note here only that (1)
$\CompileCmd$ may return $\True$ for $\failed$ to reject the input program,
such as when it detects a race condition (described further in \autoref{sec:stability-vs-data-races}),
or if expression depth exceeds the limit assumed by the
register allocation scheme model
\ifPreprint (elided to \autoref{sec:reg-alloc-model});
\else (elided to Appendix B);
\fi
also, (2) relative to the label allocation scheme
\ifPreprint (elided to \autoref{sec:labels-seq})
\else (elided to Appendix A)
\fi
we proved that
the control flow of each program fragment compiled by the \WRCompiler remains self-contained even when composed sequentially with other such fragments.

\iffalse
% robs: Version for if we do an appendectomy.
Details on label allocation (e.g.~$l,l',nl,nl'$)
and register allocation are relegated to
\citet{robs-phd} and
our Isabelle/HOL formalisation.
We note here only that (1)
$\CompileCmd$ may return $\True$ for $\failed$ to reject the input program,
such as when it detects a race condition (described further in \autoref{sec:stability-vs-data-races}),
or if expression depth exceeds a limit
assumed by our register allocation scheme model;
also, (2)
relative to our label allocation scheme
we proved that
the control flow of each program fragment compiled by the \WRCompiler remains self-contained even when composed sequentially with other such fragments.
\fi

%\subsection{Enforcement of shared variable stability to rule out data races}
\subsubsection{Requirements on inputs to the \WRCompiler}
\label{sec:stability-vs-data-races}

We define a shared variable $v$ to be recorded as assumed \emph{stable} if it
and all its control variables (i.e.~$\Cvars\ v$)
cannot presently be written to by another thread---that is,
if they are recorded %by an $\AsmrecType$
as having either of \textbf{AsmNoW} or \textbf{AsmNoRW} active on them.
Formally:

\begin{definition} [Stability of variable $v$ according to assumption record $\stableS$]
\[
    \VarStable\ \mathcal{S}\ v\ \defineq \ 
    v \in (\fst\ \mathcal{S} \cup \snd\ \mathcal{S})\ \land \ 
    (\forall v' \in \Cvars\ v.\ 
    v' \in (\fst\ \mathcal{S} \cup \snd\ \mathcal{S}))
\]
\end{definition}

For register record entries to be of any help in ensuring consistency of \WhileLang and \RISCLang expression evaluation, we exclude expression evaluation on
race-prone variables by lifting the concept of stability to register records.
The following predicate asserts internal consistency of the compilation record $C$ created by $\CompileCmd$, in the sense that the register record may only map to expressions that mention variables that are recorded as \textsf{stable} by the assumption record accompanying it.
(Here, $\defined{ran}$ denotes the \emph{range} of a map.)

\begin{definition} [Stability of the register record in compilation record $C$]
\label{def:regrec-stable}
\[
    \RegrecStable\ C\ \defineq \ 
    \forall e \in \defined{ran}\ (\Regrec\ C). \ 
    (\forall v \in \mathsf{exp{\text -}vars}\ e.\ 
    \VarStable\ (\Asmrec\ C)\ v)
\]
\end{definition}

We then implement a collection of
$\StabilityChecks \oftype \cmd \times \Comprec \Rightarrow \bool$
(called $\NoUnstableExprsOld$ in \citet{Sison_Murray_19}) as a
recursive function on the structure of \WhileLang programs,
that $\CompileCmd$ will use to ensure the following requirements of the given $\cmd$
if started with a configuration consistent with the given $\Comprec$:

\begin{itemize}
    \item The first two requirements ensure that programs comply with the locking discipline:
    \begin{itemize}
    \item The requirement regarding reading \emph{from} shared variables
        establishes the main outcome of freedom from race conditions
        we described at the beginning of \autoref{sec:tracking-regs-stability}:
        The program must not refer to expressions on any unstable variables.

        As a matter of convenience, instead of introducing a dedicated
        primitive to the \WhileLang language for reading
        atomically from a single (otherwise-unstable) device memory location,
        in our case study model of \autoref{chp:cddc} we model such
        interactions using a simple assignment $\Assign{x}{y}$
        protected by a ``read-atomicity'' lock on a shared memory location $y$
        that models the hardware interface.%
        \footnote{
        When such atomic hardware primitives exist on a given architecture, we
        expect it would be straightforward for source languages to expose them
        and oblige their architecture-specific compilers to compile them to that
        single atomic instruction in the target language's semantics,
        which would eliminate the need for such locks.}

%Note that this means that even a simple assignment $\Assign{x}{y}$, of a single variable to another, must be lock-protected.
%this is stricter than what is enforced by the local compliance check of \autoref{sec:local-compliance-locks},
%Without such a dedicated primitive, this is needed as a means of requiring the
%developer of the \WhileLang-language program to specify explicitly
%that an assignment from an otherwise-unstable variable
%should still execute atomically in the \RISCLang output program.
%Here, we will call locks introduced solely for this purpose ``read-atomicity'' locks,
%a practice we use for the case study model of \autoref{chp:cddc}.

    \item If the program assigns \emph{to} an unstable shared variable, then
        it must not be a lock-governed one according to the locking discipline.
        This prevents the violation of any guarantees not to write to the variable (due to not holding its lock).
    \end{itemize}

    \item The remaining two requirements
        follow some simplifying assertions,
        originally made by the security type system
        of \citet{Murray_SPR_16},
        that ensure mode state remains consistent
        after conditional branching and looping:

    \begin{itemize}
    \item The two sides of any \IfKw-conditional branches in the program must both end with, effectively, the same set of locks held---to be precise, judging by their effect on the mode state, as captured by the assumption record.

    \item For similar reasons, we require any \WhileKw-loops in the program to restore the original set of locks held on loop entry (again, as captured by the assumption record) on loop termination.
    \end{itemize}

        We believe these to be reasonable simplifications
        given that, in our setting, the set of variables governed by each lock
        does not change at runtime in such a way that would require access
        to them to be lock protected (or not) in a conditional manner.

\end{itemize}

\ifPreprint % (The publisher probably wouldn't find this acceptable.)
{\par \fussy % Prevent an overzealous linebreak that causes an orphaned line.
\fi
Together, $\RegrecStable\ C$ and $\StabilityChecks\ c\ C$ make up the main two requirements of a predicate $\CompilerInputReqs\ C\ l\ \nl\ c$ imposed on the input arguments to $\CompileCmd$.
(Its other two requirements reflect that the terminated \WhileLang program $\Stop$ has no valid compilation,
and that the initial label, if provided, must be valid---%
\ifPreprint
see \autoref{sec:labels-seq}.%
\else
details are relegated to supplementary Appendix A.%
\fi)
If any of these requirements are violated, $\CompileCmd$ rejects the program with $\failed=\True$:
\ifPreprint
\par} % \fussy
\fi

\begin{definition} [Requirements on inputs to $\CompileCmd$]
\label{def:compiler-input-reqs}
\[
\begin{aligned}
    \CompilerInputReqs\ C\ l\ \nl\ c\ \defineq \ \ 
    & \StabilityChecks\ c\ C\ \land \ 
      \RegrecStable\ C\ \land \\
    & c \ne \Stop\ \land \ 
      (\forall x. \ l = \Some\ x \longrightarrow x < \nl)
\end{aligned}
\]
\end{definition}

%\subsection{Proof that registers track stable expressions on stable variables}
%\subsection{Proof that expression evaluation is race free}
\subsubsection{Proof that all tracked register contents are stable}
\label{sec:proof-regrecs-stable}

Imposing the predicate $\CompilerInputReqs$ (\autoref{def:compiler-input-reqs}) gives us enough information to prove a lemma that $\CompileCmd$ only ever outputs stable register records, that attest to the fact that registers contain the results of evaluating expressions on stable variables.

Stated more precisely, every
\RISCLang program returned by a successful invocation of $\CompileCmd$
is annotated by $\Comprec$s all with
stable register records, and furthermore that the final $\Comprec$'s register record is also stable:

\begin{lemma} [Successful compilations output only stable register records] \label{thm:compiled-cmd-regrec-stable}
\[
  \mprset{vskip=0.5ex}
  \inferrule{
    (\PCs, l', \nl', C', \False) = \CompileCmd\ C\ l\ \nl\ c \and
      \CompilerInputReqs\ C\ l\ \nl\ c
  } {
      (\forall \pc < \length\ \PCs.\ \RegrecStable\ \listderef{(\mapsnd\ \PCs)}{\pc})\,\land\,
      \RegrecStable\ C'
  }
\]
\end{lemma}
\begin{proof}
    By induction on the structure of the \WhileLang language program $c$, making reference to the implementation of $\CompileCmd$.

    For cases that must compile expressions, we furthermore prove and make use of a lemma by induction on the structure of expressions, making reference to the implementation of the expression compiler function $\CompileExpr$ called by $\CompileCmd$.
    In essence, we prove that (sub)expressions appearing in register records must be stable, for two reasons:

    First, they are always only ever subexpressions over variables that must have been stable in the input program when their contents were first loaded into registers.

    Second, when compiling an $\LockRel{k}$, the \WRCompiler will always flush all register records that make reference to any variables that the $\LockRel{k}$ makes unstable.
\end{proof}
 % from thesis

\subsection{Preserving a ban on secret-dependent control flow}
\label{sec:wr-compiler-no-hbranch}
The \WRCompiler assumes that
input \WhileLang programs have no \emph{conditional branches on $\High$-sensitivity values} (\emph{$\High$-branching}), and therefore no secret-dependent control flow.
This is a restriction
% Draws too much attention to a type system that is out of scope. -robs.
%applied by the \WhileLang-language
%security type system of
%%\autoref{chp:add-locks}
%\citet{Murray_SPR_16}
%as adapted to support mutex locks by \citet{robs-phd}, and
commonly applied
% Revised: I make the case in the intro that this is a
% reasonable restriction for a practical class of security-critical programs.
as a means to prevent all implicit flows, including timing leaks.
%note that we will revisit this in \autoref{chp:h-branching}.
%
This restriction will then be preserved by the \WRCompiler for its output \RISCLang programs, reflected primarily in the design of the concrete coupling invariant $\CouplInvWR$ (see \autoref{sec:couplinv_wr}).

Specifically, the \WRCompiler assumes that the confidentiality of input \WhileLang programs is witnessed by a strong low-bisimulation modulo modes
with an extra requirement
(supplied as a parameter, as in \autoref{sec:init-extra-params})
that effectively
disallows any present or past $\High$-branching.
Relying on the fact that a low-bisimulation
already asserts $\Low$-equivalence of memories,
the extra requirement
asserts that
it furthermore pairs only configurations
at the same program location,
and that any $\IfKw$-conditional expressions
must evaluate to the same value in both configurations' memories.
Here, the helper function \textsf{leftmost-cmd} gives the leftmost in a sequence of ;-separated \WhileLang-language commands:

\begin{definition} [An extra requirement for low-bisimulations $\mathcal{B}$ to ban $\High$-branching]
\label{def:no-h-branching}
\begin{align*}
    & \NoHighBranching\ \mathcal{B} \defineq \\
    & \forall c\ c'\ \mds\ \mem\ \mem'.\ 
      (\LocalConfWhile{c}{\mds}{\mem},
       \LocalConfWhile{c'}{\mds}{\mem'}) \in \mathcal{B}\ 
       \longrightarrow \ c = c'\ \land \\
    & \qquad (\forall e\ c_1\ c_2.\ 
              \mathsf{leftmost{\text -}cmd}\ c = \ITEg{e}{c_1}{c_2} \longrightarrow \ 
              \mathsf{ev_{exp}}\,\mem\ e = \mathsf{ev_{exp}}\,\mem'\,e)
\end{align*}
\end{definition}

Then, in \autoref{sec:compiler-outputs}, we will prove that
the \WRCompiler produces confidential \RISCLang programs
with no secret-dependent control flow,
as witnessed by a low-bisimulation that asserts a similar extra requirement for \RISCLang programs.
In effect, this is the \emph{pc-security} notion of \citet{Molnar05},
but also explicitly equating the program text:

\begin{definition} [A pc-security--like requirement for \RISCLang bisimulations $\mathcal{B}$]
\label{def:pc-security}
\begin{align*}
    & \PCSecurity\ \mathcal{B} \defineq \ 
      \forall \pc\ \pc'\ P\ P'\ \regs\ \regs'\ \mds\ \mem\ \mem'. \\
    & (\LocalConfRISC{pc}{P}{\regs}{\mds}{\mem},
       \LocalConfRISC{pc'}{P'}{\regs'}{\mds}{\mem'}) \in \mathcal{B}\ 
       \longrightarrow \ pc = pc'\ \land P = P'
\end{align*}
\end{definition}
 % from thesis

\subsection{Use of the decomposition principle}
\label{sec:proof-cvdni-compilation}
Having covered the most relevant aspects of the \WRCompiler's
implementation, we now present the refinement relation $\RefRelWR$ (in \autoref{sec:refrel_wr}), pacing function $\AbsStepsWR$ (in \autoref{sec:abs-steps_wr}), and concrete coupling invariant $\CouplInvWR$ (in \autoref{sec:couplinv_wr}),
parameters we use to
apply the decomposition principle we presented in \autoref{sec:just-decomp-reqs}
to prove (in \autoref{sec:successful-compilations}) that successful compilations are legitimised by $\SecureRefinement$ (\autoref{def:secure-refinement})---the desired confidentiality-preserving notion of refinement for mixed-sensitivity concurrent programs.

The strategy laid out by the decomposition principle will be to prove that these parameters satisfy $\SimplerRefinementSafe$ (\autoref{def:simpler-refinement-safe}) for a targeted class of input \WhileLang-language programs---%
ones with no secret-dependent control flow,
as we specified in \autoref{sec:wr-compiler-no-hbranch}---%
meaning (for such programs) we can use the parameters to enforce that \WRCompiler introduces no secret-dependent inconsistencies in termination, timing behaviour, or assume--guarantee modes.

In doing so we
avoid a direct proof of the cube-shaped refinement diagram (\autoref{fig:coupling-inv-pres}) of \citet{Murray_SPR_16}---which would have involved reasoning about both $\RefRelWR$ and $\CouplInvWR$ at once---%
and instead prove (with the assistance of $\AbsStepsWR$)
a square-shaped refinement diagram for $\RefRelWR$ (\autoref{fig:decomp-R}) more typically found in compiler verification.
 % from thesis
\subsubsection{Refinement relation $\RefRelWR$ and its invariants}
\label{sec:refrel_wr}
In this section we introduce the refinement relation $\RefRelWR$ that characterises compilation of programs from \WhileLang to \RISCLang using the \WRCompiler,
and prove it satisfies
the two properties demanded of $\RefRelWR$ (alone) by
formal $\SecureRefinement$ (\autoref{def:secure-refinement}):
\begin{enumerate}
    \item Preservation of modes and all contents of shared memory ($\PreservesMM$, \autoref{def:preserves_modes_mem}), and
    \item Closedness under changes by other threads ($\ClosedOthers$, \autoref{def:closed-others}).
\end{enumerate}

\noindent An actual proof of refinement (using the square-shaped diagram of \autoref{fig:decomp-R}) for $\RefRelWR$ will be deferred to \autoref{sec:abs-steps_wr}, which introduces the $\AbsStepsWR$ function pacing it.

Just like the earlier example of a secure refinement relation (in \autoref{fig:refinement-example}), the refinement relation $\RefRelWR$
%---that we will nominate here, for the \WRCompiler---
pairs abstract (here, \WhileLang-language) with concrete (here, \RISCLang-language) program configurations.
%Here, we will focus on one of $\RefRelWR$'s cases in order to highlight its most notable characteristics for understanding why it is suitable to describe a security-preserving compilation.
For example, the $\mathtt{if\_expr}$ case of $\RefRelWR$
relates the expression-evaluation part of the \WhileLang command $\ITEg{e}{c_1}{c_2}$,
with the corresponding part
of the \texttt{RISC} program obtained by running $\CompileCmd$ on it,
including the conditional jump $\Jz$ after expression evaluation.
(This case is depicted in
\ifPreprint \autoref{fig:Rwr-if-expr},
\else Figure C.1,
\fi
and a relevant excerpt of the $\CompileCmd$ implementation provided in
\ifPreprint \autoref{fig:impl-if-expr}
\else Figure C.2
\fi
for comparison, both
\ifPreprint on page~\pageref{fig:Rwr-if-expr} of \autoref{sec:R_wr-informal}.
\else in supplementary Appendix C.
\fi
An informal description of all the cases of $\RefRelWR$, their purpose, and the invariants they maintain,
can also be found in
\ifPreprint \autoref{sec:R_wr-informal}.%
\else Appendix C.%
\fi)

We define almost all the cases
of $\RefRelWR$ to
assert at least one successful run of $\CompileCmd$
(where $\failed=\False$).
We then define a guard that we impose to
restrict the scope of $\RefRelWR$ only to consider
local program configurations
consistent with the relevant compilation record produced by $\CompileCmd$.
In short, this ensures the actual values in the register bank $\regs$ equal any expression the register record says they should have,
as evaluated under the current $\mem$;
and furthermore, that the assumption record is consistent with the $\AsmNoW$ and $\AsmNoRW$ modes in the actual $\mds$.
Formally:

\begin{definition} [Configuration consistency requirements for compiled commands] \label{def:compiled-cmd-config-consistent}
\begin{align*}
    & \CompiledCmdConfigConsistent\ C\ \regs\ \mds\ \mem\ \defineq \\
    & \quad \RegrecMemConsistent\ (\Regrec\ C)\ \regs\ \mem\ \land \ 
      \AsmrecMdsConsistent\ (\Asmrec\ C)\ \mds \\
    & \textnormal{where} \\
    & \quad
    \RegrecMemConsistent\ \Phi\ \regs\ \mem\ \defineq \ 
    \forall r\ e.\ \Phi\ r = \Some\ e \longrightarrow \regs\ r = \mathsf{ev_{exp}}\ \mem\ e \\
    & \qquad \text{(Consistency between register record, register bank, and shared memory)} \\
    & \quad
     \AsmrecMdsConsistent\ \mathcal{S}\ \mds\ \defineq \ 
     \mathcal{S} = (\mds\ \mathbf{AsmNoW},\ \mds\ \mathbf{AsmNoRW}) \\
    & \qquad \text{(Consistency between an assumption record and a mode state)}
\end{align*}
\end{definition}

Apart from using
$\CompiledCmdConfigConsistent$
to restrict the scope of $\RefRelWR$ in this manner,
we will also impose it
in \autoref{sec:successful-compilations} as
\emph{initial configuration requirements} for compiled programs:
Only configurations obeying them may be used to initialise a \texttt{RISC} program compiled by the \WRCompiler with initial $\Comprec$ $C$.

The cases of $\RefRelWR$ also tend to assert
$\RegrecStable$ (\autoref{def:regrec-stable}),
which we already proved holds for all compilation records produced by the \WRCompiler (\autoref{thm:compiled-cmd-regrec-stable}).

Finally, whenever a case of $\RefRelWR$ is inductive
(e.g.~the $\mathtt{if\_expr}$ case,
for its nested calls to $\CompileCmd$ for each of its ``then'' and ``else'' branches)
it quantifies over all configurations that obey
$\CompiledCmdConfigConsistent$
(\autoref{def:compiled-cmd-config-consistent}) and
$\RegrecStable$
(\autoref{def:regrec-stable})
relative to the initial compilation record given to each nested
call to $\CompileCmd$.

With $\RefRelWR$ thus specified, we can now prove the two requirements for $\SecureRefinement$ that pertain to $\RefRelWR$ alone:
$\PreservesMM$ (\autoref{def:preserves_modes_mem}), and $\ClosedOthers$ (\autoref{def:closed-others}).
In short, $\PreservesMM$ is largely enforced by the definition of $\RefRelWR$, but $\ClosedOthers$
relies in part on $\RefRelWR$ only ever talking about stable register records:

\begin{lemma} [$\RefRelWR$ preserves modes and memory] \label{thm:preserves-modes-mem-R_wr}
\[
    \PreservesMM\ \RefRelWR
\]
\end{lemma}
\begin{proof}
By induction on the structure of $\RefRelWR$.

For all cases of $(\lc_{w}, \lc_{r}) \in \RefRelWR$, $\LCSameMdsMem{\lc_w}{\lc_r}$ is either asserted directly by the guards or obtainable from the inductive hypothesis.
\end{proof}

\begin{lemma} [$\RefRelWR$ is closed under changes by others] \label{thm:closed-others-R_wr}
\[
    \ClosedOthers\ \RefRelWR
\]
\end{lemma}
\begin{proof}
    By induction on the structure of $\RefRelWR$.

    Changes by others (\autoref{def:closed-others}) only modify $\Writable$ variables the same way for both configurations, so preservation of $\LCSameMdsMemOp$ is immediate.
    Also, $\RegrecMemConsistent$ is unaffected because by \autoref{thm:compiled-cmd-regrec-stable}, $\CompileCmd$ only creates $\RegrecStable$ records---i.e.~referring to no $\Writable$ variables.
    No other $\RefRelWR$ guards mention shared memory.
\end{proof}
 % from thesis
\subsubsection{Refinement pacing function $\AbsStepsWR$}
\label{sec:abs-steps_wr}
In this section we nominate a pacing function, $\AbsStepsWR$, specifying the number of evaluation steps with which a \WhileLang program should simulate
each step of the \RISCLang program to which the \WRCompiler compiled it.
Using the square-shaped ``refinement preservation'' diagram of \autoref{fig:decomp-R} (part of \autoref{def:secure-refinement-simpler}),
we then prove that the $\RefRelWR$ relation
we introduced in \autoref{sec:refrel_wr}
is a refinement when ``paced'' by $\AbsStepsWR$ in this manner.

Here we define $\AbsStepsWR$ to depend only on the current program location;
consequently,
as long as the $\WRCompiler$ introduces no secret-dependent control flow,
it will also introduce no timing leaks---that is, no secret-dependent variations to the pacing of the program, as disallowed by \autoref{fig:decomp-abs-steps} (part of \autoref{def:simpler-refinement-safe})---which we will be obliged to prove in \autoref{sec:successful-compilations}.
To this end, $\AbsStepsWR$ primarily looks at the form of the \RISCLang instruction (sometimes \WhileLang command) about to be executed,
dividing them into three categories:
\begin{itemize}
    \item Instructions output by $\CompileExpr$: $\Load$, $\Op$, and $\MoveK$.
        For these, $\AbsStepsWR$ returns 1 if the
        \textsf{leftmost-cmd}
        (the leftmost in a sequence of ;-separated commands)
        of the \texttt{While} program is ``$\Whileg{e}{c}$'', to allow it to step to ``$\ITEg{e}{(\Seqg{c}{\Whileg{e}{c}})}{\Stop}$'' concurrently with the first \texttt{RISC} step of the compiled expression itself.
        Otherwise, $\AbsStepsWR$ returns 0, to indicate the \texttt{While} program standing still while the \texttt{RISC} program takes \emph{new} steps to evaluate the expression.
    \item ``Epilogue'' steps: $\Jmp$ and $\Nop$ when used for control flow at the end of a smaller compiled program in the context of a larger one.
        For these, $\AbsStepsWR$ returns 0.
    \item All other \texttt{RISC} instructions are assumed to proceed at a lockstep pace with the \texttt{While} command they were compiled from, and for these $\AbsStepsWR$ returns 1.
\end{itemize}

Having nominated $\AbsStepsWR$ and $\RefRelWR$, we now have the parameters over which we are obliged,
by $\SecureRefineSimpler$ (\autoref{def:secure-refinement-simpler}),
to prove refinement preservation (\autoref{fig:decomp-R}).
To this end, we prove firstly
that every step of execution of a \texttt{RISC} program, produced by the \WRCompiler from a \texttt{While} program, maintains the consistency demanded by $\CompiledCmdConfigConsistent$ between configurations and compilation records:

\begin{lemma} [Successfully compiled programs maintain config consistency requirements]\label{thm:compiled-cmd-eval-maintains-comprec-consistency}
\[
  \mprset{vskip=0.5ex}
  \inferrule {
    (\PCs, l', \nl', C', \failed) = \CompileCmd\ C\ l\ \nl\ c \and
      \CompilerInputReqs\ C\ l\ \nl\ c \and \\
    \failed = \False \and
      \pc < \length\ \PCs \and
      P = \mapfst\ \PCs \and
      \Cs = \mapsnd\ \PCs \and \\
    \CompiledCmdConfigConsistent\ \listderef{\Cs}{\pc}\ \regs\ \mds\ \mem \and \\
    \LocalConfRISC{\pc}{P}{\regs}{\mds}{\mem}
      \EvalStepRISC
      \LocalConfRISC{\pc'}{P}{\regs'}{\mds'}{\mem'})
  } {
    \CompiledCmdConfigConsistent\ 
      (\mathtt{if}\ \pc' < \length\:P\ 
       \mathtt{then}\ \listderef{\Cs}{\pc'}\ 
       \mathtt{else}\ C')\:
      \regs'\:\mds'\:\mem'
  }
\]
\end{lemma}
\begin{proof}
    Unfolding \autoref{def:compiled-cmd-config-consistent}, we in fact prove it separately for $\RegrecMemConsistent$ and $\AsmrecMdsConsistent$, both times by induction on the structure of \texttt{While} program $c$.

    In each case, we use the simplifiers for the $\CompileCmd$ implementation to yield the corresponding \texttt{RISC} program fragment in question, and then prove the lemma for each of the possible locations of $\pc$ in the compiled program.
    For both proofs, there is some trickiness in accounting for (and ruling out) which destination $\pc'$ must be considered for each of these cases of $\pc$, particularly for those \texttt{While} programs that compile to \texttt{RISC} programs that may have jumps in them.

    Control flow trickiness aside, the intuition for $\RegrecMemConsistent$ is that it tests the correctness of the compilation of expressions.
    For this we prove a sublemma for maintenance of $\CompiledCmdConfigConsistent$, by induction on the structure of expressions $e$ that are encountered in the \texttt{While} programs $\ITEg{e}{c_1}{c_2}$, $\Whileg{e}{c'}$, and $\Assign{v}{e}$.
    Additionally, $\LockRel{k}$ flushes register record entries mentioning variables that are to become unstable, and $\Whileg{e}{c'}$ conservatively flushes entries to force evaluation of the loop condition expression.
    This is safe trivially because flushing entries can never make a consistent register record inconsistent.
    The rest of the cases for $c$ are straightforward because they do not touch the register record.

    Then for $\AsmrecMdsConsistent$, the substantial part of the proof is as a test of the correctness of the compiler's bookkeeping of assumptions being consistent with the semantics of $\LockAcq{k}$ and $\LockRel{k}$.
    The other cases for $c$ do not touch the mode state.
\end{proof}

Also, we must prove a correctness lemma for the expression compiler:
\begin{lemma} [Correctness of the expression compiler] \label{thm:compile-expr-correct}
\[
    (\PCs, r, C', \False) = \CompileExpr\ C\ A\ l\ e\ 
    \implies (\Regrec\ C')\ r = \Some\ e
\]
\end{lemma}
\begin{proof}
    By induction on the structure of expressions $e$, using the simplification rules for the implementation of $\CompileExpr$, and also relying on assumptions of correctness of the register allocation scheme supplied by the instantiator of the theory.
\end{proof}

Armed with these facts, we can now prove the main refinement preservation result:
\begin{lemma} [$\RefRelWR$ is a refinement paced by $\AbsStepsWR$] \label{thm:R_wr-refinement-abs-steps_wr}
% Isabelle name: R_preservation
\begin{align*}
    & \forall \lc_w\ \lc_r.\ (\lc_w, \lc_r) \in \RefRelWR \longrightarrow \ 
      (\forall \lc_r'.\ \lc_r \EvalStepRISC \lc_r' \longrightarrow \\
    & \qquad (\exists \lc_w'.\ \lc_w \NEvalStepWhile{(\AbsStepsWR\ \lc_w\ \lc_r)} \lc_w'\ \land \ (\lc_w', \lc_r') \in \RefRelWR))
\end{align*}
\end{lemma}
\begin{proof}
    By induction on the structure of $\RefRelWR$.
    (Refer to
    \ifPreprint \autoref{sec:R_wr-informal}
    \else supplementary Appendix C
    \fi
    for an informal description of all cases of $\RefRelWR$.)

    The base case \texttt{stop} is immediate, as it pertains to a terminated \texttt{While} and \texttt{RISC} program.
    The base cases that proceed in one step to a terminating program configuration (\texttt{skip\_nop}, \texttt{assign\_store}, \texttt{lock\_acq}, \texttt{lock\_rel}) are fairly straightforward because after dealing with the single step, the resulting obligation can then be handled by the \texttt{stop} case.
    This leaves the last remaining base case \texttt{assign\_expr}, which proceeds in one step either to itself, or to \texttt{assign\_store}.
    In all these cases, we use \autoref{thm:compiled-cmd-eval-maintains-comprec-consistency} to obtain the preservation of the guards demanded by the $\RefRelWR$ introduction rule for the destination configuration of the step.
    Particularly, the \texttt{assign\_store} case must make use of $\RegrecMemConsistent$ and the correctness of $\CompileExpr$ (\autoref{thm:compile-expr-correct}) to ensure that once the evaluated expression is written back to shared memory, $\LCSameMdsMem{\lc_w'}{\lc_r'}$ holds as demanded by the \texttt{stop} case.

    The inductive cases that concern expression evaluation (\texttt{if\_expr}, \texttt{while\_expr}) are much like \texttt{assign\_expr} in that they have the possibility of progressing in one step to themselves.
    Unlike \texttt{assign\_expr} however, their other possibility is a conditional jump based on the result of that expression.
    Again we use \autoref{thm:compile-expr-correct} to obtain that the result is an accurate calculation of the expression, and this time we prove by the two different cases whether \texttt{if\_expr} ends up in \texttt{if\_c1} or \texttt{if\_c2}, or if \texttt{while\_expr} ends up in \texttt{while\_inner} or at \texttt{stop} (having jumped to the exit label).
    In these cases, the guards over which the inductive references to $\RefRelWR$ have been quantified are versatile enough to discharge themselves (when \texttt{*\_expr} steps to itself), or to discharge any reachable initial starting state for the nested compiled \texttt{RISC} program, given that \autoref{thm:compiled-cmd-eval-maintains-comprec-consistency} ensures the invariance of these guards.

    This just leaves the inductive cases that pertain to configurations inside a nested compiled \texttt{RISC} program (\texttt{if\_c1}, \texttt{if\_c2}, \texttt{while\_inner}), or at the end of one (\texttt{epilogue\_step}, \texttt{while\_loop}).
    In these cases, the inductive hypotheses obtained from the inductive reference to $\RefRelWR$ are always enough to satisfy the guards demanded by the possible destination cases.
    Like in the proof of \autoref{thm:compiled-cmd-eval-maintains-comprec-consistency}, the trickiness mostly comes from accounting for all the possible cases of control flow (ruling out spurious destinations) that need to be considered.
\end{proof}
 % from thesis
\subsubsection{Concrete coupling invariant $\CouplInvWR$}
\label{sec:couplinv_wr}
The next element needed is the concrete coupling invariant $\CouplInvWR$.
Recall from \autoref{sec:wr-compiler-no-hbranch} that
the $\NoHighBranching$ requirement (\autoref{def:no-h-branching})
ensures that
input \WhileLang programs
have no
secret-dependent control flow;
here we choose $\CouplInvWR$ to ensure that the \WRCompiler has not introduced any \emph{new}
secret-dependent control flow
in the output \RISCLang program.

We define $\CouplInvWR$ formally
to assert that the witness strong low-bisimulation (modulo modes) to be derived for the output program only pairs local configurations that are at the same location $\pc = \pc'$ of the same \texttt{RISC} program $P = P'$:
\begin{definition} [Concrete coupling invariant $\CouplInvWR$ for compiled programs]
\label{def:couplinv-wr}
\[
\CouplInvWR \defineq
\{(\LocalConfRISC{\pc}{P}{\regs}{\mds}{\mem},
   \LocalConfRISC{\pc'}{P'}{\regs'}{\mds'}{\mem'})\ |\ 
   (\pc, P) = (\pc', P')\}
\]
\end{definition}

From this definition, $\PCSecurity$ (\autoref{def:pc-security}) is clearly immediate for any concrete bisimulation
$\BCof~\mathcal{B}~\mathcal{R}~\mathcal{\CouplInvWR}$ (\autoref{def:concrete-bisim})
derived using $\CouplInvWR$.
 % from thesis
\subsubsection{Proof of CVDNI-preserving refinement}
\label{sec:successful-compilations}
With $\RefRelWR$, $\AbsStepsWR$, and $\CouplInvWR$ nominated, we are ready to prove confidentiality-preserving refinement using the decomposition principle $\SecureRefineSimpler$ (\autoref{def:secure-refinement-simpler}).

To this end, we now prove the suitability of these three parameters,
for \WhileLang programs that do not branch on $\High$-sensitivity values (as we specified earlier, in \autoref{sec:wr-compiler-no-hbranch}):

% Isabelle name: simpler_refinement_safe_R_simple_evB_eq_B
\begin{lemma} [$\RefRelWR,\AbsStepsWR,\CouplInvWR$ are safe for $\SecureRefinement$ decomposition]
\label{thm:simpler-refinement-safe_wr}
\begin{align*}
\inferrule{
    \StrongLowBisimMM\ \mathcal{B} \and
    \NoHighBranching\ \mathcal{B}
} {
    \SimplerRefinementSafe\ \mathcal{B}\ \RefRelWR\ \CouplInvWR\ \AbsStepsWR
}
\end{align*}
\end{lemma}
\begin{proof}
    Unfolding \autoref{def:simpler-refinement-safe} gives us the following obligations. (See also \autoref{fig:decomp}.)

    For consistent stopping behaviour, we prove a lemma that \texttt{RISC} programs stop if and only if their $\pc$ is outside the program text $P$, i.e. $\pc > \length\ P$.
    Because $\CouplInvWR$ equates $\pc$ and $P$ for the two configurations, then clearly both have identical stopping behaviour.

    For consistency of change in timing behaviour, $\AbsStepsWR$ depends only on \texttt{While} and \texttt{RISC} program locations, and $\NoHighBranching$ and $\CouplInvWR$ forces them (respectively) to be equal for the local configurations under consideration.

    For closedness of $\CouplInvWR$ under lockstep execution, the only non-straightforward cases to consider are conditional branching, and the locking primitives.
    For conditional branching, we use $\NoHighBranching$ for $\mathcal{B}$ with memory preservation via $\RefRelWR$ (\autoref{thm:preserves-modes-mem-R_wr}) to ensure that the conditional branching outcome is the same on both sides.

    Finally, as the only operations that touch mode state, the locking primitives are the only non-straightforward cases for modes-equality maintenance under lockstep execution.
    As all lock memory is classified $\Low$ (\autoref{thm:no-H-locks}),
    we use $\StrongLowBisimMM$ for $\mathcal{B}$ with memory preservation via $\RefRelWR$ to ensure the \texttt{RISC} configurations behave consistently.
\end{proof}

\begin{lemma} [$\RefRelWR,\AbsStepsWR,\CouplInvWR$ meet decomposed $\SecureRefinement$ requirements]
\begin{align*}
\inferrule{
    \StrongLowBisimMM\ \mathcal{B} \and
    \NoHighBranching\ \mathcal{B}
} {
    \SecureRefineSimpler\ \mathcal{B}\ \RefRelWR\ \CouplInvWR\ \AbsStepsWR
}
\end{align*}
\end{lemma}
\begin{proof}
    Unfolding \autoref{def:secure-refinement-simpler}, the obligations pertaining only to $\RefRelWR$ and $\AbsStepsWR$ are discharged by \autoref{thm:R_wr-refinement-abs-steps_wr}, \autoref{thm:closed-others-R_wr}, and \autoref{thm:preserves-modes-mem-R_wr}.
    Pertaining to $\CouplInvWR$: Clearly $\CouplInvWR$ is symmetric, and furthermore it is $\CgConsistent$ (\autoref{def:cg-consistent}) because the actions over which $\CouplInvWR$ must be closed modify only the shared memory, and $\CouplInvWR$ places only restrictions on the program text and current location.
    The final obligation (regarding $\SimplerRefinementSafe$) is discharged by \autoref{thm:simpler-refinement-safe_wr}.
\end{proof}

From this it follows immediately via \autoref{thm:secure-refinement-simpler-sound} that $\RefRelWR$ with the help of $\CouplInvWR$ describes a confidentiality-preserving refinement for non-$\High$-branching \texttt{While} programs:

% Isabelle name: secure_refinement_R_simple_evB_eq_B
\begin{corollary} [$\RefRelWR$ is a secure refinement for non-High-branching programs] \label{thm:secure-refinement-R_wr}
\[
\inferrule{
   \StrongLowBisimMM\ \mathcal{B} \and
   \NoHighBranching\ \mathcal{B}
} {
   \SecureRefinement\ \mathcal{B}\ \RefRelWR\ \CouplInvWR
 }
\]
\end{corollary}

Finally we prove that successful compilation produces a \texttt{RISC} program related by $\RefRelWR$ to its input \texttt{While} program, when started with corresponding (same $\mds,\mem$) and reasonable (according to $\CompiledCmdConfigConsistent$) initial configurations:

\begin{theorem} [Successful compilations are refinements in $\RefRelWR$] \label{thm:compile-cmd_correctness_R_wr}
\[
\mprset{vskip=0.5ex}
\inferrule{
    (\PCs, l', \nl', C', \failed) = \CompileCmd\ C\ l\ \nl\ c \and
      \CompilerInputReqs\ C\ l\ \nl\ c \\
    \failed = \False \and
      \CompiledCmdConfigConsistent\ C\ \regs\ \mds\ \mem \and
      P = \mapfst\ \PCs
} {
    (\LocalConfWhile{c}{\mds}{\mem},
     \LocalConfRISC{0}{P}{\regs}{\mds}{\mem}) \in \RefRelWR
}
\]
\end{theorem}
\begin{proof}
    By induction on the structure of the \WhileLang-language.

    The compiler input and initial configuration conditions we impose allow us to have each of
    $\Skip$, $\Seqg{\cmd}{\cmd}$, $\ITEg{exp}{\cmd}{\cmd}$,
    $\Whileg{exp}{\cmd}$, $\Assign{v}{exp}$,
    $\LockAcq{k}$, and $\LockRel{k}$
    and their compiled output meet the guards of the introduction rules for the cases
    \texttt{skip}, \texttt{seq}, \texttt{if\_expr}, \texttt{while\_expr}, \texttt{assign\_expr}, \texttt{lock\_acq}, and \texttt{lock\_rel} of $\RefRelWR$
    (described further in
    \ifPreprint \autoref{sec:R_wr-informal}%
    \else supplementary Appendix C%
    \fi)
    that we designed for them, respectively.
\end{proof}
 % from thesis

% Toby: Pitch wr-compiler theorem as instance of whole-system refinement?
\subsection{Proof of compositional noninterference preservation}
\label{sec:compiler-outputs}
% NB: there are subsection headings hidden in compiler-outputs
Going beyond the level of detail of our presentation in \citet{Sison_Murray_19},
we now present the final few steps
to obtain preservation of
whole-system security for concurrent compositions of \RISCLang threads
when all are obtained via compilation by the \WRCompiler
(\autoref{sec:whole-system_wr}).
In addition to this, we obtain preservation of
per-thread compositional security for each program thread compiled,
and other properties that may be useful for their composition with
\RISCLang threads proved secure directly at the \RISCLang level
(\autoref{sec:per-thread_wr}).

\subsubsection{Whole-system security preservation}
\label{sec:whole-system_wr}

To use the whole-system refinement theorem (\autoref{thm:sys-refine}),
we are obliged to show that, in addition
to establishing a $\SecureRefinement$
(\autoref{def:secure-refinement},
which we just showed in \autoref{sec:proof-cvdni-compilation}),
the \WRCompiler also preserves
$\LocalModeUse$
%(\autoref{def:local-mode-compliance})
as demanded by $\SysRefineReqs$ (\autoref{def:sys-refine-reqs}).
Then, as we noted in \autoref{sec:sys-refine},
there is no need for us to prove preservation of the
non-compositional
$\GlobalModeUse$ condition---% (\autoref{def:global-modes-compatibility})---%
the whole-system refinement theorem takes care of that.
%As noted, preservation of the non-compositional global compatibility
%condition down to RISC level is automatically ensured by the theorem.

The local compliance preservation
result follows from a property of the refinement relation, $\RefRelWR$.
Here, ``$\RespectsOwnGuars$''
%$\PreservesLocalCompliance$
is from \autoref{def:local-mode-compliance}:

% Isabelle name: establishes_local_guarantee_compliance_R
% NB: Remark on the fact that it does not use the While-level local compliance?
\begin{lemma} [Each step from a \RISCLang configuration in $\RefRelWR$ respects its own guarantees]
\label{thm:Rwr-respects-own-guarantees}
\[
\mprset{vskip=0.5ex}
\inferrule{
    (\LocalConfWhile{c}{\mds}{\mem},
     \LocalConfRISC{\pc}{P}{\regs}{\mds}{\mem}) \in \RefRelWR
} {
    \RespectsOwnGuars\ (((\pc, P), \regs), \mds)
}
\]
\end{lemma}
\begin{proof}
    By induction on the structure of $\RefRelWR$.

    %The only cases that matter are ones for \WhileLang commands (and their corresponding \RISCLang instructions) that involve any reading and writing of memory.
    %As discussed in \autoref{sec:compiler-impl}, the $\NoUnstableExprs$ static check ensures that the source \WhileLang program never accesses a lock-governed variable unless the relevant lock is held (as reflected by the mode state in accordance with the locking discipline as specified in \autoref{sec:locking-discipline}).
    %This check is asserted directly by $\CompilerInputReqs$, which is imposed directly by every relevant case of $\RefRelWR$ in ensuring that it only relates valid inputs and outputs of the \WRCompiler.
    Knowing that the \WhileLang command does not access lock-governed variables without holding the relevant lock (via the $\StabilityChecks$ asserted as part of $\CompilerInputReqs$ by every relevant case of $\RefRelWR$),
    %, in ensuring that it only relates valid inputs and outputs of the \WRCompiler),
    we are obliged to show that the \RISCLang instruction paired to it by $\RefRelWR$ similarly respects the guarantee modes implied by the locking discipline (as specified in \autoref{sec:locking-discipline}).
    We do so with a mixed Isar/``apply''-style proof that exercises the relevant cases of the \RISCLang semantics, using lemmas about control flow under sequential composition (mentioned in \autoref{sec:compiler-impl}; see also
    \ifPreprint \autoref{sec:labels-seq}%
    \else supplementary Appendix A%
    \fi).
    Propositions \ref{thm:no-locks-in-C}
    and \ref{thm:lock-interp-C-vars} also play a role in
    excluding certain cases from consideration.
\end{proof}

% Isabelle name: establishes_locally_sound_mode_use_R
\begin{lemma} [Refinements in $\RefRelWR$ ensure local mode compliance]
\label{thm:Rwr-local-mode-compliance}
\[
\mprset{vskip=0.5ex}
\inferrule{
    (\LocalConfWhile{c}{\mds}{\mem},
     \LocalConfRISC{\pc}{P}{\regs}{\mds}{\mem}) \in \RefRelWR
} {
     \LocalModeUse\ \LocalConfRISC{\pc}{P}{\regs}{\mds}{\mem}
}
\]
\end{lemma}
\begin{proof}
    Unfolding \autoref{def:local-mode-compliance}, we must show that what was proved by \autoref{thm:Rwr-respects-own-guarantees} holds for every \RISCLang configuration reachable from $\LocalConfRISC{\pc}{P}{\regs}{\mds}{\mem}$.
    %(according to $\LocReach$, a notion of reachability defined in \citet{Murray16-Dependent-SIFUM-AFP}).

    First, we prove a lemma that establishes that every such reachable \RISCLang configuration is also paired by $\RefRelWR$ to some \WhileLang configuration.
    Specifically, we prove that $\RefRelWR$ is closed under a notion of ``pairwise reachability under mode-permitted havoc'', wherein:
    \begin{enumerate}
        \item Every one step by the \RISCLang program is matched by either zero or one step by the \WhileLang program, as specified by $\AbsStepsWR$ (\autoref{sec:abs-steps_wr}).
        \item Between each evaluation step, arbitrary changes are allowed to occur to the memory locations judged by the mode state to be $\Writable$ (\autoref{def:writable}).
    \end{enumerate}

    Because all such \RISCLang configurations reachable from the initial one are in $\RefRelWR$, it then follows from \autoref{thm:Rwr-respects-own-guarantees} that they respect their own guarantees, as required.
\end{proof}

We then initialise the compiler with an empty $\InitComprec \oftype \Comprec$
that knows nothing about the register contents,
and does not assume any variables to be stable:

\begin{definition} [Empty compilation record $\InitComprec$]
\label{def:init-comprec}
\[
    \InitComprec\ \defineq\ (\AlwaysNone, (\emptyset, \emptyset))
\]
\end{definition}

With these definitions we have the desired consistency result:

% Isabelle name: no_locks_acquired_C0_mds0_init_reqs
\begin{lemma} [Initial $\InitComprec,\InitMds$ are consistent with $\NoLocksHeld$]
\label{thm:init-conds-consistent}
\[
    \NoLocksHeld\ \mem\ \Longrightarrow\
    \CompiledCmdConfigConsistent\ \InitComprec\ \regs\ \InitMds\ \mem
\]
\end{lemma}
\begin{proof}
    This is straightforward by unfolding Definitions \ref{def:compiled-cmd-config-consistent}, \ref{def:no-locks-held}, \ref{def:init-mds}, and \ref{def:init-comprec}, also relying on the cleanliness conditions
    \autoref{thm:lone-lock-per-var} and
    \autoref{thm:lock-interp-no-overlap}
    on locking disciplines
    specified in \autoref{sec:locking-restrictions}.
\end{proof}

% robs: The other 'per-thread' option will be relegated to the next section.
We now have enough information to derive a whole-system security result,
for concurrent \RISCLang programs
obtained by running the \WRCompiler
on any list ``$\cs$'' of secure \WhileLang commands
(one for each thread in the program).
As we explained in \autoref{sec:wr-compiler-no-hbranch},
the \Covern \WRCompiler's preservation of security is only for programs with $\NoHighBranching$ (\autoref{def:no-h-branching}); furthermore, so that we can derive global compatibility for multiple of these programs run concurrently as threads (as per \autoref{sec:init-conds-global-compat}), we will impose $\NoLocksHeld$ (\autoref{def:no-locks-held}) as an initial condition.
Therefore, the security preservation theorem we choose to prove here demands
that every thread of the input \WhileLang program
be $\ComSecureParam{\NoLocksHeld}{\NoHighBranching}$
(\autoref{def:com-secure}, with additional requirements as specified).
It then promises that the output program is
$\SysSecureParam{\NoLocksHeld}$:

% Isabelle name: whole_system_compilation_sys_refine
\begin{theorem} [Secure threads compiled by the \WRCompiler form a secure system]
\label{thm:compile-cmd-sys-refine}
\begin{gather*}
\begin{align*}
    & \length\ \cms_r = \length\ \cs
    \ \land \\
    & \forall i < \length\ \cms_r.\ \exists l\ \nl\ \PCs\ l'\ \nl'\ C'\ \regs. \\
    & \quad \ComSecureParam{\NoLocksHeld}{\NoHighBranching}\ (\listderef{\cs}{i},\InitMds)
    \ \land\ \\
    & \quad (\forall \mem.\ \NoLocksHeld\ \mem \longrightarrow \LocalModeUse\ \LocalConfWhile{c}{\InitMds}{\mem})
    \ \land\ \\
    & \quad (\PCs, l', \nl', C', \False) = \CompileCmd\ \InitComprec\ l\ \nl\ \listderef{\cs}{i}
    \ \land\ 
    \CompilerInputReqs\ \InitComprec\ l\ \nl\ \listderef{cs}{i}
    \ \land\ \\
    & \quad \listderef{\cms_r}{i} = (((0, \mapfst\ \PCs), \regs), \InitMds)
\end{align*} \\
\cline{1-2}
\SysSecureParam{\NoLocksHeld}\ \cms_r
\end{gather*}
\end{theorem}
\begin{proof}
    We invoke \autoref{thm:sys-refine},
    supplying:
    \begin{itemize}
        \item $\NoLocksHeld$ for the $\INITparam$ parameter at both
            \WhileLang and \RISCLang level.
        \item $\BAll$, $\RefRelWR$, $\CouplInvWR$ to be respectively the
            witness bisimulation, %$\mathcal{B}$,
            refinement relation, %$\mathcal{R}$,
            and coupling invariant %$\mathcal{I}$
            for all compiled threads,
            where we define $\BAll$ to be the arbitrary union of
            all strong low-bisimulations modulo modes
            that disallow high-branching:
            \[
                \BAll\ \defineq\ 
                \bigcup\ \{\ \mathcal{B}\ |\ \StrongLowBisimMM\ \mathcal{B} \land \NoHighBranching\ \mathcal{B}\ \}
            \]
        \item $\InitMds$ to be the initial mode state for all \WhileLang threads in $\cs$.
            %by setting
            %$\cms_A\ \defineq\ \map (\lambda c.\ (c, \InitMds)) \cs$.
    \end{itemize}
    The first thing we must prove is that the original program satisfies
    $\SoundModeUse$ (\autoref{def:sound-mode-use})
    %global modes compatibility
    %$\GlobalModeUse$
    when initialised with $\InitMds$ and
    $\NoLocksHeld$;
    we have the local part from this theorem's
    $\LocalModeUse$ assumption,
    and the global part from \autoref{thm:init-conds-global-compat}.

    We then discharge the demands of $\SysRefineReqs\ \BAll\ \RefRelWR\ \CouplInvWR$
    (\autoref{def:sys-refine-reqs})
    using
    \autoref{thm:secure-refinement-R_wr},
    \autoref{thm:Rwr-local-mode-compliance},
    and by unfolding \autoref{def:couplinv-wr}.

    It only remains for us to show that the
    initial \RISCLang--\WhileLang and \WhileLang--\WhileLang
    configuration pairs of interest
    are captured respectively by $\RefRelWR$ and $\BAll$.
    We obtain the former using this theorem's assumptions and
    \autoref{thm:init-conds-consistent}
    to discharge the guards of
    \autoref{thm:compile-cmd_correctness_R_wr}.
    Finally, we use the assumption
    that the original program is $\ComSecureParam{\NoLocksHeld}{\NoHighBranching}$
    and unfold \autoref{def:com-secure}
    to obtain that
    there exists some $\StrongLowBisimMM\ \mathcal{B}$ that
    enforces $\NoHighBranching$
    for every configuration pair with
    low-equal memories (modulo $\InitMds$) and $\NoLocksHeld$ initially;
    therefore, these state pairs must all be captured by $\BAll$.
\end{proof}

\subsubsection{Per-thread compositional security preservation}
\label{sec:per-thread_wr}

For system developers who may want to run programs
compiled from \WhileLang to \RISCLang
concurrently with other programs
written directly in \RISCLang,
per-thread security preservation results may be useful.
To compose the security proofs for those threads,
direct \RISCLang-level lemmas for the ``$\SoundModeUse$'' side conditions
of the compositionality theorem (\autoref{thm:com-secure-composes})
will also be needed.
We therefore present these
as an alternative method to obtain compositional security
results for \RISCLang programs,
applicable when only partially produced
by compilation from \WhileLang by the \WRCompiler.
% robs: Hmm.... not really 'duplicates', more like does things
%       that might be considered unnecessary.
%(Note that, to some extent,
%this section duplicates some effort of the preceding section.)

Given the facts we established in
%the preceding sections,
\autoref{sec:proof-cvdni-compilation},
we have straightforwardly that such programs' executions are captured by the bisimulation derived from $\mathcal{B},\RefRelWR,\CouplInvWR$, when started with reasonable initial configurations corresponding to those paired by $\mathcal{B}$:

% Isabelle name: compile_cmd_correctness_Bc_of
\begin{lemma} [Programs witnessed by $\mathcal{B}$ are captured by $\BCofApplyTo{\mathcal{B}}{\RefRelWR}{\CouplInvWR}$ once compiled]
\label{thm:compile-cmd_correctness_Bcof_wr}
\[
\mprset{vskip=0.5ex}
\inferrule{
    \StrongLowBisimMM\ \mathcal{B} \and
    (\LocalConfWhile{c}{\mds}{\mem_1},
     \LocalConfWhile{c}{\mds}{\mem_2}) \in \mathcal{B} \\
    (\PCs, l', \nl', C', \failed) = \CompileCmd\ C\ l\ \nl\ c \and
      \CompilerInputReqs\ C\ l\ \nl\ c \\
    \failed = \False \and
      \CompiledCmdConfigConsistent\ C\ \regs\ \mds\ \mem_1 \and
      P = \mapfst\ \PCs \\
    \CompiledCmdConfigConsistent\ C\ \regs\ \mds\ \mem_2
} {
    (\LocalConfRISC{0}{P}{\regs}{\mds}{\mem_1},
     \LocalConfRISC{0}{P}{\regs}{\mds}{\mem_2})
    \in \BCofApplyTo{\mathcal{B}}{\RefRelWR}{\CouplInvWR}
}
\]
\end{lemma}
\begin{proof}
    Straightforward from the definition of $\BCof$ (\autoref{def:concrete-bisim}), using \autoref{thm:compile-cmd_correctness_R_wr} to show membership of $\RefRelWR$, and the definition of $\StrongLowBisimMM$ (\autoref{def:strong-low-bisim-mm}) to show that the memories are low-equal modulo modes, as required by $\BCof$.
    Finally, membership of $\CouplInvWR$ (\autoref{def:couplinv-wr}) follows from the fact that the paired configurations are at the same location (program counter 0) of the same program $P$.
\end{proof}

We are ready to state the per-thread security preservation result formally.
Given an input \WhileLang command that satisfies
$\ComSecureParam{\NoLocksHeld}{\NoHighBranching}$ with $\InitMds$ initially,
it promises that the \RISCLang program
output by the \WRCompiler
is $\ComSecureParam{\NoLocksHeld}{\PCSecurity}$ with $\InitMds$:

\begin{restatable} [Preservation of per-thread confidentiality by the \WRCompiler] {theorem} {thmcompilecmdcomsecure}
\label{thm:compile-cmd-com-secure}%
% Isabelle name: per_thread_compilation_secure_thesis
\[
\mprset{vskip=0.5ex}
\inferrule{
    \ComSecureParam{\NoLocksHeld}{\NoHighBranching}\ (c, \InitMds) \\\\
    (\PCs, l', \nl', C', \False) = \CompileCmd\ \InitComprec\ l\ \nl\ c \and
    \CompilerInputReqs\ \InitComprec\ l\ \nl\ c
} {
    \ComSecureParam{\NoLocksHeld}{\PCSecurity}\ (((0, \mapfst\ \PCs), \regs), \InitMds)
}
\]
\end{restatable}
\begin{proof}
    We are given by $\ComSecureParam{\NoLocksHeld}{\NoHighBranching}$ (\autoref{def:com-secure}) that for low-equal starting configurations (modulo modes) of $c$ with no locks held, there exists some witness $\mathcal{B}$ satisfying both
    $\StrongLowBisimMM$ and
    $\NoHighBranching$.

    From this and \autoref{thm:compile-cmd_correctness_Bcof_wr} we have that the output program's corresponding execution is captured by a \RISCLang semantics-level relation
    $\BCofApplyTo{\mathcal{B}}{\RefRelWR}{\CouplInvWR}$
    derived from this $\mathcal{B}$, with
    \autoref{thm:init-conds-consistent} discharging
    the $\CompiledCmdConfigConsistent$ requirements.

    \autoref{thm:secure-refinement-R_wr} then gives us that
    $\SecureRefinement\ \mathcal{B}\ \RefRelWR\ \CouplInvWR$ holds, and
    from this and $\StrongLowBisimMM\ \mathcal{B}$ using \autoref{thm:strong-low-bisim-mm-preserved}
    we have
    $\StrongLowBisimMM\ (\BCofApplyTo{\mathcal{B}}{\RefRelWR}{\CouplInvWR})$.
    This is enough to show $\ComSecureParam{\NoLocksHeld}{\PCSecurity}$ for the \RISCLang program, by \autoref{def:com-secure};
    as \autoref{sec:couplinv_wr} noted,
    $\PCSecurity$ (\autoref{def:pc-security}) is immediate
    from the definition of $\CouplInvWR$. %\autoref{def:couplinv-wr}
\end{proof}

To prove a whole-system security result at the \RISCLang level for the compiled program, we must also prove $\SoundModeUse$ (\autoref{def:sound-mode-use}).
To that end, we prove a local and global result for \RISCLang programs output by the \WRCompiler when given a secure \WhileLang program.
The former follows
from the local compliance result in the preceding section:

% Isabelle name: per_thread_compilation_locally_sound
\begin{lemma} [Threads compiled by the \WRCompiler obey local compliance]
\label{thm:compile-cmd-local-compliance}
\[
\mprset{vskip=0.5ex}
\inferrule{
    (\PCs, l', \nl', C', \False) = \CompileCmd\ \InitComprec\ l\ \nl\ c \and
    \CompilerInputReqs\ \InitComprec\ l\ \nl\ c \\\\
    \NoLocksHeld\ \mem
} {
    \LocalModeUse\ \LocalConfRISC{0}{\mapfst\ \PCs}{\regs}{\InitMds}{\mem}
}
\]
\end{lemma}
\begin{proof}
    We use \autoref{thm:compile-cmd_correctness_R_wr} and
    \autoref{thm:init-conds-consistent}
    to obtain membership in $\RefRelWR$,
    which then allows us to use
    \autoref{thm:Rwr-local-mode-compliance}.
\end{proof}

% robs: I briefly considered moving this to sec:target-language,
%       and remarking strongly up front that there exists an option
%       that doesn't require using it.
%       But I think it's too much of a distraction if presented before the
%       whole-system preservation result in the preceding subsection.
Then we prove invariance of global modes compatibility (as in \autoref{sec:global-compatibility-locks}) for compiled \RISCLang programs, due to \RISCLang's identical semantics to \WhileLang regarding locking and modes:

% Isabelle name: whole_system_global_soundness
% Isabelle name: no_locks_empty_mds_global_soundness
\begin{lemma} [Initialising \RISCLang with $\NoLocksHeld, \InitMds$ ensures global compatibility]
\label{thm:init-risc-global-compat}
\[
\inferrule{
    \NoLocksHeld\ \mem \and
    \forall (((\pc, P), \regs), \mds) \in \toSet\ \cms_r.\ \mds = \InitMds
} {
    \GlobalModeUse\ (\cms_r, \mem)
}
\]
\end{lemma}
\begin{proof}
    We firstly prove versions of
    \autoref{thm:single-step-managed},
    \autoref{thm:single-step-unmanaged}, and
    \autoref{thm:soundness-mode-management}
    for \RISCLang,
    following exactly the same reasoning as we did in
    \autoref{sec:global-compatibility-locks}
    for \WhileLang.
    This is because the \RISCLang instructions $\RISCLockAcq\ k$ and $\RISCLockRel\ k$ are (like $\LockAcq{k}$ and $\LockRel{k}$ in \WhileLang) the only ones in their language that modify mode state, and their semantics regarding mode state and lock memory are identical to those of the $\LockAcq{k}$ and $\LockRel{k}$ commands.
    The present result then follows for the same reason that \autoref{thm:init-conds-global-compat} did for \WhileLang.
\end{proof}

With this result,
%\autoref{thm:compile-cmd-local-compliance} and \autoref{thm:init-risc-global-compat},
it is now possible to invoke
\autoref{thm:com-secure-composes}
to compose \RISCLang-level
per-thread security and mode compliance, % properties,
%into a whole-system security property,
whether they were obtained via the \WRCompiler
(using \autoref{thm:compile-cmd-com-secure} and
\autoref{thm:compile-cmd-local-compliance}, respectively),
or proved directly at \RISCLang level.

We remark that,
for programs wholly compiled by the \WRCompiler,
\autoref{thm:compile-cmd-sys-refine}
can be subsumed by
a whole-system preservation result
%with less assumptions than
that no longer demands %for which
%the per-thread compositionality side condition
$\LocalModeUse$ for each thread,
%is no longer needed
due to our ability to obtain it
directly at \RISCLang level:
%(using \autoref{thm:compile-cmd-local-compliance}):

% Isabelle name: whole_system_compilation_secure_thesis
\begin{theorem} [Secure threads compiled by the \WRCompiler form a secure system]
\label{thm:compile-cmd-sys-secure}
\begin{gather*}
\begin{align*}
    & \forall i < \length\ \cms_r.\ \exists c\ l\ \nl\ \PCs\ l'\ \nl'\ C'\ \regs. \\
    & \quad \ComSecureParam{\NoLocksHeld}{\NoHighBranching}\ (c,\InitMds)
    \ \land\ \\
    % Actually it's not a preservation result, but establishment
    % independent of whether the original program complies.
    %& \quad (\forall \mem.\ \NoLocksHeld\ \mem \longrightarrow \LocalModeUse\ \LocalConfWhile{c}{\InitMds}{\mem})
    %\ \land\ \\
    & \quad (\PCs, l', \nl', C', \False) = \CompileCmd\ \InitComprec\ l\ \nl\ c
    \ \land\ 
    \CompilerInputReqs\ \InitComprec\ l\ \nl\ c
    \ \land\ \\
    & \quad \listderef{\cms_r}{i} = (((0, \mapfst\ \PCs), \regs), \InitMds)
\end{align*} \\
\cline{1-2}
\SysSecureParam{\NoLocksHeld}\ \cms_r
\end{gather*}
\end{theorem}
\begin{proof}
    By \autoref{thm:com-secure-composes} and unfolding \autoref{def:sound-mode-use}, we are required to prove security and local mode compliance for every thread of the compiled \RISCLang program, and global modes compatibility between them all as a whole, assuming $\NoLocksHeld$ and using $\InitMds$ initially.
    These requirements are immediate using \autoref{thm:compile-cmd-com-secure}, \autoref{thm:compile-cmd-local-compliance}, and \autoref{thm:init-risc-global-compat}.
\end{proof}

 % from thesis

\section{Case study: Cross Domain Desktop Compositor input handler}
\label{chp:cddc}

This section presents---as the main case study for the \Covern \WRCompiler---%
a mixed-sensitivity concurrent program whose source-level
noninterference properties are
preserved by verified secure compilation down to an assembly-level model.

The Cross Domain Desktop Compositor (CDDC) of \citet{Beaumont_MM_16}
is a desktop device that gives
trusted users the option of
replacing multiple monitors, keyboards, and mice %mouse setups
with a single multi-level secure %(MLS)
user interface (via a single monitor, keyboard, and mouse,
as depicted in \autoref{fig:cddc-user})
when using several desktop computers simultaneously.

Here we present as case study
a program (replacing customised hardware)
that handles the incoming mouse and keyboard inputs to the CDDC.
This program has served as a particularly good case study,
because it features both of the characteristics
for which proving information-flow security is this
work's main focus:
\begin{itemize}
    \item Concurrency---%
        here, between \emph{software components} whose execution
        is interleaved
        (by the seL4 operating-system microkernel \citep{Klein_AEMSKH_14}),
        and that interact via shared memory.
    \item
        Mixed-sensitivity reuse---here, of system resources
        (notably the input devices) and memory locations,
        for input whose sensitivity level can be different at different times.
\end{itemize}

By exercising the \Covern \WRCompiler on
%\emph{compiler verification} result of
%\autoref{chp:wr-compiler}
a \WhileLang model of this case study,
we show this compiler verification-based approach to be feasible
for obtaining the preservation of noninterference properties
proved at \WhileLang level,
straightforwardly and for little extra effort,
down to a \RISCLang model of the program.

The section will proceed as follows.
Following an overview
in \autoref{sec:cddc-overview}
of the main characteristics of the case study,
\autoref{sec:cddc-results-formally}
presents the formal security properties
proved about its \WhileLang model---%
as our focus is its compilation, further details on this model and
the proof techniques used to prove
these properties at \WhileLang level are left to \citet{robs-phd}.
\autoref{sec:case-study-wr} then presents the formal preservation
of security properties down to a \RISCLang model,
obtained from running the verified \WRCompiler of \autoref{chp:wr-compiler}
on the \WhileLang model.

% NB: several subsection headings are hidden in cddc-model-fresh
\subsection{Overview of the case study}
\label{sec:cddc-overview}
\lstdefinelanguage{While}{
    morekeywords={if, then, else, fi, skip, while, do, od, lock, unlock},
    morecomment=[s]{/*}{*/},
    % Mixing tt with normalfont is a bit of a hack, but I really don't
    % want to mess with the fact that I've already referred to
    % constants and variables names as tt in the text,
    % and the default tt doesn't support bold. So, whatever.
    basicstyle=\normalfont, % No bold.
    identifierstyle=\ttfamily,
    %keywordstyle=\color{black}\normalfont\bfseries,
    keywordstyle=\color{black}\bfseries,
}
% https://tex.stackexchange.com/q/33685
\lstset{% https://tex.stackexchange.com/a/33021
        %basicstyle=\ttfamily,
        language=While,
        % Customise listing to wrap, and a red arrow to indicate wrap.
        %   https://tex.stackexchange.com/a/116572
        breaklines=true,
        postbreak=\mbox{\textcolor{red}{$\hookrightarrow$}\space}}
% https://tex.stackexchange.com/a/33021
\lstset{basicstyle=\scriptsize}

\begin{figure}[t]
    \begin{subfigure}[b]{0.49\textwidth}
        \centering
        \ifPreprint
        \includegraphics[width=\textwidth]{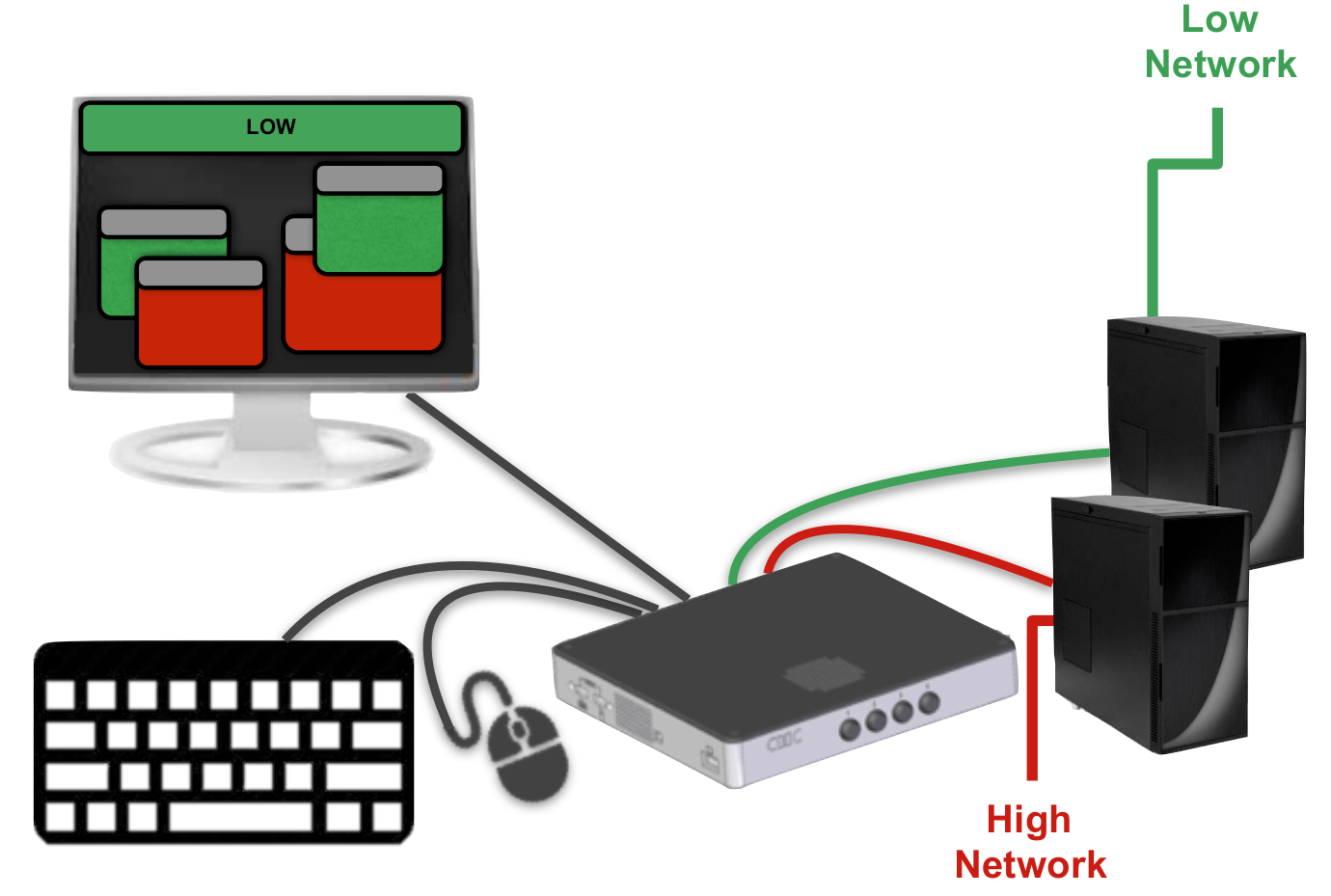}
        \else % !ifPreprint
        \includegraphics[width=\textwidth]{cddc_hardware_user_setup.eps}
        \fi % !ifPreprint
        \caption{CDDC hardware use-case setup. \smallskip \\
        The bar painted at the top of the screen
        %by the video compositor
        indicates the computer set to receive all keyboard events.
        Mouse events are delivered to the owner of the topmost window
        underneath the mouse cursor.}
        \label{fig:cddc-user}
    \end{subfigure}
    \hfill
    \begin{subfigure}[b]{0.48\textwidth}
        \centering
        \ifPreprint
        \includegraphics[width=\textwidth]{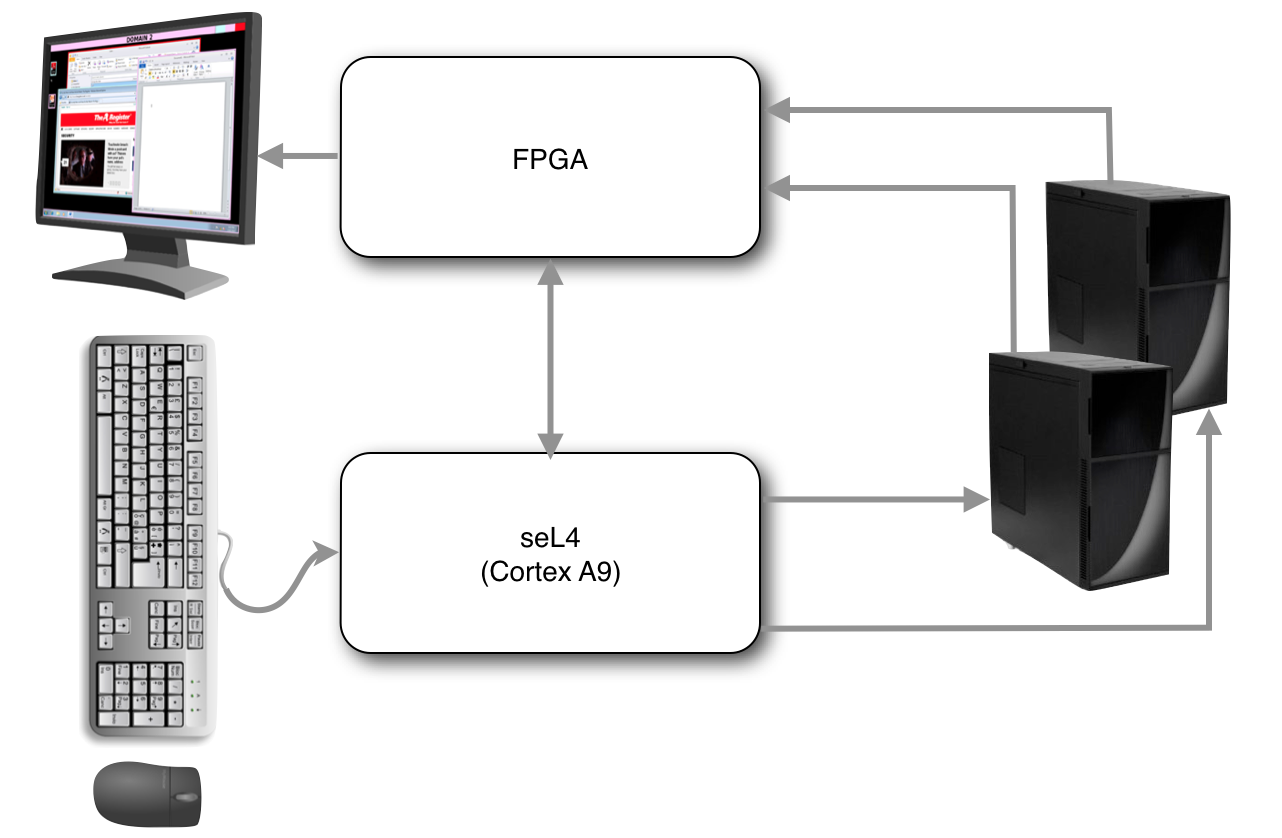}
        \else % !ifPreprint
        \includegraphics[width=\textwidth]{FPGACDDCSplit.eps}
        \fi % !ifPreprint
        \caption{CDDC hardware architecture. \smallskip \\
        The HID switch---implemented in software on top of seL4---runs
        %The software implementation of the HID switch runs
        on an ARM Cortex A9 core,
        and operates a compositor device implemented (as in \citet{Beaumont_MM_16}) using an FPGA.}
        \label{fig:cddc-hw}
    \end{subfigure}
    \newcommand{\figCDDChw}{Functional schematics for Cross Domain Desktop Compositor hardware}
    \caption[\figCDDChw]{\figCDDChw. \\
    Reproduced from \citet{Murray_SE_18}.}
    \label{fig:cddc-setup}
\end{figure}

The case study is a software implementation of the \emph{human interface device (HID) switch} in the CDDC (see \autoref{fig:cddc-hw}).
In short, this part of the CDDC is
responsible for determining
the destination of all \emph{HID input} (\emph{keyboard} and \emph{mouse} device) events,
and ensuring that the user
%suffers no confusion about that destination.
remains informed
of that destination
(by operating a \emph{video compositor} device, which renders display elements for that purpose on a shared monitor, as depicted in \autoref{fig:cddc-user}).
%(by keeping the compositor-device hardware updated accordingly).

\subsubsection{Information-flow security}

The HID switch's responsibilities are security critical, as the CDDC is intended % for use
%The security story for this program arises from the expectation that
to provide an
interface
%in situations where
to multiple desktop computers
%resides in
belonging to
different security domains;
hence,
the user of the CDDC is expected to choose the sensitivity of the data they
%type into the keyboard
input,
based on
%which security domain is ``active''---i.e.~
the computer to which they expect it to be delivered.
Furthermore, part of the CDDC's functionality is to allow users to choose which computer they are interacting with,
%where the input events are to be delivered,
by clicking on (accordingly responsive) display elements using the mouse.
Thus, the desired information-flow security property
for the HID switch
is that, in providing this functionality,
%it manages the user's expectations such that
it never
delivers inputs to a destination contrary to the user's expectations.
%to be proved for the input-processing switch is that keyboard events not be delivered to the ``wrong'' destination.

We simplify analysis to the classic $\High \not\rightarrow \Low$ security policy over the basic two-point $\{\High, \Low\}$ security lattice, and model the HID switch to service only two potential destination computers.%
\footnote{Aside from presenting a more minimal case study,
any verification for an arbitrary security lattice can be reduced to multiple applications of verification to the basic $\High \not\rightarrow \Low$ policy, with the locations reclassified appropriately.
Furthermore, the design of the CDDC's HID switch program is symmetrical for each user.}
One computer is designated as belonging to the $\High$ security domain, and is the only legitimate destination for $\High$-sensitivity input events;
the other is designated as belonging to the $\Low$ security domain.
The hardware and connections that the \CDDCswitch component uses
to forward events to these computers are modelled as shared variables
classified statically: one $\High$, the other $\Low$
(as depicted in \autoref{fig:cddc-kb-outputs}).
The attacker is then considered to be an entity that can read at any time from
%the output-device interface variable that is classified
the $\Low$-classified one.

\begin{figure}[t]
    \ifPreprint
    \includegraphics[width=\textwidth]{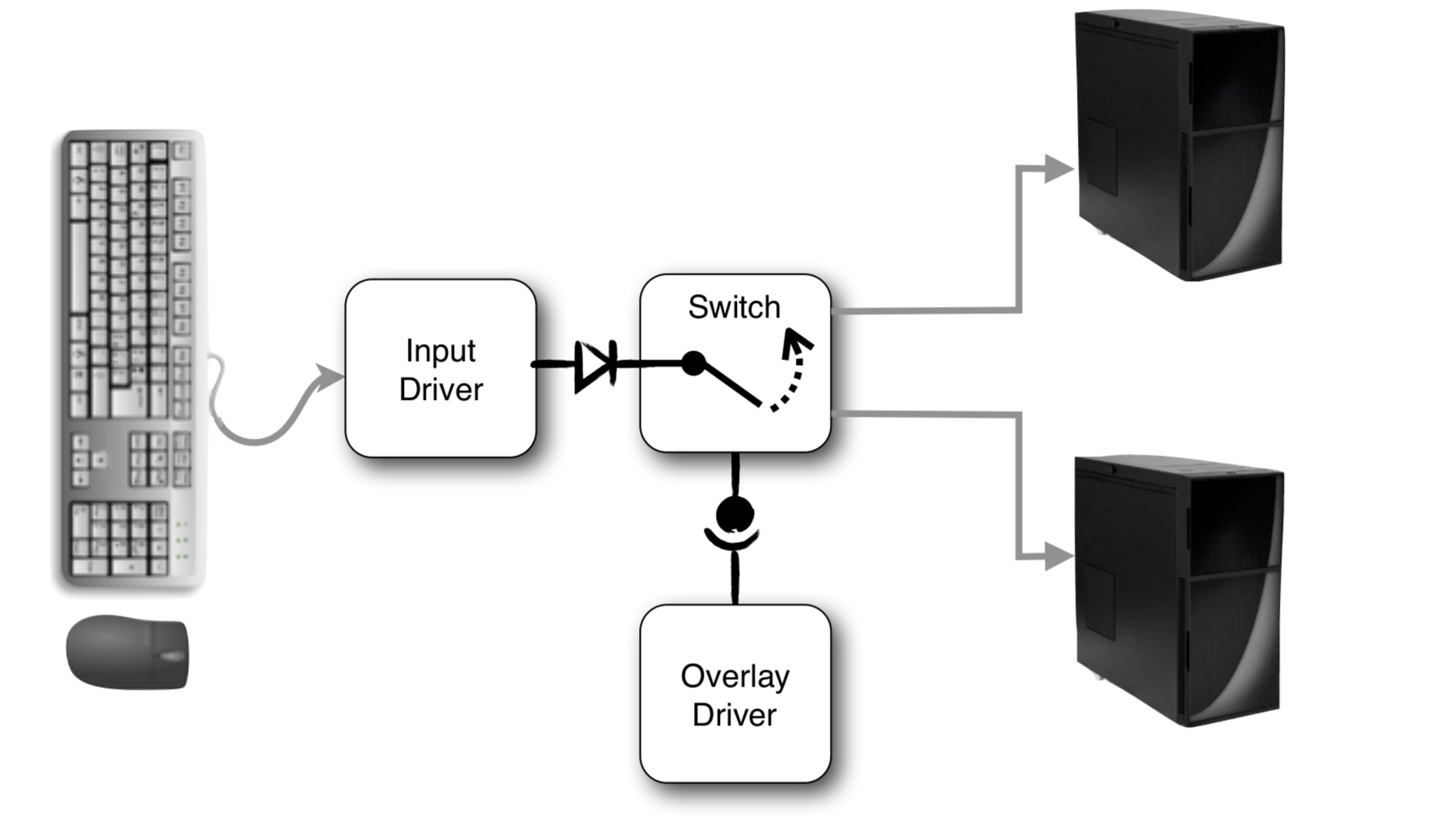}
    \else % !ifPreprint
    \includegraphics[width=\textwidth]{cddc_seL4_component_arch.eps}
    \fi % !ifPreprint
    %\caption{3-component model architecture.}
    %\caption{Component architecture for \WhileLang model of the CDDC's HID switch.}
    \newcommand{\figCDDCsw}{Functional schematic of seL4 component architecture for CDDC HID switch}
    \caption[\figCDDCsw]{\figCDDCsw. \\
    Reproduced from \citet{Murray_SE_18}.}
    \label{fig:cddc-3comp-arch}
\end{figure}

\subsubsection{Shared-variable concurrency}

\begin{figure}[h]
\begin{subfigure}[l]{0.48\textwidth}
\input{graphics/fig-cddc-kb-inputs}
\end{subfigure}
\hfill
\begin{subfigure}[l]{0.48\textwidth}
\input{graphics/fig-cddc-kb-outputs}
\end{subfigure}
\newcommand{\figCDDCexternal}{Examples of external device interactions by the CDDC HID switch, as modelled in \WhileLang}
\caption[\figCDDCexternal]{\figCDDCexternal---here, for the keyboard events. The full model is in the Isabelle/HOL supplement.}
\label{fig:cddc-uses-external}
\end{figure}

\begin{figure}[h]
\begin{subfigure}[l]{0.47\textwidth}
\input{graphics/fig-cddc-query-top}
\end{subfigure}
\hfill
\begin{subfigure}[l]{0.49\textwidth}
\input{graphics/fig-cddc-switch-top}
\end{subfigure}
\newcommand{\figCDDCcompositor}{Excerpts of the \CDDCswitch component interfacing with the compositor device}
\caption[\figCDDCcompositor]{\figCDDCcompositor.}
\label{fig:cddc-uses-compositor}
\end{figure}

\begin{figure}[h]
\begin{subfigure}[l]{0.49\textwidth}
\input{graphics/fig-cddc-buffer-init}
\end{subfigure}
\hfill
\begin{subfigure}[l]{0.47\textwidth}
\input{graphics/fig-cddc-buffer-copy}
\end{subfigure}
\newcommand{\figCDDCbuffer}{Excerpts of the \CDDCswitch component interacting with the input-event buffer}
\caption{\figCDDCbuffer.}
\label{fig:cddc-uses-buffer}
\end{figure}

The software implementation
(replacing the original FPGA-based implementation \citep{Beaumont_MM_16})
of the CDDC's HID switch
is a system of software components written in C, that all run in user mode on top of the seL4
%operating-system
microkernel \citep{Klein_AEMSKH_14}.

Here, we have abstracted from the seL4-based C implementation's details, to model in the \WhileLang language the basic functionality of its three main software components (as depicted in \autoref{fig:cddc-3comp-arch}) as a shared-variable concurrent program of three threads:
%Their roles can be summarised as follows:
\begin{itemize}
    \item The \CDDCinput driver is responsible for taking events from input-device interfaces and placing them on an input-event buffer for consumption by the \CDDCswitch (\autoref{fig:cddc-kb-inputs}).

    \item The \CDDCswitch is responsible for
        inspecting all input events on the buffer
        from the \CDDCinput driver, %(\autoref{fig:cddc-buffer-copy}),
        querying the compositor device (as modelled in
        \autoref{fig:cddc-query-top})
        and \CDDCoverlay driver
        to determine
        if any constitute a user-directed change to
        the destination of subsequent events,
        and if so,
        updating the compositor device
        to display that change (as modelled in \autoref{fig:cddc-switch-top}).
        %to display the new destination,
        %if so.
        %It fulfils these responsibilities both
        %by operating compositor-device interfaces directly, and
        %by making remote procedure calls to the \CDDCoverlay driver.
        Finally, it is responsible for delivering all events to their destination computer via the appropriate \emph{output-device} interface
        (\autoref{fig:cddc-kb-outputs}).%

    \item The \CDDCoverlay driver is responsible for servicing remote procedure calls (RPCs, made by the \CDDCswitch) that query a subset of the compositor-device interface, regarding the position of certain mouse-clickable elements the compositor is rendering as part of a visual overlay on the trusted user's video monitor.
        (As no mixed-sensitivity reuse occurs in this part of the model,
        we leave its details to the Isabelle/HOL supplement.)
\end{itemize}

% NB: At this point, the reader may be wondering in what way
%     the device interfaces are modelled.
\noindent
The device interfaces, shared buffers (for input events and RPC mechanisms), and local variables used by each component are all modelled as program variables in shared memory.
Consequently in the \WhileLang model, mutex locks are used to model all synchronisation and restriction of concurrent access by the components to those variables.

So that we do not need to add separate \WhileLang semantics for interacting with
private as opposed to shared memory,
we model thread-private memory as shared program variables
protected by a permanently held lock acquired at initialisation time
(e.g.~as in \autoref{fig:cddc-buffer-init}).
%memory, regardless of whether it is private or shared,
We consider this to be a stand-in for the memory isolation properties
established by the underlying operating system between the
program threads that it hosts.

% You just talked about the presence of concurrency in the program.
%     You put it down to "that's the way the actual implementation is".
%     You could possibly also explain it as a ubiquitous consequence
%       of seL4's system design philosophy
%       (perhaps here, or with a brief remark pointing to a later section).

\subsubsection{Mixed-sensitivity reuse}

% Now you need to talk about the presence (and broadly the purpose)
% of mixed-sensitivity reuse in the program.
%     It's an unavoidable fact of having info of differing sensitivity
%     travel through the one set of locations, rather than duplicating
%     (input devices, shared buffer, and switch-internal variables).
% Perhaps as a single paragraph.

Inherently to the CDDC's role as a multi-level secure user interface,
its HID switch receives data
of differing sensitivity levels (at different times)
from a single set of input device memory locations
(e.g.~as modelled in \autoref{fig:cddc-kb-inputs}, for the keyboard events),
%(via device memory),
rather than from those of distinct device sets for each sensitivity level.

Furthermore,
the HID switch propagates all input event data (regardless of sensitivity)
%That data is propagated
through a single set of memory locations
(the input-event buffer and \CDDCswitch-internal copies of its contents,
as modelled in \autoref{fig:cddc-uses-buffer}),
rather than % propagating it
%statically allocating
duplicating those memory locations
for each security domain.
Consequently in the \WhileLang model, all of these memory locations that are subject to mixed-sensitivity reuse are assigned value-dependent classifications, reflecting the trusted user's expectation of the sensitivity level of the data they contain: %, as at the time they originally entered that data.

\begin{itemize}
    \item
        To model a user that we trust to type sensitive information into
        the keyboard only when the compositor device indicates
        the $\High$ domain computer is \emph{active}
        (i.e.~set to receive all keyboard events),
        we have the \CDDCinput driver draw keyboard events
        from a shared variable named \HidKeyboardSource
        (as depicted in \autoref{fig:cddc-kb-inputs})
        that has classification dependent on
        a control variable \IndicatedDomain
        modelling the relevant state of the compositor
        (here, \texttt{DOM\_HIGH} is a designated constant):
        \begin{align*}
            \begin{cases}
                \High,& \text{if } \IndicatedDomain = \texttt{DOM\_HIGH} \\
                \Low, & \text{otherwise.}
            \end{cases}
        \end{align*}

        %This models a user that is trusted to type sensitive information into
        %the keyboard only when they see that the compositor device is
        %indicating that it is
        %the $\High$ domain whose computer is \emph{active}, meaning it is
        %currently the computer that is set to receive all keyboard events.
        %
        %Following a similar assumption by \citet{Beaumont_MM_16},
        %we model the user's perception and actions as being entirely faithful
        %to what is indicated by the compositor, by using a
        %Here, the shared variable named \IndicatedDomain to model the security
        %domain that the compositor is indicating as active.

    \item In contrast,
        as clicking on composited user interface elements
        has the potential ability to change the
        future \IndicatedDomain %active domain
        (which, as a control variable, is never allowed to
        receive any $\High$-sensitivity data),
        the model trusts the user not to encode sensitive
        information into the mouse input in any way.
        %have a potential ability to influence the future value of control
        %variables, which in turn are not ever
        %permitted to receive any $\High$-sensitivity data.
        Thus, the $\CDDCinput$ driver always draws mouse events from a
        statically $\Low$-classified shared variable.
        %regardless of the current value of \IndicatedDomain.

        Consequently, as the data portion \InputEventData
        of the input-event buffer%
        \footnote{We model in \WhileLang only a single-place buffer, which could easily be extended to a buffer of arbitrary size by duplicating the same basic pattern of access, classification, and lock-protection, for multiple places.}
        between the \CDDCinput driver and \CDDCswitch
        may carry either
        keyboard data of value-dependent sensitivity
        %(dependent on whether the \IndicatedDomain was $\High$
        %when the user typed it),
        or $\Low$-sensitivity mouse data,
        we assign it a classification dependent on
        the values of both its control portion and the \IndicatedDomain:
        \begin{align*}
        \begin{cases}
            \High,& \text{if } \InputEventType = \texttt{KEYBOARD}\ \land
                    \enskip\ \IndicatedDomain = \texttt{DOM\_HIGH} \\
            \Low, & \text{otherwise.}
        \end{cases}
        \end{align*}

    \item Finally, we model the seL4-based \CDDCswitch component's copying
        of the event from the buffer into its own local variables,
        giving its data portion a classification dependent on
        its own private view of the currently active domain
        (modelled as \ActiveDomain):
        \begin{align*}
        \begin{cases}
            \High,& \text{if } \CurrentEventType = \texttt{KEYBOARD}\ \land
                    \enskip\ \ActiveDomain = \texttt{DOM\_HIGH} \\
            \Low, & \text{otherwise.}
        \end{cases}
        \end{align*}

        To ensure \ActiveDomain remains authoritative with what is composited by
        the CDDC into the display, in the \WhileLang model
        the \CDDCswitch initialises \IndicatedDomain to
        match the initial value of \ActiveDomain
        (as depicted in \autoref{fig:cddc-buffer-init}),
        updates it whenever \ActiveDomain changes
        (as depicted in \autoref{fig:cddc-switch-top}),
        and checks at runtime that $\ActiveDomain = \IndicatedDomain$
        when copying data from the buffer to its own private variables
        (as depicted in \autoref{fig:cddc-buffer-copy}).

\end{itemize}

The CVDNI properties' (1) value dependence on control variables,
(2) quantification over all initial values for the control variables
and (3) assumptions of environmental havoc on write-unprotected shared variables
between evaluation steps (\autoref{def:cg-consistent})
then ensure noninterference between $\High$ inputs and $\Low$-classified sinks,
%$\High$-classified data never arrives at
%locations at the time they are $\Low$-classified,
regardless of the initial and dynamically changing sensitivity of
all such locations subject to mixed-sensitivity reuse.
 % from thesis

\subsection{CVDNI properties of the \WhileLang model to be preserved}
\label{sec:cddc-results-formally}
%Having now presented the \WhileLang-language model for the CDDC's HID switch,
This section will now give a brief formal exposition
of the security properties of the CDDC HID switch's \WhileLang-language model
that our compiler will preserve down to \RISCLang.
%We now state formally the results of verification of the model.

As the per-thread proof techniques for \WhileLang
that we used for the case study are outside the scope of this paper,
we note only that they consist of an adaptation to mutex locks by
\citet{robs-phd}
of a security type system and local mode compliance check developed by
\citet{Murray_SPR_16,Murray16-Dependent-SIFUM-AFP}.
Nevertheless, we have provided
their full formalisation in our Isabelle/HOL supplement,
and we refer the reader
to these prior works on their design,
and particularly to \citet{robs-phd} for further discussion
on their application to this case study.
%Further details on the design and application of
%the \WhileLang-language proof techniques
%can be found in \citet{robs-phd}.
%and in the Isabelle/HOL formalisation.

In short,
from applying local type checks
%a security type system and local compliance check
on the \WhileLang-language commands for each of the three software components
(\CDDCinput, \CDDCswitch, and \CDDCoverlay)
to obtain per-thread security ($\ComSecure$, \autoref{def:com-secure})
and modes compliance
($\LocalModeUse$, \autoref{def:local-mode-compliance}),
we have from \autoref{thm:com-secure-composes}
that the concurrent program of all three components
satisfies
the whole-system security property ($\SysSecure$, \autoref{def:sys-secure})
%from \citet{Murray16-Dependent-SIFUM-AFP}
%as presented in \autoref{sec:cvdni-definition},
as instantiated to specify that no locks are held initially.

%My approach will be to use the security type system of \autoref{sec:type-system-locks} to establish the per-thread security property for each component, and then use the compositionality theorem (\autoref{thm:com-secure-composes}) to derive the whole system security property from the per-thread ones.

So that we can use the approach we gave in
\autoref{sec:source-language}
to obtain the global modes compatibility part of
the $\SoundModeUse$ side-condition (\autoref{def:sound-mode-use}),
we specify $\NoLocksHeld$ (\autoref{def:no-locks-held})
as the $\INITparam$ requirement on memory,
and use the initial mode state $\InitMds$ (\autoref{def:init-mds})
for all of the components in the system.

This $\NoLocksHeld$ predicate and $\InitMds$ are both defined relative to a lock interpretation parameter
%$(\NoWvarsFuncCDDC, \NoRWvarsFuncCDDC)$
that we supply (as required by \autoref{sec:locking-discipline})
for the CDDC model.
The locks in the CDDC model fall under the following categories:
\begin{itemize}
    \item The locks coordinating inter-component interactions grant exclusive read--write access to the shared variables they govern.

    \item There are also locks granting the \CDDCswitch and \CDDCinput components exclusive read--write access to a set of ``private'' variables each, for internal use.
        %As depicted in \autoref{fig:cddc-buffer-init} for \CDDCswitch,
        The components acquire these prior to entering their main loop, and never release them.
    \item Finally the model uses read-atomicity locks---a practice introduced in \autoref{sec:stability-vs-data-races}.
        %and depicted in Figures \ref{fig:cddc-kb-inputs} and \ref{fig:cddc-query-top}.
        These grant exclusive write access to shared variables used to model hardware interfaces,
        to make explicit an assumption (normally implicit in the atomicity of expression evaluation in the \WhileLang language) that these variables will not have their value changed by the environment during a simple assignment from those variables.

        Note that these read-atomicity locks are not needed
        to prove confidentiality for the \WhileLang model, % confidentiality result proved in this section,
        but rather we add them to satisfy the requirements demanded by
        the $\WRCompiler$ so that it can preserve confidentiality
        %the preservation of that result
        (via small-step semantic preservation) down to the \RISCLang model.
        %by the $\WRCompiler$.
        %(this will be presented later, in \autoref{sec:case-study-wr}). %via the small-step semantic preservation demanded by $\SecureRefinement$ (\autoref{def:secure-refinement}).
\end{itemize}

%It so happens %that a lemma from \citet{Murray16-Dependent-SIFUM-AFP}
%(and unaffected by the changes in \autoref{chp:add-locks})
%gives us
% Revised:
% This draws too much attention to a type system that is out of scope. -robs.
%that the \WhileLang-language security type system of \citet{Murray_SPR_16}
%(as adapted for locks by \citet{robs-phd})
%when a thread program typechecks
%with final typing environment $\Gamma',\stableS',P'$,
%the bisimulation $\TypeBisimDefault$ constructed by the security type system
The \WhileLang-language proof techniques we apply to each thread
of the program yield
%as does its adaptation for mutex lock support by \citet{robs-phd},
%Thus, applying the security type system
%so applying it in fact yields
$\ComSecureParam{ }{\NoHighBranching}$,
a stronger version of the per-thread CVDNI property that
enforces $\NoHighBranching$ (\autoref{def:no-h-branching}).
%This is enough to give us the needed per-thread properties for preservation by the \WRCompiler via \autoref{thm:compile-cmd-com-secure}:
Furthermore, we have trivially from the definition of $\ComSecure$
(\autoref{def:com-secure})
that if a program is secure without imposing any initial conditions,
then it remains secure if we impose any $\INITparam$ parameter arbitrarily.
%---note this holds \emph{regardless} of any $\BISIMREQSparam$ requirements imposed on the bisimulation witness:
Therefore, for each thread we have
$\ComSecureParam{\NoLocksHeld}{\NoHighBranching}$
(\autoref{def:com-secure}, with
$\INITparam \defineq \NoLocksHeld$ and
$\BISIMREQSparam \defineq \NoHighBranching$):

\begin{lemmas} [Per-thread confidentiality results for CDDC \WhileLang model]
\label{thm:per-thread-cddc-stronger}
\begin{align*}
    & \ComSecureParam{\NoLocksHeld}{\NoHighBranching}\ (\CDDCoverlay,\InitMds) \\
    & \ComSecureParam{\NoLocksHeld}{\NoHighBranching}\ (\CDDCinput,\InitMds) \\
    & \ComSecureParam{\NoLocksHeld}{\NoHighBranching}\ (\CDDCswitch,\InitMds)
\end{align*}
\end{lemmas}

From this and \autoref{thm:com-secure-composes},
using local compliance checks
and \autoref{thm:init-conds-global-compat}
to discharge the
$\SoundModeUse$ (\autoref{def:sound-mode-use}) side condition,
we have a whole-system confidentiality theorem
for the system of all three components running concurrently:

\begin{theorem} [Whole-system confidentiality result for the CDDC \WhileLang model]
\label{thm:whole-system-cddc}
\begin{align*}
    \SysSecureParam{\NoLocksHeld}\ [(\CDDCoverlay, \InitMds),
                 (\CDDCinput, \InitMds),
                 (\CDDCswitch, \InitMds)]
\end{align*}
\end{theorem}

 % from thesis

\subsection{Confidentiality-preserving compilation to \RISCLang model}
\label{sec:case-study-wr}
We now turn to applying the \Covern \WRCompiler of \autoref{chp:wr-compiler} to our \WhileLang-language model of the CDDC's HID switch;
we then have automatically that it preserves the security properties
presented in \autoref{sec:cddc-results-formally}
down to the compiler's \RISCLang-language output.

The \WRCompiler is \emph{executable} in the Isabelle proof assistant.
Using Isabelle's \texttt{eval} tactic,
we execute the \WRCompiler's main function, $\CompileCmd$ (whose implementation was described in \autoref{sec:compiler-impl}) on the \WhileLang-language models for
all three of the CDDC's \CDDCinput driver, \CDDCswitch, and \CDDCoverlay driver components,
to obtain their \RISCLang-language compilations.
%(together totalling about 250 \texttt{RISC} instructions)
%Now, the \RISCLang texts of the CDDC model's components are obtained by invoking the compiler on each of their \WhileLang programs.
(Recall from \autoref{sec:compiler-impl} that we obtain the \RISCLang text trivially as the $\mapfst$ of the $CompRec$-annotated \RISCLang program, which is the $\fst$ output of $\CompileCmd$.)

\begin{definition} [\RISCLang-language program texts of CDDC model's components]
\label{def:cddc-risc-texts}
\begin{align*}
    \CDDCoverlayRISC\ &\defineq\ \mapfst\ (\fst\ (\CompileCmd\ \InitComprec\ \None\ 0\ \CDDCoverlay)) \\
    \CDDCinputRISC\ &\defineq\ \mapfst\ (\fst\ (\CompileCmd\ \InitComprec\ \None\ 0\ \CDDCinput)) \\
    \CDDCswitchRISC\ &\defineq\ \mapfst\ (\fst\ (\CompileCmd\ \InitComprec\ \None\ 0\ \CDDCswitch))
\end{align*}
\end{definition}

Our approach to obtain per-thread confidentiality for each of these \RISCLang texts will be to use the theorem of its preservation by the $\WRCompiler$ (\autoref{thm:compile-cmd-com-secure}).
Recall, this was:

\thmcompilecmdcomsecure*

%Note that this only works for non--$\High$-branching programs.

Then, for $\CompileCmd$ to execute successfully (i.e.~to return $\failed = \False$), the model must pass the $\StabilityChecks$ discussed in \autoref{sec:compiler-impl}.
All three of $\CDDCoverlay$, $\CDDCinput$, and $\CDDCswitch$ pass the checks (1) because they use locks to protect the atomicity of reads from (otherwise unstable) variables used to model hardware interfaces, and (2) as a consequence of having passed the local security and mode compliance checks mentioned in \autoref{sec:cddc-results-formally}.

We are now in a position to prove a whole-system confidentiality result for the compiled \RISCLang model---here, with each thread's register bank initialised to zero: $\InitRegs\ \defineq\ \AlwaysZero$.

\begin{theorem} [Whole-system confidentiality result for the CDDC \RISCLang model]
\begin{align*}
    \SysSecureParam{\NoLocksHeld}\ [&(((0, \CDDCoverlayRISC), \InitRegs), \InitMds), \\
                 &(((0, \CDDCinputRISC), \InitRegs), \InitMds), \\
                 &(((0, \CDDCswitchRISC), \InitRegs), \InitMds)]
\end{align*}
\end{theorem}
\begin{proof}
    A few approaches are available; we obtained formal proofs of this theorem in Isabelle/HOL using all three of the following alternatives
    (unfolding \autoref{def:cddc-risc-texts}):

    Option 1.
    Use either of
    \autoref{thm:compile-cmd-sys-refine} or
    \autoref{thm:compile-cmd-sys-secure}, both of which
    established whole-system security
    for \RISCLang outputs of the \WRCompiler when executed on $\ComSecureParam{\NoLocksHeld}{\NoHighBranching}$ \WhileLang programs
    (which we have here from \autoref{thm:per-thread-cddc-stronger}).
    This is the easiest option to take
    for programs that are already verified in the \WhileLang language, and then compiled successfully to \RISCLang by the \WRCompiler.
    It is possible to take here because all of $\CDDCoverlayRISC$, $\CDDCinputRISC$, and $\CDDCswitchRISC$ were obtained in this manner.

    Option 2.
    Use \autoref{thm:compile-cmd-com-secure} and \autoref{thm:compile-cmd-local-compliance} to obtain $\ComSecureParam{\NoLocksHeld}{\PCSecurity}$ and $\LocalModeUse$ (resp.)~for each of $\CDDCoverlayRISC$, $\CDDCinputRISC$, and $\CDDCswitchRISC$, then use
    %the compositionality result of \citet{Murray_SPR_16}
    \autoref{thm:com-secure-composes} directly to obtain $\SysSecureParam{\NoLocksHeld}$.
    This option can be used for systems where some of the threads are written directly in \RISCLang;
    for such threads, $\ComSecureParam{\NoLocksHeld}{\PCSecurity}$ and $\LocalModeUse$ would need to be proved directly at \RISCLang level.
    However, \autoref{thm:init-risc-global-compat} still discharges the
    $\GlobalModeUse$ requirement for \RISCLang, provided all threads are initialised with $\InitMds$ and $\NoLocksHeld$.

    Option 3.
    Use \autoref{thm:sys-refine} directly.
    %Use the more generic whole-system refinement framework from \citet{Murray_SPR_16} (as formalised in Isabelle/HOL by \citet{Murray16-Refinement-AFP}).
    This option can be used for systems where all the \RISCLang threads
    are secure refinements (according to \autoref{def:secure-refinement})
    of the threads of some \WhileLang program that
    satisfied $\SoundModeUse$ with $\NoLocksHeld$ initially,
    but some were obtained by other means than the \WRCompiler
    (i.e.~not all via the refinement $\RefRelWR$).
\end{proof}
 % from thesis

\section{Related work}
\label{sec:related-work}
First in \autoref{sec:related-compositionality}, we describe other
recent and related works
that address concerns of noninterference proof compositionality in a concurrent
setting (of the kind we tackled in \autoref{sec:source-language}).
The remaining sections focus on related works on verified compilation:
The works in \autoref{sec:related-ni-pres} and \autoref{sec:related-conc-ni-pres},
like ours, focus on compilation preserving a form of noninterference.
In \autoref{sec:related-robust}
we describe our work's relationship with varieties of
robust property preservation, and
other compilation verification efforts in \autoref{sec:related-general}.

\subsection{Compositionality of concurrent noninterference proofs}
\label{sec:related-compositionality}

Alternative approaches exist to establishing the
non-compositional global modes compatibility condition we proved as
invariant to concurrent \WhileLang executions in \autoref{sec:source-language}.
For the precursor (non--value-dependent) notion of concurrent noninterference
to CVDNI, \citet{Mantel_SS_11} originally proposed that such
a condition be met by a non-compositional
\emph{may happen in parallel} analysis (e.g.~\citet{Masticola93}).
Then, instead of demanding the explicit declaration of the
sorts of guarantees implied by locking discipline (as we do),
\citet{Mantel_MPW_15} proposed automating their inference
and proof of the compatibility condition
using a reachability analysis making use of dynamic pushdown networks.
% Not sure this strikes as relevant, given we're talking purely static here.
%\citet{Askarov_CM_15} proposed a solution using dynamic monitoring.
We leave adapting and implementing such an approach
for our CVDNI setting to future work.

We note also that,
like the CVDNI theory
%presented in \autoref{sec:cvdni-notions}
and our work of \autoref{sec:source-language},
recent work by \citet{Frumin21} concerns
compositionality of machine-checked proof efforts
for noninterference in a concurrent setting
that are obtained potentially via a variety of proof techniques.
They model more fine-grained synchronisation than we do here,
via atomic compare-and-swap operations
that can be used to implement mutex locking primitives.
%They instantiate their underlying theory to model a greater variety
%of synchronisation primitives.
%Their separation logic for noninterference, SeLoC, permits composition of manual proofs with ones discharged via a type system they developed on top of it; like our work, theirs is fully machine checked.
However, they do not study compilation as a means of preserving such proofs, which is the focus of our work here.
%Regarding verifying compilation between languages with different synchronisation primitives
We believe that the CVDNI refinement notions we presented could support certain cases of compilation between
%languages with
different synchronisation primitives,
provided
only new thread-private state is needed
(like the registers in \RISCLang),
and the shared variable interactions can be proved as preserved.
%new state needed is added to the thread-private state of each thread.
% (e.g. hardware registers, relying on context switch by the underlying operating system to guarantee their effective thread privacy).
For example, we expect mutex locking primitives
(with slightly different semantics to ours here)
%(e.g. write a thread identifier to the lock memory to acquire, write zero to release),
could feasibly be refined to a compare-and-swap--based implementation in this way---%
this we also leave to future work.

\subsection{Noninterference-preserving compilation}
\label{sec:related-ni-pres}

\citet{Tedesco16} present a type-directed compilation scheme that preserves
a \emph{fault-resilient} noninterference property. The compilation scheme
of our \WRCompiler was inspired by theirs. Like our $\ComSecure$ CVDNI
security property that \WRCompiler preserves, \citeauthor{Tedesco16}'s security property
is also \emph{strong bisimulation}-based~\citep{Sabelfeld00}.
But where our property accounts (via mode states) for \emph{controlled interference} by other threads, theirs instead quantifies over all possible interference by the environment with the memory contents.
While this simplifies their task of proving that their security property is preserved under compilation---as it need not require the compiler to preserve the contents of memory---it means their security property cannot capture value-dependent noninterference.
In contrast, our \WRCompiler must obey our $\SecureRefinement$ notion's requirement that memory contents are preserved.
\iffalse robs: No need for this.
\footnote{Consequently, we found and fixed a bug in their expression compiler (acknowledged privately) whereby registers in use were incorrectly reallocated.
Expressions like $v + (v + 1)$ were thus compiled incorrectly to programs yielding $(v + 1) + (v + 1)$ instead, causing a violation of memory contents preservation.}
\fi

The line of work most relevant to ours
is that which was conducted (concurrently)
by \citet{Barthe20},
wherein they achieved the remarkable result of proving
that a modification of the CompCert C compiler \citep{Leroy09} preserves
the \emph{cryptographic constant-time} class of noninterference
(2-safety) properties.
Their proof approach was to use
various notions of
\emph{constant-time simulation} (CT-simulation) first presented
by \citet{Barthe18},
originally intended for application to the
Jasmin compiler \citep{Almeida17_jasmin}.
Although not targeting
programs with concurrency or mixed-sensitivity reuse (as our work does),
CT-simulation shares in common with the refinement notions used by this paper
%(from \citet{Murray_SPR_16},
%which we recall in \autoref{chp:preliminaries},
%\autoref{fig:coupling-inv-pres}),
that it in essence rests on a simulation diagram that is cube-shaped,
as it must preserve a 2-safety hyperproperty.
We submit that \citet{Barthe20} broadly validates the argument we made in
\citet{Sison_Murray_19},
that decomposing such cube-shaped diagrams into square-shaped ones
is what will make them feasible to apply
to the verification of fully fledged compilers like CompCert---%
noting that
they described the only compilation pass they proved with their non-decomposed, cube-shaped diagram as ``not especially pleasant because the diagrams are difficult to exploit'' \citep{Barthe20}.

Note that the refinement theory of \citet{Barthe18,Barthe20} preserves
security via refinement phrased in terms of \emph{forward simulation}
\citep{Leroy09}---%
that is,
each step of the abstract program must be simulated by the target program.
In contrast, our theory presented here is instead targeted towards preserving
refinement via \emph{backward simulations},%
\footnote{Again, as commonly referred to
in the compiler verification literature from \citet{Leroy09} onwards.
This is not to be confused with the ``backward simulations'' of
concurrency verification \citep{Lynch96}
and data refinement \citep{Roever98,Cavalcanti02},
where the refined program instead simulates the original, and
where simulation proceeds from the end of the program back to the beginning.}
in which each step of the
concrete (compiled) program must be simulated by the abstract program.
This difference arises because in our setting we need to account for
leakage that might occur and be visible only in intermediate states. In
their setting, in contrast, leakage that occurs in intermediate states
remains visible forever in the concrete program semantics via a
\emph{leakage trace}. % Terminology looks consistent with Barthe et al. -robs.
It remains unclear whether we could have adopted a similar approach in our
work, thereby enabling a (simpler) forward simulation argument. In particular,
it is not clear what the semantics of leakage traces should be for a language
that supports both value-dependent classification and shared-memory concurrency
as ours does.

\subsection{Concurrency-compositional noninterference-preserving compilation}
\label{sec:related-conc-ni-pres}

Neither of the above consider per-thread compositional compilation of concurrent, shared memory programs, nor value-dependent noninterference policies -- the focus of our theory and compiler.
\citet{Barthe10,Barthe07_multi} however did aim to preserve noninterference of multithreaded programs by compilation, extending a prior \emph{(security) type-preserving} compilation approach~\citep{Barthe04,Barthe07_single}.
Their noninterference property however was termination- and timing-\emph{insensitive}, so preventing internal timing leaks relied on the scheduler disallowing certain interleavings between threads.
Also, their type-preservation argument was derived from a big-step semantics preservation property for their compiler.
Here we instead rely on preservation of a small-step semantics (specifically memory contents), which is necessary for us to preserve value-dependent security under compilation, as well as to avoid imposing non-standard requirements on the scheduler.

\subsection{Robust property preservation}
\label{sec:related-robust}

Other recent works have improved on \emph{fully abstract compilation} (surveyed by \citet{PatrignaniAC19}) by mapping out the spectrum \citep{Abate19} or developing specific forms \citep{PatrignaniG19} of \emph{robust property preservation}, concerned with \emph{robustness} of source program (hyper)properties to concrete \emph{adversarial} contexts.
Like \citet{Tedesco16}, these works differ from ours in quantifying over a wider range of hostile interference.
They also focus prominently on changes to data types, which we do not support.
Thus, as a 2-safety hyperproperty quantifying over a lesser range of interference, we expect CVDNI-preservation to be implied by R2HSP (robust 2-hypersafety preservation), but do not expect it to imply any other secure compilation criterion on \citeauthor{Abate19}'s spectrum.

While recently \citet{PatrignaniG19} instantiated their \emph{robustly safe compilation} for shared-memory fork-join concurrent programs, it only preserves (1-)safety properties.
Previously however, \citet{Patrignani17} proved their
\emph{trace-preserving compilation} preserves $k$-safety
hyperproperties~\citep{Clarkson10}, including
noninterference properties.
However, it disallows the removal or addition of trace entries, which would be necessary to change the passage of time as seen in the observable trace events.
Thus it excludes the sorts of changes to pacing carried out by our compiler (regulated by $\abssteps$) and studied as optimisations by the two other works \citep{Tedesco16,Barthe20} on timing-sensitive security-preserving compilation mentioned above.

\subsection{Compiler verification in general}
\label{sec:related-general}

Finally, there has been much work on large-scale verified compilation~\citep{Leroy09,Kumar_MNO_14} some of which has also treated compilation of shared-memory concurrent
programs~\citep{Lochbihler18} including taking weak-memory consistency into
account~\citep{Podkopaev19}. Our work here does not consider the effects of weak-memory models.
In particular, such models are often defined axiomatically
rather than operationally. Our notion of secure refinement and our decomposition
principle
(Definitions \ref{def:secure-refinement}
and \ref{def:secure-refinement-simpler}, respectively)
are defined assuming an operational
semantics for the source and target languages.

Our work differs to prior work on verified concurrent compilation, in that it formalises and proves a compiler's ability to use information about the application's locking protocol, both to exclude unsafe access to shared variables,
and conversely to know when it is safe to allow optimisations on shared variables that would typically be excluded. %(see \autoref{sec:comprec}).
 % from itp19

\section{Conclusion}
\label{sec:conclusion}
To our knowledge, we have presented the first mechanised verification that a compiler
preserves concurrent, value-dependent noninterference.
To this end, we provided a general
decomposition principle for compositional, secure refinement. Although our compiler is a proof-of-concept targeting simple source and target languages, we nevertheless
applied it to produce a verified assembly-level model of
an input-handling system for the CDDC \citep{Beaumont_MM_16},
a nontrivial mixed-sensitivity concurrent program.

We expect this decomposition principle to remain applicable in reducing noninterference-refinement proof efforts for compilers that overcome the specific limitations of ours here.
For example, a compiler that inserts padding to equalise the time taken on either side of a $\High$-branch---%
which may change when it expands expressions into multiple instructions---%
may instantiate the decomposition principle
with a more sophisticated concrete coupling invariant that does not require $\PCSecurity$.

This work
serves to demonstrate that verified security-preserving compilation for mixed-sensitivity concurrent programs is now
within reach, by augmenting traditional proof obligations for
verified compilation (e.g.\ square-shaped semantics preservation) with those specific
to security (e.g.\ absence of termination- and timing-leaks) as depicted in
\autoref{fig:decomp}. We hope that this work paves the way for future large-scale verified security-preserving compilation efforts.
 % from itp19

\subsection*{Acknowledgements}
We would like to thank our anonymous referees, and
to thank again all those who provided feedback on
the conference version of this paper
\citep{Sison_Murray_19}
and on Robert Sison's PhD thesis \citep{robs-phd}.
This paper describes research that was conducted during
Robert's PhD candidature at UNSW Sydney and CSIRO's Data61,
which was funded by
an Australian Government Research Training Program (RTP) Scholarship
and a CSIRO Data61 Research Project Award.
We thank the Trustworthy Systems group at CSIRO's Data61 for cultivating
an excellent working and learning environment.

\subsection*{Conflicts of Interest}
None

\bibliography{references}

\ifPreprint
\appendix
\renewcommand{\thesection}{\Alph{section}}

\section{Label allocation and sequential composability}
\label{sec:labels-seq}
The \WRCompiler fixes the label type $\Lab \defineq \nat$
%allocating natural numbers to use as labels for \texttt{RISC} instructions.
to allow it to ensure freshness merely by using the highest natural number reached so far on a ``next label'' counter
(the argument $\nl$%
\ifPreprint
\ in \autoref{eg:compile-cmd-invocation}%
\fi);
it then increments the counter before passing it to subsequent calls, and outputs the next available unused label on return (the return value $\nl'$%
\ifPreprint
\ in the example%
\fi).

Relative to this scheme, we prove that two \emph{consecutively compiled} \RISCLang programs---in the sense that the relevant outputs from the first call are fed directly into the second call---only ever jump to locations within themselves (and not in the other).

Specifically, we define two \texttt{RISC} programs $P_1, P_2$ to be $\Joinable$ if they are both:
\begin{itemize}
    \item $\JoinableFwd$: $P_1$ only ever jumps to labels that are either
        \begin{itemize}
            \item labelling an instruction in $P_1$ itself, or
            \item the label of the very first instruction in $P_2$.
        \end{itemize}
    \item $\JoinableBwd$: $P_2$ does not jump to any of the labels of instructions in $P_1$.
\end{itemize}
The lemma we prove
%(whose details are left to the Isabelle/HOL formalisation)
then says that two \texttt{RISC} programs output by consecutive invocations of the \WRCompiler are $\Joinable$.

Proving that the control flow of programs compiled by the \WRCompiler always remains self-contained in this manner facilitates reasoning about their sequential composition.
 % from thesis

\section{Register allocation scheme model}
\label{sec:reg-alloc-model}
Like \citet{Tedesco16} we generalise over the (user-supplied) register allocation scheme, and assume there are enough registers to service the maximum depth of expressions in the source program.
We leave for future work the modelling and analysis of a compiler phase that spills register contents to memory, in order to make this assumption unnecessary.

Here we model the (user-supplied) register allocation scheme with two functions $reg\_alloc$ and $reg\_alloc\_cached$ on the \emph{register record} $\Phi$
\ifPreprint
(see \autoref{sec:compiler-impl})
\fi
and the set $A$ of registers whose contents are needed to evaluate the current expression.
To avoid loading from memory unnecessarily, the compiler may first call $reg\_alloc\_cached\ \Phi\ A\ v$ to identify a register that $\Phi$ records as already containing the variable $v$.
When the compiler needs a fresh register, it will call $reg\_alloc\ \Phi\ A$.
Neither function is allowed to allocate a register in $A$, so the allocator is permitted to fail if it cannot find any suitable register.
However, registers typically become available again as expression evaluation is resolved.
 % from thesis

\section{Informal descriptions of cases of refinement relation $\RefRelWR$}
\label{sec:R_wr-informal}
\subsection{Base cases}

\begin{itemize}
    \item \texttt{stop}: This case relates a terminated \texttt{While} program with a terminated \texttt{RISC} program (i.e.~one where the program counter is at the length of the program text).

    \item \texttt{skip\_nop}: This case relates the \texttt{While} program $\Skip$ with the configuration where the program counter is at the start of the \texttt{RISC} program $[\Nop]$.

    \item \texttt{assign\_expr}: This case relates the expression evaluation part (for the expression $e$) of the \texttt{While} program $\Assign{v}{e}$ with the corresponding part of the \texttt{RISC} program obtained by compiling it with the \WRCompiler.

    \item \texttt{assign\_store}: As for \texttt{assign\_expr}, but for the very last $\Store$ instruction that commits the result of the expression evaluation back to shared memory variable $v$.

        It asserts additionally that $v$ must be stable if lock-governed, and non-lock-governed otherwise.
        This prevents threads from violating the locking discipline%
        \ifPreprint
        \ (see \autoref{sec:locking-discipline})%
        \fi.

    \item \texttt{lock\_acq}: This case relates $\LockAcq{k}$ with $\RISCLockAcq\ k$.
    \item \texttt{lock\_rel}: This case relates $\LockRel{k}$ with $\RISCLockRel\ k$.
\end{itemize}

\subsection{Inductive cases}

\begin{itemize}
    \item \texttt{seq}: This case relates the \texttt{While} program $\Seqg{c_1}{c_2}$ with the \emph{concatenation} $P_1 @ P_2$ of the \texttt{RISC} programs $P_1$ and $P_2$ that are respectively the outputs of successful consecutive compilation (see \autoref{sec:labels-seq}) of $c_1$ and $c_2$ by the \WRCompiler.
        It is intended for cases where the \texttt{While} (resp.~\texttt{RISC}) program is currently in $c_1$ (resp.~$P_1$).

        It is an inductive case of $\RefRelWR$, in that:
        \begin{itemize}
            \item $c_1$ is required to be related by $\RefRelWR$ to the present location in $P_1$.
            \item For all local configurations that obey the $\CompiledCmdConfigConsistent$ requirements,
                $c_2$ is required to be related by $\RefRelWR$ to the first instruction of $P_2$.
                This quantification ensures that $\RefRelWR$ remains closed when execution progresses from the first program to the second program.
        \end{itemize}

        It asserts that $P_1$ and $P_2$ are $\Joinable$ (\autoref{sec:labels-seq}), which is particularly relevant here to ensure that $P_1$ can only jump to locations within or at the end of itself (i.e.~the start of $P_2$).

    \item \texttt{join}: This case relates a \texttt{While} program $c$ with an offset $\pc > \length\ P_1$ into a \texttt{RISC} program $P_1 @ P_2$, assuming the inductive hypothesis that $c$ is related by $\RefRelWR$ with the offset $\pc - \length\ P_1$ into the \texttt{RISC} program $P_2$ alone.

        It is intended primarily for cases where the \texttt{While} (resp.~\texttt{RISC}) program is currently in the $c_2$ (resp.~$P_2$) of some consecutively compiled $\Seqg{c_1}{c_2}$ (resp.~$P_1$ concatenated with $P_2$) but applies more broadly to allow any prepend of dead, unreachable instructions onto the front of a \texttt{RISC} program without breaking $\RefRelWR$.

        It also asserts that $P_1$ and $P_2$ are $\Joinable$, which is important here to ensure that $P_2$ cannot jump back into $P_1$.

    \item \texttt{if\_expr}: This case relates the expression evaluation part (for the expression $e$) of the \texttt{While} program $\ITEg{e}{c_1}{c_2}$ with the corresponding part (including the conditional jump $\Jz$ at the end of expression evaluation) of the \texttt{RISC} program obtained by compiling it with the \WRCompiler.

        It relies on both $c_1$ and $c_2$ being related by $\RefRelWR$ to its compiled \texttt{RISC} counterparts when started with initialisation states judged valid by $\CompiledCmdConfigConsistent$.

        This case is depicted in full in \autoref{fig:Rwr-if-expr}, on page~\pageref{fig:Rwr-if-expr};
        for comparison, \autoref{fig:impl-if-expr} depicts the relevant part of the $\CompileCmd$ implementation.

    \item \texttt{if\_c1}: This case relates some \texttt{While} program $c_1'$ reachable from $c_1$ with the corresponding part within the $c_1$ part of the \texttt{RISC} program obtained by compiling $\ITEg{e}{c_1}{c_2}$ with the \WRCompiler.

        It relies on $c_1$ being related by $\RefRelWR$ to its compiled \texttt{RISC} counterpart at the appropriate program counter offset.

    \item \texttt{if\_c2}: As for \texttt{if\_c1}, but for $c_2$.

    \item \texttt{epilogue\_step}: This case relates a terminated \texttt{While} program to the silent control flow steps navigating to the end of a \texttt{RISC} program from the end of the ``then'' and ``else'' branches of a compiled if-conditional.

        It works only for the ``epilogue'' step forms: $\Jmp$ and $\Nop$%
        \ifPreprint
        \ (see \autoref{sec:abs-steps_wr})%
        \fi.

        It is inductive in that it asserts closedness of $\RefRelWR$ over pairwise reachability from the pair currently under consideration---the only case to do so directly.

    \item \texttt{while\_expr}: This case relates the \texttt{While} program ($\Whileg{e}{c}$)'s initial intermediate step to $\ITEg{e}{(\Seqg{c}{\ \Whileg{e}{c}})}{\Stop}$, and its expression evaluation part, with the expression evaluation and conditional jump of the \texttt{RISC} program that $\Whileg{e}{c}$ was compiled to by $\CompileCmd$.

        It relies on $c$ being related by $\RefRelWR$ to its compiled \texttt{RISC} counterpart when started with initialisation states judged valid by $\CompiledCmdConfigConsistent$.

    \item \texttt{while\_inner}: This case relates some program $\Seqg{c_I}{\Whileg{e}{c}}$ reachable from $\Seqg{c}{\Whileg{e}{c}}$ to the loop body part of the \texttt{RISC} program compiled from $\Whileg{e}{c}$.

        It relies on $c_I$ being related by $\RefRelWR$ to its compiled \texttt{RISC} counterpart at the appropriate program counter offset.

        It also carries around the same reliance on $c$ being related by $\RefRelWR$ to its compiled \texttt{RISC} counterpart for all initialisation states judged valid by $\CompiledCmdConfigConsistent$.

    \item \texttt{while\_loop}: This case handles epilogue steps for the inner loop body program, and the final jump back to the beginning of the While-loop.

        It requires $\RefRelWR$ to relate the terminated \texttt{While} program to the end of the compiled loop body, and furthermore also carries around the same reliance on $c$ being related by $\RefRelWR$ to its compiled \texttt{RISC} counterpart for all initialisation states judged valid by $\CompiledCmdConfigConsistent$.

\end{itemize}
 % from thesis

\begin{figure}[h]
%\begin{definition} [Introduction rule for case \texttt{if\_expr} of $\RefRelWR$]
\[
{\small
\mprset{vskip=0.5ex}
\inferrule{
    c = \ITEg{e}{c_1}{c_2} \and
    \CompilerInputReqs\ C\ l\ \nl\ c \and \\
    (\PCs, \unused{l'}, \nl_2, \unused{C'}, \False) = \CompileCmd\ C\ l\ \nl\ c \and
    \qquad(P_e, \unused{r}, C_1, \False) = \CompileExpr\ C\ \varnothing\ l\ e \and \\
    % Some manual spacing to get compile-cmd assumptions to line up nicely
    (P_1, \unused{l_1}, \nl_1, \unused{C_2}, \False) = \CompileCmd\ C_1\ \None\ (\Suc\ (\Suc\ \nl))\ c_1 \and
        \quad \pc \leq \length\ P_e\ \ \ \ \ \  \and \\
    \,\, (P_2, \unused{l_2}, \nl_2, \unused{C_3}, \False) = \CompileCmd\ C_1\ (\Some\ \nl)\ \nl_1\ c_2 \and
        \quad C_{\pc} = \listderef{(\mapsnd\ \PCs)}{\pc} \and \\
    \CompiledCmdConfigConsistent\ C_{\pc}\ \regs\ \mds\ \mem \and
    \RegrecStable\ C_{\pc} \and \\
    \forall \mds'\ \mem'\ \regs'. \ 
    \CompiledCmdConfigConsistent\ C_1\ \regs'\ \mds'\ \mem'\, \land \,
    \RegrecStable\ C_1 \\
    \quad\  \longrightarrow ((\LocalConfWhile{c_1}{\mds'}{\mem'}, \LocalConfRISC{0}{\mapfst\ P_1}{\regs'}{\mds'}{\mem'}) \in \RefRelWR\ \land \\
    \qquad \quad (\LocalConfWhile{c_2}{\mds'}{\mem'}, \LocalConfRISC{0}{\mapfst\ P_2}{\regs'}{\mds'}{\mem'}) \in \RefRelWR) 
} {
    (\LocalConfWhile{c}{\mds}{\mem}, \LocalConfRISC{\pc}{\mapfst\ \PCs}{\regs}{\mds}{\mem}) \in \RefRelWR
} }
\]
    \newcommand{\figRefRelIfExpr}{Introduction rule for case \texttt{if\_expr} of refinement relation $\RefRelWR$}
    \caption[\figRefRelIfExpr]{\figRefRelIfExpr. \\
This case pertains to the expression-evaluation part of an
\IfKw-conditional compiled by $\CompileCmd$ (see \autoref{fig:impl-if-expr}).
Variables ignored are in gray.}
    \label{fig:Rwr-if-expr}
\end{figure}

\lstdefinelanguage{HackyIsabelle}{
    morekeywords={let, in, if, then, else},
    morecomment=[s]{(*}{*)},
    basicstyle=\normalfont\small, % No bold.
    identifierstyle=\ttfamily,
    %keywordstyle=\color{black}\normalfont\bfseries,
    keywordstyle=\color{black}\bfseries,
}
\begin{figure}[h]
% https://tex.stackexchange.com/a/33021
\lstset{mathescape}
\begin{lstlisting} [language = HackyIsabelle, columns=fullflexible]
compile_cmd C l nl (If e c$_1$ c$_2$) =
  (let (P$_e$, r, C$_1$, fail$_e$) = (compile_expr C {} l e);
      (br, nl') = (nl, Suc nl); (ex, nl'') = (nl', Suc nl');
      (P$_1$, l$_1$, nl$_1$, C$_2$, fail$_1$) = (compile_cmd C$_1$ None nl'' c$_1$);
      (P$_2$, l$_2$, nl$_2$, C$_3$, fail$_2$) = (compile_cmd C$_1$ (Some br) nl$_1$ c$_2$);
      (* Pre-compilation check ensures asmrec C$_2$ = asmrec C$_3$ *)
      C' = (regrec C$_2$ $\sqcap_R$ regrec C$_3$, asmrec C$_2$)
   in (P$_e$ @ [((if P$_e$ = [] then l else None, Jz br r), C$_1$)] @
      P$_1$ @ [((l$_1$, Jmp ex), C$_2$)] @ P$_2$ @ [((l$_2$, Nop'), C$_3$)],
      Some ex, nl$_2$, C', fail$_e$ $\lor$ fail$_1$ $\lor$ fail$_2$))
\end{lstlisting}
\newcommand{\figImplIfExpr}{Excerpt of \WRCompiler implementation: case for $\IfKw$-conditionals}
\caption[\figImplIfExpr]{\figImplIfExpr. \\
    This case of the Isabelle/HOL function $\CompileCmd$ compiles the \WhileLang command $\ITEg{e}{c_1}{c_2}$.
    Here, $@$ denotes concatenation between two \RISCLang program texts,
and $\Phi \sqcap_R \Phi'$ denotes the subset of mappings on which the register records $\Phi$ and $\Phi'$ agree.}
\label{fig:impl-if-expr}
\end{figure}

\fi % ifPreprint

\phantomsection
\label{lastpage01}

\end{document}